\documentclass[transmag]{IEEEtran}
\usepackage{latexsym}
\usepackage{graphicx}
\usepackage{amsfonts,amssymb,amsmath}
\usepackage{hyperref}
\usepackage{cite}
\usepackage{amsmath,amssymb,amsfonts}
\usepackage{algorithmic}
\usepackage{graphicx}
\usepackage{textcomp}
\usepackage{xcolor}

\DeclareMathOperator*{\argmin}{arg\,min}
\usepackage{mathtools}

\usepackage{steinmetz}
\usepackage{authblk}
\newcommand{\minus}{\scalebox{0.75}[1.0]{$-$}}
\usepackage{array}

\DeclareMathOperator{\tr}{tr}
\usepackage[ruled,norelsize]{algorithm2e}
\makeatletter
\newcommand{\removelatexerror}{\let\@latex@error\@gobble}
\makeatother

\usepackage{caption}
\usepackage{subcaption}
\usepackage{blindtext}
\usepackage{threeparttable}
\usepackage{array}

\def\BibTeX{{\rm B\kern-.05em{\sc i\kern-.025em b}\kern-.08em T\kern-.1667em\lower.7ex\hbox{E}\kern-.125emX}}
\begin{document}

\title{Energy Efficient Dual-Functional Radar-Communication: Rate-Splitting Multiple Access, Low-Resolution DACs, \\and RF Chain Selection}

\author{Onur Dizdar, \IEEEmembership{Member,~IEEE,} Aryan Kaushik, \IEEEmembership{Member,~IEEE,}
Bruno Clerckx, \IEEEmembership{Fellow,~IEEE,} \\ and Christos Masouros, \IEEEmembership{Senior Member,~IEEE}
\thanks{This work was supported by the UK Engineering and Physical Sciences Research Council (EPSRC) Grant numbers EP/S026622/1 and EP/S026657/1, and the UK MOD University Defence Research Collaboration (UDRC) in Signal Processing.}
\thanks{Onur Dizdar and Bruno Clerckx are with the Department of Electrical and Electronic Engineering, Imperial College London, U.K. (e-mail: \{o.dizdar, b.clerckx\}@imperial.ac.uk).}
\thanks{Aryan Kaushik is with the School of Engineering and Informatics, University of Sussex, U.K. (e-mail: aryan.kaushik@sussex.ac.uk).} 
\thanks{Christos Masouros is with the Department of Electronic and Electrical Engineering, University College London, U.K. (e-mail: c.masouros@ucl.ac.uk).}  
}

\IEEEtitleabstractindextext{\begin{abstract}Dual-Functional Radar-Communication systems enhance the benefits of communications and radar sensing by jointly implementing these on the same hardware platform and using the common RF resources. An important and latest concern to be addressed in designing such systems is maximizing the energy-efficiency. In this paper, we consider a Dual-Functional Radar-Communication system performing simultaneous multi-user communications and radar sensing, and investigate the energy-efficiency behaviour with respect to active transmission elements. Specifically, we formulate a problem to find the optimal precoders and the number of active RF chains for maximum energy-efficiency by taking into consideration the power consumption of low-resolution Digital-to-Analog Converters on each RF chain under communications and radar performance constraints.  
We consider Rate-Splitting Multiple Access to perform multi-user communications with perfect and imperfect Channel State Information at Transmitter. 
The formulated non-convex optimization problem is solved by means of a novel algorithm. 
We demonstrate by numerical results that Rate Splitting Multiple Access achieves an improved energy-efficiency by employing a smaller number of RF chains compared to Space Division Multiple Access, owing to its generalized structure and improved interference management capabilities.\end{abstract}

}

\maketitle

\section{INTRODUCTION}

\IEEEPARstart{T}{he} increasing user demand for cellular services and wireless communication systems has congested the existing sub 6-GHz frequency spectrum. In Cisco’s annual report, it is expected that there will be over 70\% mobile users of the world population by 2023 with 5.3 billion internet users \cite{cisco}. As per the Ericsson mobility report \cite{ericsson}, video traffic will increase to 77\% in 2026 and the number of IMS voice subscriptions will reach to 6.8 billion by 2026. This leads the demand to seek for either widely unused spectrum such as millimeter wave for the fifth generation (5G) wireless systems \cite{andrewsJSAC2014} or reusing/sharing of the spectrum used for other applications and systems such as radar technology. For instance, the sub 6-GHz spectrum allocated to radar systems can be made available for sharing between radar and wireless systems \cite{darpa2016, griffithsPROC2015}. 

The increased demand for connectivity, data and spectrum also lead to an increased demand for energy required to enable such massive networks. 
An approach to tackle the energy demand problem is to prioritize and investigate energy-efficient transmission and networking algorithms in next generation communications standards \cite{zappone_2017, masoudi_2019}. Achieving high energy-efficiency (EE) in communications is significantly important for operators as well as end-users. 
Consequently, EE has become a highly important metric in transceiver and network architecture design for the next generation standards \cite{aryanTGCN2020, vlachosVTC2018, akGCOM2019}.

Integrated Sensing and Communication (ISAC) systems have been studied in recent literature to counter the issue of spectrum congestion problem and improve EE by having a single device transmitting instead of two devices. It was demonstrated that such joint systems have advantages over the individually operating radar and communication systems \cite{hassanienSPM2019, blissACCESS2017, barneto2021, zhang_2021, liu_2021}. Integration of communications and radar functionalities is important in commercial systems as well as military systems, with applications in areas such as automotive radars and vehicular communications \cite{ozkaptan2020, maSP2020}. There are two types of ISAC systems: the radar-communication co-existence (RCC) requires effective coordination between the communication and radar units \cite{liuTWC2018}, whereas the Dual Functional Radar-Communication (DFRC) systems share the same hardware and signal for conducting both the operations \cite{maSP2020, akJCNS2022, chenCL2021}.   

Managing the interference in DFRC systems is important for an improved system performance, which may otherwise suffer from performance degradation in terms of both radar and communications functionalities. 
The most common type of interference considered in DFRC system design is the interference between the communications and radar signals, which occurs as a result of combining two separate systems with different operational purposes. On the other hand, multi-user interference can also occur for communication users when the multiple antennas at transmitter are exploited for improved spatial-multiplexing gain by means of multi-user transmission without perfect Channel State Information at Transmitter (CSIT). 
Consequently, a communication user in a DFRC system is susceptible to interference from both radar signals and communications signals for other users.


In this work, we consider a multi-antenna DFRC system which performs simultaneous multi-user communications and radar sensing with low-resolution Digital-to-Analog Converters (DACs). 
Our aim is to perform energy-efficient joint communications and sensing by means of low-resolution DACs and only activating the optimal number of RF chains at each transmission.
We consider Rate-Splitting Multiple Access (RSMA) for multi-user communications in our proposed DFRC system. RSMA is a multiple access scheme based on the concept of Rate-Splitting (RS) and linear precoding for multi-antenna multi-user communications. RSMA splits user messages into common and private parts, and encodes the common parts into one or several common streams while encoding the private parts into separate streams. The streams are precoded using the available (perfect or imperfect) Channel State Information at the Transmitter (CSIT), superposed and transmitted via the Multi-Input Multi-Output (MIMO) or Multi-Input Single-Output (MISO) channel \cite{clerckxCM2016}. 

RSMA manages multi-user interference by allowing the interference to be partially decoded and partially treated as noise at the receivers. RSMA has been shown to encompass and outperform existing multiple access schemes, i.e., Space Division Multiple Access (SDMA), Non-Orthogonal Multiple Access (NOMA), Orthogonal Multiple Access (OMA) and multicasting. RSMA has been shown to be flexible and robust by its capability to adapt to any interference level and surpass the performance of SDMA and NOMA under perfect and imperfect Channel State Information at Transmitter (CSIT)
\cite{clerckxCM2016,clerckx2018,clerckxWCL2020,clerckxTC2016,maoTCOM2019}. This implies that RSMA can play a big role in emerging ISAC in a congested environment subject to not only interference among communications users but also among radar and communications.

\subsection{Literature Review}
The sharing of RF spectrum and hardware resources by multiple systems has increased the importance of interference management among the communication users and radar targets in an energy-efficient manner. 
The works on precoder design with interference management for co-existing and co-located DFRC systems, such as \cite{liuTWC2018, cuiSPAWC2018, zhao_2021}, consider design metrics in terms of communications rate, Signal-to-Noise Power Ratio (SNR) or multi-user interference, without any discussions on hardware power consumption or EE. 

For energy-efficient MIMO DFRC systems, one needs to take into account the hardware power consumption and reducing the RF transceiver components such as RF chains and associated DAC bits, which are power hungry components, in an optimal way.
Hybrid analog-digital (HAD) beamforming is proposed for DFRC sytems in \cite{liu_2020} to reduce the number of RF chains. HAD beamforming reduces the hardware costs and improves EE by creating a trade-off with the system performance. 
It has been shown in \cite{aryanTGCN2019, aryanComNet2020, vlachosRS2020} that optimally selecting the active RF chains for HAD beamforming in communications only systems can further improve EE while maintaining good spectral efficiency performance. Beside RF chain selection, taking into account the DAC-bit optimization is also beneficial for EE maximization \cite{kaushikICC2019}. For DFRC systems, it has been shown in \cite{aryanICC2021} that RF chain selection can provide advantages in terms of the hardware complexity with favourable radar beampattern performance. However, the investigated scenarios in \cite{aryanICC2021} consider full-bit resolution sampling and does not tackle the interference management problem directly. Furthermore, \cite{akICC2022} addresses the EE maximization in DFRC systems while considering low DAC-bit resolution sampling, however the issue of interference management is missing. 



Recent works \cite{xuICCW2020, xu2021} have tackled the interference management problem in MIMO DFRC systems by employing RSMA, however, without any considerations for EE or interference due to imperfect CSIT. In \cite{loli2021, loli2022}, the authors deal with the sum-rate (SR) maximization of a DFRC system employing RSMA under imperfect CSIT, without considering the EE performance, low-resolution DACs or the problem of RF chain selection.
In \cite{dizdarICC2021}, the authors investigate a DFRC system with low-resolution DACs under a total transmit power constraint and compare the performance of RSMA with that of SDMA, again with perfect CSIT assumption. The results in \cite{dizdarICC2021} demonstrate the advantage of RSMA in systems considering power consumption by DACs, which provides a motivation to study RSMA in more advanced setups with RF chain selection, which can further reduce power consumption and increase EE.


\subsection{Contributions}
The main contributions of this work can be listed as follows:
\begin{itemize}
	\item We propose a DFRC system model to perform multi-user communications and radar sensing simultaneously. The proposed system model considers the practical impairments resulting from low-resolution DACs and imperfect CSIT. We consider RSMA as the enabling technology in the proposed system. Our aim is to find the optimal precoders, message split into common and private streams, and active RF chains to operate with maximum EE. To the best of our knowledge, this is the first paper to consider RF chain selection for EE in a DFRC system with low-resolution DACs and RSMA.
	\item We formulate a non-convex optimization problem to find the optimal precoders and active RF chains. The formulated problem is an EE maximization problem under total power, constant modulus, communications performance and radar performance constraints. The communications performance is guaranteed by a Quality-of-Service (QoS) constraint, which imposes a minimum SR value. The radar performance is guaranteed by limiting the norm of the covariance matrix of the designed waveform to a reference matrix, which can be chosen to optimize performance for different radar functionalities, such as target detection and parameter estimation. We propose alternative reference matrices to account for different radar operation modes, namely, detection and tracking modes.
	\item We propose a novel Alternating Optimization (AO) based algorithm to solve the formulated non-convex problem by iterating between subproblems which find the optimal precoders and active RF chains separately. The subproblem to find the optimal precoders is solved by an algorithm based on Alternating Direction Method of Multipliers (ADMM) method. The subproblem to determine the active RF chains is solved by an algorithm based on Successive Convex Approximation (SCA) and Semi Definite Relaxation (SDR) methods. We prove the convergence of the overall proposed algorithm by analytical means and provide an algorithmic complexity analysis. 
	We perform simulations to demonstrate the effects of number of active RF chains and quantization bits on EE, target detection, target parameter estimation, and SR performance explicitly. We show that RSMA can operate with a higher EE while achieving similar communications and radar performance, due to its improved interference management capability and higher degrees-of-freedom in the precoder design space.
\end{itemize}

\textit{Notation:} Vectors are denoted by bold lowercase letters and matrices are denoted by bold uppercase letters. The operations $|.|$ and $||.||$ represent the absolute value of a scalar and $l_{2}$-norm of a vector, respectively. The notation $\mathbf{a}^{H}$ denotes the Hermitian transpose of a vector $\mathbf{a}$. $\mathcal{CN}(0,\sigma^{2})$ represents the Circularly Symmetric Complex Gaussian distribution with zero mean and variance $\sigma^{2}$. $\mathbf{I}_{n}$ represents the $n$-by-$n$ identity matrix. $\mathbf{0}_{n}$ represents the all-zero vector of dimensions $n$-by-$1$. $\mathfrak{R}(a)$ and  $\mathfrak{I}(a)$ represent the real and imaginary parts of the complex number $a$, respectively. The operator $\mathrm{Diag}(\mathbf{X}_{1}, \ldots, \mathbf{X}_{K})$ builds a matrix $\mathbf{X}$ by placing the matrices $\mathbf{X}_{1}$, $\ldots$, $\mathbf{X}_{K}$ diagonally and setting all other elements to zero. 
The operator $\mathrm{diag}(\mathbf{X})$ builds a vector $\mathbf{x}$ from the diagonal elements of $\mathbf{X}$. The operator $\mathrm{vec}(\mathbf{X})$ vectorizes the matrix $\mathbf{X}$ into a column vector by concatenating its columns. 


The organization of the paper is as follows: Section~\ref{sec:system} presents the system model. Section~\ref{sec:problem} gives the problem formulation and the proposed solution algorithm. 
Section~\ref{sec:simulation} gives the numerical results. Section~\ref{sec:conclusion} concludes the paper.

\section{System Model}
\label{sec:system}
\begin{figure}[t!]
	\centerline{\includegraphics[width=3.4in,height=3.4in,keepaspectratio]{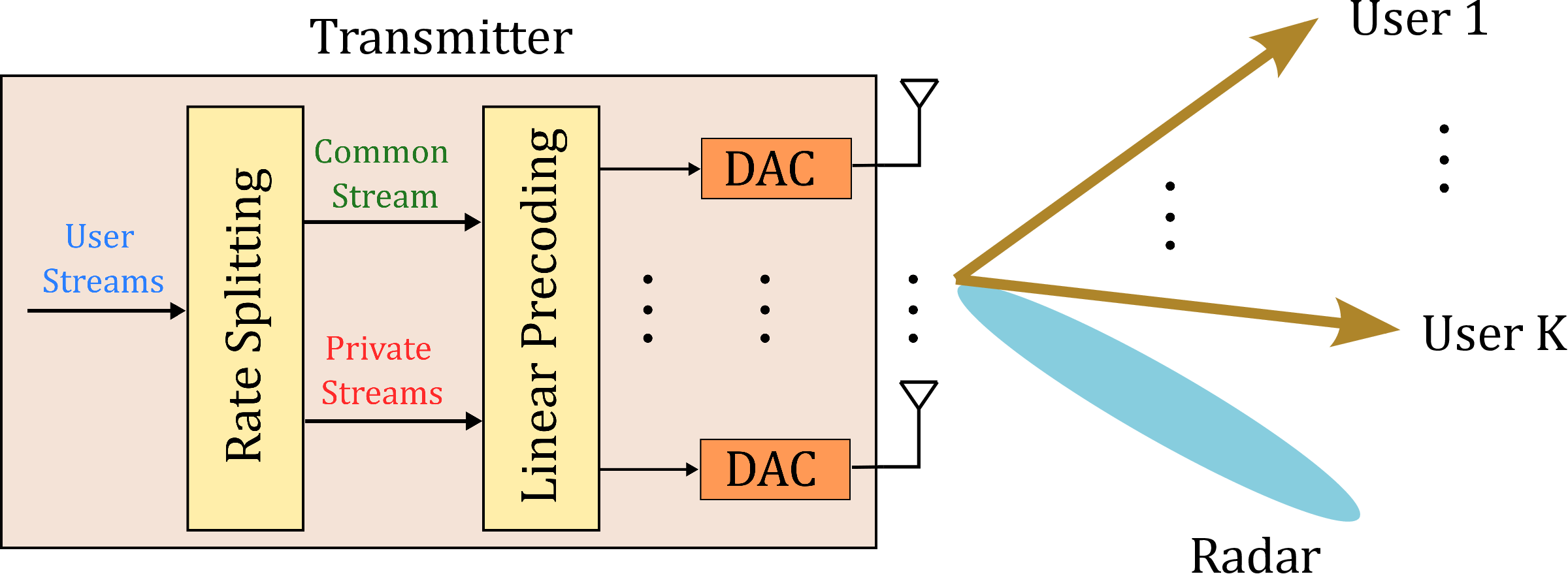}}
	\caption{System model.}
	\label{fig:system}
\end{figure}

Fig.~\ref{fig:system} shows the system model. We consider a DFRC system consisting of one transmitter with $N_{t}$ transmit antennas. The transmitter employs 1-layer RSMA \cite{clerckx2018} to serve $K$ single-antenna users indexed by $\mathcal{K}=\left\lbrace1,\ldots ,K \right\rbrace $ and perform sensing operation simultaneously. The RSMA transmit signal consists of $L$ consecutive samples in each Pulse Repetition Interval (PRI). The digitally formed transmit signal is converted to an
analog signal by means of separate DACs connected to each of the transmit antennas. 

RSMA is a multiple access technique that relies on linearly precoding at the transmitter and Successive Interference Cancellation (SIC) at the receivers. RSMA splits the user messages into common and private parts, encodes the common parts of the user messages into a common stream, encodes the private parts of the user messages into private streams and superposes them in a non-orthogonal manner.

Let us denote the messages intended for the communication users as $W_{k}$, $\forall k\in \mathcal{K}$. The transmitter splits each user message into common and private parts, i.e., $W_{c,k}$ and $W_{p,k}$. The common parts of the messages, $W_{c,k}$, are combined into the common message $W_{c}$. The common message $W_{c}$ and the private messages $W_{p,k}$ are independently encoded into streams $s_{c}$ and $s_{k}$, $\forall k\in \mathcal{K}$, respectively. Linear precoding is applied to all streams.
The transmit signal is expressed as
\begin{align}
	\mathbf{x}=\mathbf{p}_{c}s_{c}+\sum_{k=1}^{K}\mathbf{p}_{k}s_{k},
	\label{eqn:transmit_signal}	
\end{align}
where $\mathbf{p}_{c}$, $\mathbf{p}_{k} \in\mathbb{C}^{n_t}$ are the precoders for the common stream and the private stream of user-$k$, respectively. 

We assume that the streams have unit power, so that \mbox{$\mathbb{E}\left\lbrace \mathbf{s}\mathbf{s}^{H}\right\rbrace =\mathbf{I}$}, for \mbox{$\mathbf{s}=[s_{c}, s_{1}, \ldots, s_{K}]$}. We assume that $s_{c}$ and $s_{k}$, $\forall k\in\mathcal{K}$, are chosen independently from a Gaussian alphabet for theoretical analysis. The signal received by user-$k$ is written as
\begin{align}
	\mathbf{y}_{k}&=\mathbf{h}_{k}^{H}\mathbf{x}+z_{k}, \quad \forall k \in \mathcal{K}, 
	\label{eqn:receive_signal}	
\end{align}
where \mbox{$\mathbf{h}_{k} \in \mathbb{C}^{n_{t}}$} is the channel vector and \mbox{$z_{k} \sim \mathcal{CN}(0,\sigma_{n}^{2})$} is the Additive White Gaussian Noise (AWGN) term for user-$k$. 

The detection of the messages is carried out using SIC algorithm. The common stream is detected first to obtain the common message estimate $\widehat{W}_{c}$ by treating the private streams as noise. The common stream is then reconstructed using $\widehat{W}_{c}$ and subtracted from the received signal. The remaining signal is used to detect the private messages $\widehat{W}_{p,k}$. Finally, the estimated message for user-$k$, $\widehat{W}_{k}$, is obtained by combining $\widehat{W}_{c,k}$ and $\widehat{W}_{p,k}$. 
We note here that SDMA or conventional multiuser linear precoding is a subscheme of \eqref{eqn:transmit_signal} that is obtained when no power is allocated to the common stream $s_{c}$ and $W_{k}$ is directly encoded into $s_{k}$. Hence, the sequel also holds for SDMA by simply turning off the common stream.

In this work, we consider the practical case of imperfect CSIT, in which the transmitter uses the imperfect channel estimates for precoder calculation. Let $\widehat{\mathbf{h}}_{k}$ denote the CSIT for user-$k$. The instantaneous channel is written in terms of CSIT as
\begin{align}
	\mathbf{h}_{k}=\sqrt{1\minus\sigma_{ce}^{2}}\widehat{\mathbf{h}}_{k}+\sigma_{ce}\widetilde{\mathbf{h}}_{k}, 
	\label{eqn:error_model}
\end{align}
where $\widetilde{\mathbf{h}}_{k,n}$ is the channel estimation error and $0 \leq \sigma^{2}_{ce} \leq 1$ is the channel estimation error variance. The entries of $\widehat{\mathbf{h}}_{k,n}$ and $\widetilde{\mathbf{h}}_{k,n}$ are independent. 

In the following subsections, we present details on the quantization model, the quantized RSMA signal, and the performance metrics considered in the problem formulation.  
\subsection{Quantization Model}
We consider a linear model approximation to express the signal quantized by the DACs \cite{mo_2018, vlachosVTC2018, kaushikICC2019}. We define  uniform scalar quantizer function $Q(x)$ for an RF chain as
\begin{align}
	Q(u)\approx \delta u+\epsilon,
\end{align}
where the parameter $\delta$ represents the quantization resolution of $b$ bits and is expressed in terms of $b$ as
\begin{align}
	\delta=\sqrt{1-\frac{\pi\sqrt{3}}{2}2^{-2b}}.
	\label{eqn:delta}
\end{align} 
The quantization noise \mbox{$\epsilon \sim \mathcal{CN}(0,\sigma_{e}^{2})$} and the input signal $u$ are uncorrelated \cite{mo_2018}. The quantization noise variance is expressed as $\sigma^{2}_{e}=\delta^{2}(1-\delta^{2})^{2}$. Accordingly, the quantized transmitted signal is written as
\begin{align}
	Q(\mathbf{x})\approx \boldsymbol{\Delta}\mathbf{x}+\boldsymbol{\epsilon},
	\label{eqn:quantization}
\end{align}
where \mbox{$\boldsymbol{\Delta}=\mathrm{Diag}(\delta_{1}, \delta_{2}, \ldots, \delta_{N_{t}})$},   \mbox{$\boldsymbol{\epsilon} \sim \mathcal{CN}(\mathbf{0}_{N_{t}},\boldsymbol{\Sigma})$} with \mbox{$\boldsymbol{\Sigma}=\mathrm{Diag}(\sigma^{2}_{e,1}, \sigma^{2}_{e,2}, \ldots, \sigma^{2}_{e,N_{t}})$}. The power consumption of each active DAC is proportional to the number of quantization bits with the power consumption model expressed as
\begin{align}
	P(\delta)=\left[ P_\textrm{DAC}\sqrt{\frac{\pi\sqrt{3}}{2(1-\delta^{2})}} \right], 
	\label{eqn:power}
\end{align}
with $P_\textrm{DAC}$ being the power consumption coefficient. 
\subsection{RSMA with DACs and RF Chain Selection for DFRC}
\label{sec:rsma_syst}
We define the RF chain selection indicator, $\lambda_{i} \in \left\lbrace 0,1 \right\rbrace$, to represent the status of the $i$-th RF-chain, such that, $\lambda_{i}=1$ represents that the $i$-th RF-chain is active. We further define the indicator vector and matrix as $\boldsymbol{\lambda}=[ \lambda_{1}, \lambda_{2}, \ldots, \lambda_{N_{t}}]$ and $\boldsymbol{\Lambda}=\mathrm{Diag}( \lambda_{1}, \lambda_{2}, \ldots, \lambda_{N_{t}})$, respectively.

We consider a transmission model where $L$ consecutive symbols are transmitted, with the symbol indexes chosen from the set $\mathcal{L}=\left\lbrace 1,2,\ldots,L\right\rbrace$. We can express the $l$-th transmitted symbol for RSMA with RF chain selection under the quantization effects of the DACs from \eqref{eqn:transmit_signal} and \eqref{eqn:quantization} as  
\begin{align}
	\tilde{\mathbf{x}}_{l}&=Q(\boldsymbol{\Lambda}\mathbf{x}_{l})=\boldsymbol{\Delta}(\boldsymbol{\Lambda}\mathbf{p}_{c,l}s_{c,l}\hspace{-0.1cm}+\hspace{-0.1cm}\sum_{k=1}^{K}\boldsymbol{\Lambda}\mathbf{p}_{k,l}s_{k,l})+\boldsymbol{\Lambda}\boldsymbol{\epsilon}_{l} \nonumber \\
	&=\boldsymbol{\Delta}\boldsymbol{\Lambda}\mathbf{P}_{l}\mathbf{s}_{l}+\boldsymbol{\Lambda}\boldsymbol{\epsilon}_{l}, \quad \forall l \in \mathcal{L}. 
	\label{eqn:transmit_signal_quant}
\end{align}
Using \eqref{eqn:receive_signal} and \eqref{eqn:transmit_signal_quant}, the $l$-th received RSMA symbol at user-$k$ under DAC quantization error is written as
\begin{align}
	\tilde{\mathbf{y}}^{c}_{k,l}=\mathbf{h}_{k}^{H}\tilde{\mathbf{x}_{l}}+z_{k,l}=\mathbf{h}_{k}^{H}\boldsymbol{\Delta}\boldsymbol{\Lambda}\mathbf{P}_{l}\mathbf{s}_{l}+\underbrace{\mathbf{h}_{k}^{H}\boldsymbol{\Lambda}\boldsymbol{\epsilon}_{l}+z_{k,l}}_{\eta_{k,l}},
	\label{eqn:receive_signal_quant}	
\end{align}
where the term $\eta_{k,l}$ is treated as if it has the worst-case Gaussian
distribution \cite{mo_2018}, such that \mbox{$\eta_{k,l} \sim \mathcal{CN}(0,\sigma^{2}_{\eta,k})$} with $\sigma^{2}_{\eta,k}=\mathbf{h}_{k}^{H}\boldsymbol{\Lambda}\boldsymbol{\Sigma}\boldsymbol{\Lambda}\mathbf{h}_{k}+\sigma_{n}^{2}$. The channel $\mathbf{h}_{k}$, $k \in \mathcal{K}$, is assumed to be static for $L$ consecutive symbols.
As it can be observed from the expression \eqref{eqn:receive_signal_quant}, the effects of DAC quantization is reflected by a multiplicative factor $\boldsymbol{\Delta}$ on the precoders and an increased noise variance.

We express the received Signal-to-Interference-Plus-Noise Ratios (SNRs) for the common and private streams at the $l$-th symbol and user-$k$ as 
\begin{subequations}
	\begin{alignat}{3}
	\gamma_{c,k,l}(\mathbf{P}_{l}, \boldsymbol{\Lambda}, \mathbf{B})&\hspace{-0.1cm}=\hspace{-0.1cm}\frac{|\mathbf{h}_{k}^{H}\boldsymbol{\Delta}\boldsymbol{\Lambda}\mathbf{p}_{c,l}|^{2}}{\sigma_{n}^{2}\hspace{-0.1cm}+\hspace{-0.1cm}\mathbf{h}_{k}^{H}\boldsymbol{\Lambda}\boldsymbol{\Sigma}\boldsymbol{\Lambda}\mathbf{h}_{k}\hspace{-0.1cm}+\hspace{-0.1cm}\sum_{k^{\prime}\in\mathcal{K}}|\mathbf{h}_{k}^{H}\boldsymbol{\Delta}\boldsymbol{\Lambda}\mathbf{p}_{k^{\prime},l}|^{2}},   \label{eqn:sinr_1} \\
	\gamma_{k,l}(\mathbf{P}_{l}, \boldsymbol{\Lambda}, \mathbf{B})&\hspace{-0.1cm}=\hspace{-0.1cm}\frac{|\mathbf{h}_{k}^{H}\boldsymbol{\Delta}\boldsymbol{\Lambda}\mathbf{p}_{k,l}|^{2}}{\sigma_{n}^{2}\hspace{-0.1cm}+\hspace{-0.1cm}\mathbf{h}_{k}^{H}\boldsymbol{\Lambda}\boldsymbol{\Sigma}\boldsymbol{\Lambda}\mathbf{h}_{k}\hspace{-0.1cm}+\hspace{-0.1cm}\sum_{\substack{k^{\prime}\in\mathcal{K}, \\ k^{\prime} \neq k}}|\mathbf{h}_{k}^{H}\boldsymbol{\Delta}\boldsymbol{\Lambda}\mathbf{p}_{k^{\prime},l}|^{2}},  \label{eqn:sinr_2} 
	\end{alignat}
\end{subequations}
where the matrix $\mathbf{B}=\mathrm{Diag}(b_{1}, b_{2}, \ldots, b_{N_{t}})$ is used to represent the dependency on the number of quantization bits at each RF chain. 
The rates of the common and private streams are written in terms of the SINR expressions as given below.
\begin{subequations}
	\begin{alignat}{3}
	R_{c,k,l}(\mathbf{P}_{l}, \boldsymbol{\Lambda}, \mathbf{B})&=\log\left(1+\gamma_{c,k,l}(\mathbf{P}_{l}, \boldsymbol{\Lambda},\mathbf{B})\right), \label{eqn:common_rate} \\
	R_{k,l}(\mathbf{P}_{l}, \boldsymbol{\Lambda}, \mathbf{B})&=\log\left(1+\gamma_{k,l}(\mathbf{P}_{l}, \boldsymbol{\Lambda}, \mathbf{B})\right). \label{eqn:private_rate}
	\end{alignat}
\end{subequations}

According to the RSMA framework, the common stream should be decodable by all receivers. Consequently, the rate of the common stream, denoted by $C_{l}(\mathbf{P}_{l}, \boldsymbol{\Lambda}, \mathbf{B})$, should be at most the minimum of the $R_{c,k,l}(\mathbf{P}_{l}, \boldsymbol{\Lambda}, \mathbf{B})$, $\forall k \in \mathcal{K}$, {\sl i.e.,}
\begin{align}
	C_{l}(\mathbf{P}_{l}, \boldsymbol{\Lambda}, \mathbf{B})=\min_{k \in \mathcal{K}}R_{c,k,l}(\mathbf{P}_{l}, \boldsymbol{\Lambda}, \mathbf{B}).
	\label{eqn:common_rate_constr}
\end{align}

For radar processing, the $l$-th received quantized RSMA symbol is expressed as  
\begin{align}
	\tilde{\mathbf{y}}^{r}_{l}&=\alpha_{r} A(\theta)\tilde{\mathbf{x}}_{l}+\mathbf{n}_{l} \nonumber \\
	&=\alpha_{r} A(\theta)\boldsymbol{\Delta}\boldsymbol{\Lambda}\mathbf{P}_{l}\mathbf{s}_{l}+\alpha_{r} A(\theta)\boldsymbol{\Lambda}\boldsymbol{\epsilon}_{l}+\mathbf{n}_{l},  
	\label{eqn:radarrx_quant}
\end{align}
where $\alpha_{r}$ is the complex path loss including
the propagation loss and the coefficient of reflection, $A(\theta)=\mathbf{a}(\theta)\mathbf{a}^{T}(\theta)$ is the transmit-receive steering matrix, and $\mathbf{n}_{l}\sim\mathcal{CN}(\mathbf{0}, \sigma_{n}^{2}\mathbf{I}_{N_{t}})$ is the additive Gaussian noise \cite{khawar_2015}. The steering vector is defined as $\mathbf{a}(\theta)=[1,e^{j2\pi sin(\theta)d}, \ldots,e^{j2\pi (N_{t}-1)sin(\theta)d} ]^{T}$, where $d$ the normalized distance between adjacent array elements with respect to wavelength. 

The transmit power is subject to the uniform elemental power constraint, which ensures that the power amplifiers driving the antenna elements are operated at full power \cite{fuhrmann_2008}. Accordingly, we write
\begin{align}
	\mathrm{tr}(\mathbf{E}_{N_{t},i}(\boldsymbol{\Delta}\mathbf{P}_{l}\mathbf{P}_{l}^{H}\boldsymbol{\Delta}+\boldsymbol{\Sigma}))=P_{ant},  
	\label{eqn:const_pant}
\end{align}
$\forall i \in \left\lbrace 1,2,\ldots, N_{t} \right\rbrace$ and $\forall  l \in \mathcal{L}$, where $P_{ant}$ is the signal transmit power at each transmit antenna and $\mathbf{E}_{N_{t},i}$ is a $N_{t}$-by-$N_{t}$ diagonal matrix with the $i$-th diagonal element as $1$ and the rest as $0$.

\textit{Remark:} One can note from \eqref{eqn:const_pant} that the uniform elemental power constraint is considered for each antenna. In order to tackle the formulated optimization problem more easily, we write this constraint independent of the RF chain selection indicators. Consequently, the precoders are designed in such a way that the uniform elemental power constraint is satisfied at each antenna, regardless of the active status of the corresponding RF chain. It is seen from \eqref{eqn:common_rate}, \eqref{eqn:private_rate} that the SR is calculated by multiplying the precoders by the RF chain selection indicators. We will also show in the following section that the total power consumption is also calculated by the multiplication of RF chain selection indicators by the transmit power at each antenna. Therefore, the SR and total power consumption metrics in this work are dependent on the activation status of the RF chains, although the precoders are designed to satisfy uniform elemental power constraint at each antenna regardless of the corresponding status.

\subsection{Performance Metrics}
In this section, we explain the metrics to be used in the constraints and objective function of the optimization problem that will be formulated to design our system.

\subsubsection{Performance Metric for Communications}
Ergodic rate is an appropriate metric for system design under imperfect CSIT, as it captures the statistical properties of CSIT imperfections and leads to designs with enhanced performance \cite{clerckxTC2016}. The ergodic rates for the common and private streams are expressed as $\mathbb{E}_{\mathbf{h}_{k}}\left\lbrace R_{c,k,l}\right\rbrace$ and $\mathbb{E}_{\mathbf{h}_{k}}\left\lbrace R_{k,l}\right\rbrace$. Let us define the average rates for the common and private streams as $\bar{R}_{c,k,l}\triangleq\mathbb{E}_{\mathbf{h}_{k}|\widehat{\mathbf{h}}_{k}}\left\lbrace R_{c,k,l}\right\rbrace$ and $\bar{R}_{k,l}\triangleq\mathbb{E}_{\mathbf{h}_{k}|\widehat{\mathbf{h}}_{k}}\left\lbrace R_{k,l}\right\rbrace$, which represent the rates averaged over possible channel estimation errors for a given CSIT realization. It was shown in \cite{clerckxTC2016} that $\mathbb{E}_{\mathbf{h}_{k}}\left\lbrace R_{c,k,l}\right\rbrace=\mathbb{E}_{\widehat{\mathbf{h}}_{k}}\left\lbrace \bar{R}_{c,k,l}\right\rbrace$ and $\mathbb{E}_{\mathbf{h}_{k}}\left\lbrace R_{k,l}\right\rbrace=\mathbb{E}_{\widehat{\mathbf{h}}_{k}}\left\lbrace \bar{R}_{k,l}\right\rbrace$. Therefore, we can use the average rates $\bar{R}_{c,k,l}$ and $\bar{R}_{k,l}$ to obtain precoders for a given CSIT with robust performance under CSI imperfections.
Accordingly, we express the constraint \eqref{eqn:common_rate_constr} in terms of average rates as $\bar{C}_{l}(\mathbf{P}_{l}, \boldsymbol{\Lambda}, \mathbf{B})=\min_{k \in \mathcal{K}}\bar{R}_{c,k,l}(\mathbf{P}_{l}, \boldsymbol{\Lambda}, \mathbf{B})$. For a given CSIT realization, the communications performance of the system is measured in terms of the SR, which is defined as 
\begin{align}
	\bar{R}_{sum}(\mathbf{P}, \boldsymbol{\Lambda}, \mathbf{B})\hspace{-0.1cm}=\hspace{-0.1cm}\frac{1}{L}\sum_{l \in \mathcal{L}}\hspace{-0.1cm}\left(\hspace{-0.1cm}\bar{C}_{l}(\mathbf{P}_{l}, \boldsymbol{\Lambda}, \mathbf{B})\hspace{-0.1cm}+\hspace{-0.1cm}\sum_{k=1}^{K}\bar{R}_{k,l}(\mathbf{P}_{l}, \boldsymbol{\Lambda}, \mathbf{B})\hspace{-0.1cm}\right)\hspace{-0.1cm}, \nonumber
\end{align}
where $\mathbf{P}=[\mathbf{P}_{1}, \mathbf{P}_{2}, \ldots, \mathbf{P}_{L}]$.

\subsubsection{Performance Metric for Radar}
In this work, we consider a co-located system and a point target model. Such a model is valid especially for detecting targets by means of co-located radars when the target distance is large compared to inter-element distance of the radar antenna array \cite{skolnik_2008,khawar_2015}.

Define the covariance matrix of the transmitted radar signal $\mathbf{x}_{l}$ with $L$ samples as
\begin{align}
	\mathbf{R}_{\mathbf{x}}=\sum_{l \in \mathcal{L}}\mathbf{x}_{l}\mathbf{x}_{l}^{H}.
\end{align}
The covariance matrix of the transmitted signal can be used to investigate the detection probability \cite{khawar_2015} or angle/distance estimation performance for the radar. For each of the mentioned performance metrics, waveform designs with specific covariance matrix properties are known to achieve an optimal performance in terms of the metric in consideration. Therefore, we aim to design waveforms with covariance matrix properties similar to those of such radar-specific designs. Consequently, our radar performance metric is expressed as 
\begin{align}
	||\mathbf{R}_{\tilde{\mathbf{x}}}(\mathbf{P}, \boldsymbol{\Lambda}, \mathbf{B})-\mathbf{U} ||^{2}_{2} \leq \tau,
\end{align} 
where $\mathbf{U} $ represents the desired radar-signal covariance matrix and $\tau$ is the similarity threshold to the radar waveform. 

From the quantized signal model in \eqref{eqn:transmit_signal_quant}, one can notice the transmit signal is a stochastic process due to the random transmit symbols and quantization noise term $\boldsymbol{\epsilon}_{l}$. Proposition 1 gives an approximate form for the spatial covariance matrix of the quantized signal.

\textit{Proposition 1:} Assume that the quantization noise and transmit symbols are Wide-Sense Stationary (WSS) stochastic processes and ergodic in the wide sense. Then, for large $L$, the matrix $\mathbf{R}_{\tilde{\mathbf{x}}}(\mathbf{P}, \boldsymbol{\Lambda}, \mathbf{B})$ can be approximately expressed as
\begin{align}
	\mathbf{R}_{\tilde{\mathbf{x}}}(\mathbf{P}, \boldsymbol{\Lambda}, \mathbf{B})&\approx\sum_{l \in \mathcal{L}}\boldsymbol{\Delta}\boldsymbol{\Lambda}\mathbf{P}_{l}\mathbf{P}_{l}^{H}\boldsymbol{\Lambda}\boldsymbol{\Delta}+L\boldsymbol{\Lambda}\boldsymbol{\Sigma}\boldsymbol{\Lambda}. \nonumber \\
	&\triangleq\widehat{\mathbf{R}}_{\tilde{\mathbf{x}}}(\mathbf{P}, \boldsymbol{\Lambda}, \mathbf{B}).
\end{align}

\subsubsection{Performance Metric for the Objective Function}
We consider the maximization of the EE metric for the objective function, which is defined as
\begin{align}
	\mathcal{E}(\mathbf{P}, \boldsymbol{\Lambda}, \mathbf{B})=\frac{\bar{R}_{sum}(\mathbf{P}_{l}, \boldsymbol{\Lambda}, \mathbf{B})}{P_{tot}(\boldsymbol{\Lambda}, \mathbf{B})}.
\end{align}
We express the total power consumption as in \cite{rodriguez_2017, lee_2013, bjornson_2015}, by adding the RF chain selection indicators as
\begin{align}
	&P_{tot}(\boldsymbol{\Lambda},  \mathbf{B}) \nonumber \\
	&=\frac{\mathrm{tr}(\boldsymbol{\Lambda}^{2})P_{ant}}{\eta_{PA}}+P_{RF}+\mathrm{tr}(\boldsymbol{\Lambda}^{2}\mathbf{P}_{\boldsymbol{\Delta}})+P_{int}+P_{BB},
	\label{eqn:totalpower}
\end{align}
where $0 \leq \eta_{PA} \leq 1$ is the power amplifier efficiency, $\mathbf{P}_{\boldsymbol{\Delta}}=\mathrm{Diag}(P(\delta_{1}), P(\delta_{2}), \ldots, P(\delta_{N_{t}}))$ and $P_{BB}$ is the power consumption for baseband signal processing. The term $P_{RF}=\mathrm{tr}(\boldsymbol{\Lambda}^{2})P_{circ}+P_{syn}$ represents the power consumption by analog hardware components with $P_{circ}$ being the power consumed by analog circuitry in each RF chain and $P_{syn}$ is the power consumed by frequency synthesizer. 
The term $P_{int}=p_{int}2S_{DAC}\mathrm{tr}(\boldsymbol{\Lambda}^{2}\mathbf{B})$ represents the power consumed by data interface of the Digital Signal Processor (DSP) for each RF chain, where $p_{int}$ is the the power consumption per Gbps and $S_{DAC}$ represents the sampling rate per port (I-Q) in Gbps \cite{rodriguez_2017} \footnote{Since $\lambda_{i} \in \left\lbrace 0,1 \right\rbrace$, the power consumption expressions can also be written in terms of $\boldsymbol{\Lambda}$ instead of $\boldsymbol{\Lambda}^{2}$. However, we stick with the current expressions for the sake of consistency. Moreover, we make use of the quadratic $\boldsymbol{\Lambda}^{2}$ terms in our proposed algorithm to obtain the optimal RF chain selection indicators in Section~\ref{sec:scasdr}.}.

One can note from \eqref{eqn:totalpower} that the power transmitted in an RF chain, which is represented by $P_{ant}$, only affects the first term in the summation, and all terms except the last one are dependant on the RF chain selection indicator matrix. This implies that even if one allocates a small amount of power to a particular antenna, the corresponding RF chain still consumes power unless the antenna is completely deactivated. Consequently, RF chain selection is required to completely eliminate power consumption at selected chains.

\section{Problem Formulation and Proposed Algorithms}
\label{sec:problem}
In this section, we formulate an EE maximization problem for DFRC waveform design with RSMA and propose an algorithm to solve of the resulting formulation.
Our aim is to determine the optimal number of active RF chains to maximize EE under radar and communications performance metrics. 

\subsection{Problem Formulation}
We formulate the RSMA EE maximization problem for joint radar and communications for a given $\mathbf{B}$ (and $\boldsymbol{\Delta}$ and $\boldsymbol{\Sigma}$) as 
\begin{subequations}
	\begin{alignat}{3}
		&\max_{ \bar{\mathbf{C}}, \mathbf{P}, \boldsymbol{\Lambda}}   \    \mathcal{E}(\mathbf{P},  \boldsymbol{\Lambda}, \mathbf{B})    \\
		&\hspace{0.8cm}\text{s.t.}  \  \bar{C}_{l}(\mathbf{P}_{l}, \boldsymbol{\Lambda}, \mathbf{B}) \leq \bar{R}_{c,k,l}(\mathbf{P}_{l}, \boldsymbol{\Lambda}, \mathbf{B}) , \nonumber \\ 
		&\hspace{3.8cm}   k \in \mathcal{K}, \ \forall l \in \mathcal{L} \label{eqn:common_rate_2} \\
		& \hspace{0.6cm} \lambda_{i} \in \left\{0,1 \right\}, \   i \in \left\lbrace 1,2,\ldots, N_{t} \right\rbrace, \  \forall l \in \mathcal{L} \label{eqn:s} \\
		& \hspace{0.6cm} ||\widehat{\mathbf{R}}_{\tilde{\mathbf{x}}}(\mathbf{P}, \boldsymbol{\Lambda}, \mathbf{B})-\mathbf{U} ||^{2}_{2} \leq \tau \label{eqn:radar} \\
		& \hspace{0.6cm} \bar{R}_{sum}(\mathbf{P}, \boldsymbol{\Lambda}, \mathbf{B}) \geq R_{th} \label{eqn:comms} \\
		& \hspace{0.6cm} \mathrm{tr}(\mathbf{E}_{N_{t},i}(\boldsymbol{\Delta}\mathbf{P}_{l}\mathbf{P}_{l}^{H}\boldsymbol{\Delta}+\boldsymbol{\Sigma}))=P_{ant},  \nonumber \\ 
		&\hspace{2.2cm}   \forall i \in \left\lbrace 1,2,\ldots, N_{t} \right\rbrace, \  \forall l \in \mathcal{L}\label{eqn:radar_power_1},  
	\end{alignat}
	\label{eqn:problem3_v2}
\end{subequations}
where $\bar{\mathbf{C}}=[\bar{C}_{1}, \bar{C}_{2}, \ldots, \bar{C}_{L}]$, and $R_{th}$ is the lower bound for the SR. The constraint \eqref{eqn:common_rate_2} ensures the decodability of the common stream by all receivers, in accordance with the RSMA framework as described in Section~\ref{sec:system}. The radar detection performance is controlled by subjecting the radar metric to the upper bound in the constraint \eqref{eqn:radar}. A minimum SR performance is guaranteed by the constraint \eqref{eqn:comms}. The constraint \eqref{eqn:radar_power_1} is the uniform elemental power constraint as discussed in Section~\ref{sec:rsma_syst}.



The problem formulation \eqref{eqn:problem3_v2} is not convex with respect to the optimization parameters $\mathbf{P}$ and  $\boldsymbol{\Lambda}$, and is challenging to solve. In the following sections, we describe our proposed AO-based algorithm that alternates between finding the optimal precoders for given RF chain selection matrix and finding optimal RF chain selection matrix for given precoders by solving separate subproblems. 
The subproblem of finding the optimal precoders is solved by an ADMM-based algorithm and the problem of finding optimal RF chain selection matrix for given precoders is solved by an SCA/SDR-based algorithm. 

\subsection{Proposed AO-ADMM-Based Algorithm for Optimal Precoder Calculation}
\label{sec:aoadmm}
In this section, we address the problem of obtaining the optimal precoders $\mathbf{P}_{l}$, $\forall l \in \mathcal{L}$,  for a given $\boldsymbol{\Lambda}^{\prime}$.
We define $\mathbf{v}=[\mathbf{v}_{1}^{T},\mathbf{v}_{2}^{T}\ldots, \mathbf{v}_{L}^{T}]^{T}$,  $\mathbf{v}_{l}=[\bar{C}_{l}, \mathrm{vec}(\mathbf{P}_{l})^{T}]^{T}$, and $f(\mathbf{v}, \boldsymbol{\Lambda}^{\prime}, \mathbf{B})=\mathcal{E}(\mathbf{v}, \boldsymbol{\Lambda}^{\prime}, \mathbf{B})$. We reformulate the problem \eqref{eqn:problem3_v2} according to the ADMM framework \cite{boyd_2011} as 
\begin{subequations}
	\begin{alignat}{3}
		\min_{\mathbf{v},\mathbf{u}}&   \quad  -f(\mathbf{v, \boldsymbol{\Lambda}^{\prime}, \mathbf{B}})+\Pi(\mathbf{u, \boldsymbol{\Lambda}^{\prime}, \mathbf{B}})      \label{eqn:obj_3_2}   \\
		\text{s.t.}&  \quad  \mathbf{v}-\mathbf{u}=\mathbf{0},
	\end{alignat}
	\label{eqn:problem_admm_2}
\end{subequations}
\hspace{-0.2cm}where  $\mathbf{u}_{l}$, $\forall l \in \mathcal{L}$ and $\mathbf{u}=[\mathbf{u}_{1}^{T},\mathbf{u}_{2}^{T}\ldots, \mathbf{u}_{L}^{T}]^{T}$ are introduced to split the problem according to the ADMM procedure. 
The function $\Pi(\mathbf{v})$ is the indicator function that performs projection on the domain $\mathcal{D}$ defined by the constraints \eqref{eqn:common_rate_2}, \eqref{eqn:radar} and \eqref{eqn:radar_power_1}, {\sl i.e.}, $\Pi(\mathbf{u}, \boldsymbol{\Lambda}^{\prime}, \mathbf{B})=0$ if $\mathbf{u} \in \mathcal{D}$ and $\Pi(\mathbf{u}, \boldsymbol{\Lambda}^{\prime}, \mathbf{B})=\infty$ if $\mathbf{u} \notin \mathcal{D}$. 
We also define the real-valued vectors \mbox{$\mathbf{v}_{r,l}=[\mathfrak{R}(\mathbf{v}_{l}^{T}), \mathfrak{I}(\mathbf{v}_{l}^{T})]^{T}$}, \mbox{$\mathbf{u}_{r,l}=[\mathfrak{R}(\mathbf{u}_{l}^{T}), \mathfrak{I}(\mathbf{u}_{l}^{T})]^{T}$}, \mbox{$\mathbf{v}_{r}=[\mathbf{v}_{r,1}^{T},\mathbf{v}_{r,2}^{T}\ldots, \mathbf{v}_{r,L}^{T}]^{T}$} and \mbox{$\mathbf{u}_{r}=[\mathbf{u}_{r,1}^{T},\mathbf{u}_{r,2}^{T}\ldots, \mathbf{u}_{r,L}^{T}]^{T}$} 
in accordance with the ADMM framework, where the functions $\mathfrak{R}(.)$ and $\mathfrak{I}(.)$ return the real and imaginary parts of their inputs, respectively. 

The augmented Lagrangian function for the optimization problem \eqref{eqn:problem_admm_2} is written as 
\begin{align}
	\mathcal{L}_{\zeta}&=-f(\mathbf{v_{r}}, \boldsymbol{\Lambda}^{\prime}, \mathbf{B})+\Pi(\mathbf{u}_{r}, \boldsymbol{\Lambda}^{\prime}, \mathbf{B}) \nonumber \\
	&+\sum_{l \in \mathcal{L}} \mathbf{d}_{r,l}^{T}(\mathbf{v}_{r,l}-\mathbf{u}_{r,l})+\sum_{l \in \mathcal{L}} (\zeta/2)||\mathbf{v}_{r,l}-\mathbf{u}_{r,l}||^{2}_{2}, \nonumber
\end{align}
where $ \zeta > 0$ is called the penalty parameter, $\mathbf{d}_{r}=[\mathbf{d}_{r,1}^{T},\mathbf{d}_{r,2}^{T}\ldots, \mathbf{d}_{r,L}^{T}]^{T}$ and $\mathbf{d}_{r,n}=[\mathfrak{R}(\mathbf{d}_{n}^{T}), \mathfrak{I}(\mathbf{d}_{l}^{T})]^{T}$ and $\mathbf{d}_{l} \in \mathbb{C}^{(1+N_{t}(K+1))}$, $\forall l \in \mathcal{L}$ being the dual variables. 

The updates of iterative ADMM procedure can be written in the scaled form as
\begin{align}
	&\mathbf{v}^{t+1}_{r}\hspace{-0.1cm}=\hspace{-0.1cm}\argmin_{\mathbf{v}_{r}}(-f(\mathbf{v}_{r}, \boldsymbol{\Lambda}^{\prime}, \mathbf{B}))\hspace{-0.1cm}+\hspace{-0.1cm}\frac{\zeta}{2}\hspace{-0.1cm}\sum_{l \in \mathcal{L}}||\mathbf{v}_{r,l}-\mathbf{u}^{t}_{r,l}+\mathbf{w}^{t}_{r,l}||^{2}_{2} )  \label{eqn:v_update} \\
	&\mathbf{u}^{t+1}_{r}\hspace{-0.1cm}=\hspace{-0.1cm}\argmin_{\mathbf{u}}(\Pi(\mathbf{u}_{r}, \boldsymbol{\Lambda}^{\prime}, \mathbf{B})\hspace{-0.1cm}+\hspace{-0.1cm}\frac{\zeta}{2}\hspace{-0.1cm}\sum_{l \in \mathcal{L}}||\mathbf{v}^{t+1}_{r,l}-\mathbf{u}_{r,l}+\mathbf{w}^{t}_{r,l}||^{2}_{2} )  \label{eqn:u_update}  \\
	&\mathbf{w}^{t+1}_{r,n}\hspace{-0.1cm}=\hspace{-0.1cm}\mathbf{w}^{t}_{r,l}+\mathbf{v}^{t+1}_{r,l}-\mathbf{u}^{t+1}_{r,l}, \quad \forall l \in \mathcal{L}, \label{eqn:w_update}
\end{align}
where $\mathbf{w}_{r}=\mathbf{d}_{r}/\zeta$. The update step for $\mathbf{v}_{r}$ in \eqref{eqn:v_update} deals with the EE maximization problem. The update step for $\mathbf{u}_{r}$ in \eqref{eqn:u_update} involves projection onto the domain $\mathcal{D}$. In the following subsections, we explain our proposed solution on how to perform each update step.

\subsubsection{Update Step for $\mathbf{v}$ : }
We note that for given $\mathbf{B}$ and $\boldsymbol{\Lambda}^{\prime}$, the objective function $f(\mathbf{v}, \boldsymbol{\Lambda}^{\prime}, \mathbf{B})$ is not convex with respect to $\mathbf{v}$. In \cite{clerckxTC2016}, Rate-Mean Square Error (MSE) transformations are used to transform the non-convex SR function for RSMA into a convex one.
Accordingly, we first obtain the optimal receive filters, $g_{c,k,l}$ and $g_{k,l}$, which minimize the MSEs $\epsilon_{c,k,l}=\mathbb{E}\left\lbrace|g_{c,k,l}y^{c}_{k,l}-s_{c,l}|^{2} \right\rbrace$ and $\epsilon_{k,l}=\mathbb{E}\left\lbrace|g_{k,l}(y^{c}_{k,l}-\mathbf{h}_{k}^{H}\boldsymbol{\Delta}\boldsymbol{\Lambda}\mathbf{p}_{c,l}s_{c,l})-s_{k,l}|^{2} \right\rbrace$, respectively,  $\forall k \in \mathcal{K}$, $\forall l \in \mathcal{L}$. The expressions for $\epsilon_{c,k,l}$ and $\epsilon_{k,l}$ are given as
\begin{subequations}
\begin{align}
	\epsilon_{c,k,l}&=|g_{c,k,l}|^{2}\left(|\mathbf{h}_{k}^{H}\boldsymbol{\Delta}\boldsymbol{\Lambda}\mathbf{p}_{c,l}|^{2}+\sum_{k \in \mathcal{K}}|\mathbf{h}_{k}^{H}\boldsymbol{\Delta}\boldsymbol{\Lambda}\mathbf{p}_{k,l}|^{2}+\sigma^{2}_{\eta,k}\right) \nonumber\\
	&-2\mathfrak{R}\{g_{c,k,l}\mathbf{h}_{k}^{H}\boldsymbol{\Delta}\boldsymbol{\Lambda}\mathbf{p}_{c,l}\}+1,  \\
	\epsilon_{k,l}&=|g_{k,l}|^{2}\left(\sum_{k \in \mathcal{K}}|\mathbf{h}_{k}^{H}\boldsymbol{\Delta}\boldsymbol{\Lambda}\mathbf{p}_{k,l}|^{2}+\sigma^{2}_{\eta,k}\right)\nonumber \\
	&-2\mathfrak{R}\{g_{k,l}\mathbf{h}_{k}^{H}\boldsymbol{\Delta}\boldsymbol{\Lambda}\mathbf{p}_{k,l}\}+1.\label{eqn:mse_1}
\end{align}
\end{subequations}

It is well known that the optimal receive filter for the abovementioned problem is a Minimum MSE (MMSE) filter expressed as
\begin{align}
	g^{\mathrm{opt}}_{k,l}&\hspace{-0.1cm}=\hspace{-0.1cm}\frac{\mathbf{p}^{H}_{k,l}\boldsymbol{\Lambda}^{\prime}\boldsymbol{\Delta}\mathbf{h}_{k}}{\sigma_{n}^{2}+\mathbf{h}_{k}^{H}\boldsymbol{\Lambda}^{\prime}\boldsymbol{\Sigma}\boldsymbol{\Lambda}^{\prime}\mathbf{h}_{k}+\sum_{k^{\prime}\in\mathcal{K}}|\mathbf{h}_{k}^{H}\boldsymbol{\Delta}\boldsymbol{\Lambda}^{\prime}\mathbf{p}_{k^{\prime},l}|^{2}}, \nonumber \\
	g^{\mathrm{opt}}_{c,k,l}&\hspace{-0.1cm}=\hspace{-0.1cm}\frac{\mathbf{p}^{H}_{c,k,l}\boldsymbol{\Lambda}^{\prime}\boldsymbol{\Delta}\mathbf{h}_{k}}{\sigma_{n}^{2}+\mathbf{h}_{k}^{H}\boldsymbol{\Lambda}^{\prime}\boldsymbol{\Sigma}\boldsymbol{\Lambda}^{\prime}\mathbf{h}_{k}+\sum_{k^{\prime}\in \left\lbrace c,\mathcal{K}\right\rbrace}|\mathbf{h}_{k}^{H}\boldsymbol{\Delta}\boldsymbol{\Lambda}^{\prime}\mathbf{p}_{k^{\prime},l}|^{2}}.
	\label{eqn:mmse}
\end{align} 
The resulting MSEs are written as
\begin{align}
	\epsilon^{\mathrm{opt}}_{k,l}\hspace{-0.1cm}=&\frac{\sigma_{n}^{2}+\mathbf{h}_{k}^{H}\boldsymbol{\Lambda}^{\prime}\boldsymbol{\Sigma}\boldsymbol{\Lambda}^{\prime}\mathbf{h}_{k}+\sum_{\substack{k^{\prime}\in\mathcal{K}, \\ k^{\prime} \neq k}}|\mathbf{h}_{k}^{H}\boldsymbol{\Delta}\boldsymbol{\Lambda}^{\prime}\mathbf{p}_{k^{\prime},l}|^{2}}{\sigma_{n}^{2}+\mathbf{h}_{k}^{H}\boldsymbol{\Lambda}^{\prime}\boldsymbol{\Sigma}\boldsymbol{\Lambda}^{\prime}\mathbf{h}_{k}+\sum_{k^{\prime}\in\mathcal{K}}|\mathbf{h}_{k}^{H}\boldsymbol{\Delta}\boldsymbol{\Lambda}^{\prime}\mathbf{p}_{k^{\prime},l}|^{2}}, \nonumber \\
	\epsilon^{\mathrm{opt}}_{c,k,l}\hspace{-0.1cm}=&\frac{\sigma_{n}^{2}+\mathbf{h}_{k}^{H}\boldsymbol{\Lambda}^{\prime}\boldsymbol{\Sigma}\boldsymbol{\Lambda}^{\prime}\mathbf{h}_{k}+\sum_{k^{\prime}\in\mathcal{K}}|\mathbf{h}_{k}^{H}\boldsymbol{\Delta}\boldsymbol{\Lambda}^{\prime}\mathbf{p}_{k^{\prime},l}|^{2}}{\sigma_{n}^{2}+\mathbf{h}_{k}^{H}\boldsymbol{\Lambda}^{\prime}\boldsymbol{\Sigma}\boldsymbol{\Lambda}^{\prime}\mathbf{h}_{k}+\sum_{k^{\prime}\in \left\lbrace c,\mathcal{K}\right\rbrace}|\mathbf{h}_{k}^{H}\boldsymbol{\Delta}\boldsymbol{\Lambda}^{\prime}\mathbf{p}_{k^{\prime},l}|^{2}}.
	\label{eqn:mse}
\end{align} 

We define the augmented weighted MSEs (WMSEs) as
\begin{subequations}
\begin{align}
	\xi_{c,k,l}&=\omega_{c,k,l}\epsilon_{c,k,l}-\log_{2}(\omega_{c,k,l}),   \\
	\xi_{k,l}&=\omega_{k,l}\epsilon_{k,l}-\log_{2}(\omega_{k,l}),\label{eqn:wmse_1} 
\end{align}
\label{eqn:wmse}
\end{subequations}
where $\omega_{c,k,l}$ and $\omega_{k,l}$ are the weights for the MSEs of the common and private streams at user-$k$ and symbol-$l$. We also define the average augmented WMSEs as $\bar{\xi}_{c,k,l}=\mathbb{E}_{\mathbf{h}_{k}|\widehat{\mathbf{h}}_{k}}\left\lbrace \xi_{c,k,l}\right\rbrace$ and $\bar{\xi}_{k,l}=\mathbb{E}_{\mathbf{h}_{k}|\widehat{\mathbf{h}}_{k}}\left\lbrace \xi_{k,l}\right\rbrace$.
It can be shown that the optimum weights are given by $\omega^{opt}_{c,k,l}=(\epsilon^{\mathrm{opt}}_{c,k,l})^{-1}$ and $\omega^{opt}_{k,l}=(\epsilon^{\mathrm{opt}}_{k,l})^{-1}$, for which the MSE-rate relations are obtained as $\xi^{\mathrm{opt}}_{c,k,l}=1\minus R_{c,k,l}$ and
$\xi^{\mathrm{opt}}_{k,l}=1\minus R_{k,l}$. Accordingly, the average augmented MSE-rate transformations are written as  
\begin{subequations}
		\begin{alignat}{3}
	\bar{\xi}^{\mathrm{opt}}_{c,k,l}&=\mathbb{E}_{\mathbf{h}_{k}|\widehat{\mathbf{h}}_{k}}\left\lbrace \xi^{\mathrm{opt}}_{c,k,l}\right\rbrace=1\minus \bar{R}_{c,k,l}, \\ \bar{\xi}^{\mathrm{opt}}_{k,l}&=\mathbb{E}_{\mathbf{h}_{k}|\widehat{\mathbf{h}}_{k}}\left\lbrace \xi^{\mathrm{opt}}_{k,l}\right\rbrace=1\minus \bar{R}_{k,l}.
	\label{eqn:augMSErate}
\end{alignat}
\end{subequations}

For given $\boldsymbol{\Lambda}^{\prime}$, it is straightforward to show that maximizing $f(\mathbf{v}, \boldsymbol{\Lambda}^{\prime}, \mathbf{B})=\mathcal{E}(\mathbf{v}, \boldsymbol{\Lambda}^{\prime}, \mathbf{B})$ is equivalent to minimizing $\bar{f}(\mathbf{v}, \boldsymbol{\Lambda}^{\prime}, \mathbf{B})=\frac{1}{L}\sum_{l \in \mathcal{L}} (-\bar{C}_{l}+\sum_{k \in \mathcal{K}}\bar{\xi}_{k,l})$. 
Then, the update operation in \eqref{eqn:v_update} is rewritten in terms of $\bar{f}(\mathbf{v}_{r}, \boldsymbol{\Lambda}^{\prime}, \mathbf{B})$ as 
\begin{align}
	\mathbf{v}^{t+1}_{r}\hspace{-0.1cm}=&\hspace{-0.1cm}\argmin_{\mathbf{v}_{r}}(\bar{f}(\mathbf{v}_{r}, \boldsymbol{\Lambda}^{\prime}, \mathbf{B})\hspace{-0.1cm}+\hspace{-0.1cm}\frac{\zeta}{2}\hspace{-0.1cm}\sum_{l \in \mathcal{L}}\hspace{-0.05cm}||\mathbf{v}_{r,l}-\mathbf{u}^{t}_{r,l}+\mathbf{w}^{t}_{r,l}||^{2}_{2} ).  \label{eqn:v_update_2} 
\end{align}
We note the stochastic terms in \eqref{eqn:v_update_2}. In order to solve the resulting optimization problem, we first write deterministic equivalents for the stochastic expressions using the Sample Average Approximation (SAA) method \cite{shapiro_2009}. We define the sample set for given $\widehat{\mathbf{h}}_{k}$ and $\mathcal{M}=\left\lbrace 1,2,\ldots,M\right\rbrace$ as
	\begin{align}
		\mathbb{H}^{(M)}\triangleq\left\lbrace\mathbf{h}^{(m)}_{k}\hspace{-0.1cm}=\widehat{\mathbf{h}}_{k}+\tilde{\mathbf{h}}^{(m)}_{k}|\hspace{-0.1cm} \ \widehat{\mathbf{h}}_{k},\  \forall m\in \mathcal{M}, \forall k\in \mathcal{K} \right\rbrace\hspace{-0.1cm},
	\end{align}
	where the elements are independent and identically distributed with $f_{\mathbf{h}|\widehat{\mathbf{h}}}(\mathbf{h}|\widehat{\mathbf{h}})$.
	This set is
	used to approximate the average rates by their Sample Average
	Functions (SAFs) as
	\begin{align}
		\bar{R}^{(M)}_{c,k,l}(\mathbf{P}_{l}, \boldsymbol{\Lambda}^{\prime}, \mathbf{B})&=\frac{1}{M}\sum_{m \in \mathcal{M}}R^{(m)}_{c,k,l}(\mathbf{P}_{l}, \boldsymbol{\Lambda}^{\prime}, \mathbf{B}, \mathbf{h}^{(m)}_{k}), \nonumber \\
		\bar{R}^{(M)}_{k,l}(\mathbf{P}_{l}, \boldsymbol{\Lambda}^{\prime}, \mathbf{B})&=\frac{1}{M}\sum_{m \in \mathcal{M}}R^{(m)}_{k,l}(\mathbf{P}_{l}, \boldsymbol{\Lambda}^{\prime}, \mathbf{B}, \mathbf{h}^{(m)}_{k}),
	\end{align}
	where $R^{(m)}_{c,k,l}(\mathbf{P}_{l}, \boldsymbol{\Lambda}, \mathbf{B}, \mathbf{h}^{(m)}_{k})$ and $R^{(m)}_{k,l}(\mathbf{P}_{l}, \boldsymbol{\Lambda}, \mathbf{B}, \mathbf{h}^{(m)}_{k})$ are the rates obtained by the $m$-th channel realization $\mathbf{h}^{(m)}_{k}$. Accordingly, we define $\xi^{(m)}_{c,k,l}=\omega^{(m)}_{c,k,l}\epsilon^{(m)}_{c,k,l}-\log_{2}(\omega^{(m)}_{c,k,l})$ and $\xi^{(m)}_{k,l}=\omega^{(m)}_{k,l}\epsilon^{(m)}_{k,l}-\log_{2}(\omega^{(m)}_{k,l})$ to write
	\begin{subequations}
		\begin{alignat}{3}
			\bar{\xi}^{\mathrm{opt}(M)}_{c,k,l}&=\frac{1}{M}\sum_{m \in \mathcal{M}} \xi^{\mathrm{opt}(m)}_{c,k,l}=1\minus \bar{R}^{(M)}_{c,k,l}, \\ \bar{\xi}^{\mathrm{opt}(M)}_{k,l}&=\frac{1}{M}\sum_{m \in \mathcal{M}}\xi^{\mathrm{opt}(m)}_{k,l}=1\minus \bar{R}^{(M)}_{k,l}.
			\label{eqn:augMSErate_saa}
		\end{alignat}
	\end{subequations}
	Next, we express $\bar{\mathbf{f}}(\mathbf{v, \boldsymbol{\Lambda}^{\prime}, \mathbf{B}})$ and the constraints \eqref{eqn:common_rate_1_f} and \eqref{eqn:comms_1_f} in terms of $\omega^{(m)}_{c,k,l}$, $\omega^{(m)}_{k,l}$, $g^{(m)}_{c,k,l}$, and $g^{(m)}_{k,l}$  to describe the proposed algorithm. For that purpose, we first define
	\begin{align}
		\bar{d}_{c,k,l}&\hspace{-0.05cm}=\hspace{-0.05cm}\frac{1}{M}\hspace{-0.15cm}\sum_{m \in \mathcal{M}}\hspace{-0.15cm}\log(\omega^{(m)}_{c,k,l}), \quad \bar{d}_{k,l}\hspace{-0.05cm}=\hspace{-0.05cm}\frac{1}{M}\hspace{-0.15cm}\sum_{m \in \mathcal{M}}\hspace{-0.15cm}\log(\omega^{(m)}_{k,l}),  \nonumber \\ 
		\bar{q}_{c,k,l}&\hspace{-0.05cm}=\hspace{-0.05cm}\frac{1}{M}\hspace{-0.15cm}\sum_{m \in \mathcal{M}}\hspace{-0.15cm}\omega^{(m)}_{c,k,l}|g^{(m)}_{c,k,l}|^{2}, \quad \bar{q}_{k,l}\hspace{-0.05cm}=\hspace{-0.05cm}\frac{1}{M}\hspace{-0.15cm}\sum_{m \in \mathcal{M}}\hspace{-0.15cm}\omega^{(m)}_{k,l}|g^{(m)}_{k,l}|^{2}, \nonumber \\ 
		\bar{\mathbf{f}}_{c,k,l}&\hspace{-0.05cm}=\hspace{-0.05cm}\frac{1}{M}\hspace{-0.15cm}\sum_{m \in \mathcal{M}}\hspace{-0.15cm}\omega^{(m)}_{c,k,l}(g^{(m)}_{c,k,l})^{*}\mathbf{h}_{k}^{(m)}, \nonumber \\ 	\bar{\mathbf{f}}_{k,l}&\hspace{-0.05cm}=\hspace{-0.05cm}\frac{1}{M}\hspace{-0.15cm}\sum_{m \in \mathcal{M}}\hspace{-0.15cm}\omega^{(m)}_{k,l}(g^{(m)}_{k,l})^{*}\mathbf{h}_{k}^{(m)}, \nonumber \\ 
		\bar{\mathbf{A}}_{c,k,l}&\hspace{-0.05cm}=\hspace{-0.05cm}\frac{1}{M}\hspace{-0.15cm}\sum_{m \in \mathcal{M}}\hspace{-0.15cm}\omega^{(m)}_{c,k,l}|g^{(m)}_{c,k,l}|^{2}\mathbf{h}_{k}^{(m)}(\mathbf{h}_{k}^{(m)})^{H}, \nonumber \\ 
		\bar{\mathbf{A}}_{k,l}&\hspace{-0.05cm}=\hspace{-0.05cm}\frac{1}{M}\hspace{-0.15cm}\sum_{m \in \mathcal{M}}\hspace{-0.15cm}\omega^{(m)}_{k,l}|g^{(m)}_{k,l}|^{2}\mathbf{h}_{k}^{(m)}(\mathbf{h}_{k}^{(m)})^{H}. 
		\label{eqn:allsafs}
	\end{align}
	The average augmented MSEs for the private streams can be written by substituting \eqref{eqn:mse_1} into \eqref{eqn:wmse_1} and using \eqref{eqn:augMSErate_saa} as follows:
	\begin{align}
		\bar{\xi}_{k,l}&=\sum_{i \in \mathcal{K}}\mathbf{p}_{i,l}^{H}\boldsymbol{\Lambda}^{\prime}\boldsymbol{\Delta}\bar{\mathbf{A}}_{k,l}\boldsymbol{\Delta}\boldsymbol{\Lambda}^{\prime}\mathbf{p}_{i,l}+\bar{q}_{k,l}\sigma_{n}^{2}+\tr(\bar{\mathbf{A}}_{k,l}\boldsymbol{\Lambda}^{\prime}\boldsymbol{\Sigma}\boldsymbol{\Lambda}^{\prime})\nonumber\\
		&-2\mathcal{R}\lbrace\bar{\mathbf{f}}^{H}_{k,l}\boldsymbol{\Delta}\boldsymbol{\Lambda}^{\prime}\mathbf{p}_{k,l}\rbrace+\bar{\omega}_{k,l}-\bar{d}_{k,l}.
		\label{eqn:xibar}
	\end{align}
\begin{figure}[t!]
	\removelatexerror
	\begin{algorithm}[H]
			\caption{ADMM-Based Algorithm}
			\label{alg:admm}
			$t \gets 0$, $\mathbf{v}_{r}^{0}$, $\mathbf{u}_{r}^{0}$,  $\mathbf{w}_{r}^{0}$, $\boldsymbol{\omega}$, $\mathbf{g}$, $\boldsymbol{\Lambda}^{\prime}$, $\mathbf{B}$ \\
			\While{$\sum_{l \in \mathcal{L}}\hspace{-0.08cm}||\mathbf{r}_{r,l}^{t}||\hspace{-0.08cm}>\hspace{-0.08cm}\epsilon_{admm}$ $\mathrm{\mathbf{or}}$ $\sum_{l \in \mathcal{L}}\hspace{-0.08cm}||\mathbf{q}_{r,l}^{t}||\hspace{-0.08cm}>\hspace{-0.08cm}\epsilon_{admm}$}{
				$\mathbf{v}_{r}^{t+1} \gets \argmin_{\mathbf{v}_{r}}(\bar{f}(\mathbf{v}_{r},\boldsymbol{\Lambda}^{\prime},\boldsymbol{\omega},\mathbf{g},\mathbf{B})+\frac{\zeta}{2}\sum_{l \in \mathcal{L}}||\mathbf{v}_{r,l}-\mathbf{u}^{t}_{r,l}+\mathbf{w}^{t}_{r,l}||^{2} )$ by solving \eqref{eqn:v_update_3} via interior-point methods\\
				$\mathbf{u}^{t+1}_{r} \gets \argmin_{\mathbf{u}_{r}}(\bar{\Pi}(\mathbf{u}_{r},\boldsymbol{\Lambda}^{\prime},\boldsymbol{\omega},\mathbf{g},\mathbf{B}) +\frac{\zeta}{2}\sum_{l \in \mathcal{L}}||\mathbf{v}^{t+1}_{r,l}-\mathbf{u}_{r,l}+\mathbf{w}^{t}_{r,l}||^{2} )$ by solving \eqref{eqn:problem_umin_sdr} via interior-point methods\\
				$\mathbf{w}^{t+1}_{r,l}=\mathbf{w}^{t}_{r,l}+\mathbf{v}^{t+1}_{r,l}-\mathbf{u}^{t+1}_{r,l}, \quad \forall l \in \mathcal{L}$\\
				$\mathbf{r}^{t+1}_{r,l} \gets  \mathbf{v}^{t+1}_{r,l}-\mathbf{u}^{t+1}_{r,l}, \quad \forall l \in \mathcal{L}$\\
				$\mathbf{q}^{t+1}_{r,l} \gets \mathbf{u}^{t+1}_{r,l}-\mathbf{u}^{t}_{r,l}, \quad \forall l \in \mathcal{L}$\\
				$t \gets t + 1$\\
			}
			\Return $\mathbf{u}_{r}^{t}$
	\end{algorithm}
	\vspace{-0.5cm}
\end{figure}
Then, we substitute \eqref{eqn:xibar} into $\bar{f}(\mathbf{v}, \boldsymbol{\Lambda}^{\prime}, \mathbf{B})=\frac{1}{L}\sum_{l \in \mathcal{L}} (-\bar{C}_{l}+\sum_{k \in \mathcal{K}}\bar{\xi}_{k,l})$ to write 
	\begin{align}
		\bar{f}&(\mathbf{v}, \boldsymbol{\Lambda}^{\prime}, \boldsymbol{\omega},\mathbf{g}, \mathbf{B})\hspace{-0.1cm}=\nonumber \\
		&\frac{1}{L}\sum_{l \in \mathcal{L}}\left[ -\bar{C}_{l}+\sum_{k \in \mathcal{K}}\left(\sum_{i \in \mathcal{K}}\mathbf{p}_{i,l}^{H}\boldsymbol{\Lambda}^{\prime}\boldsymbol{\Delta}\bar{\mathbf{A}}_{k,l}\boldsymbol{\Delta}\boldsymbol{\Lambda}^{\prime}\mathbf{p}_{i,l}+\bar{q}_{k,l}\sigma_{n}^{2}\right.\right.\nonumber \\
		&\left.\left.+\tr(\bar{\mathbf{A}}_{k,l}\boldsymbol{\Lambda}^{\prime}\boldsymbol{\Sigma}\boldsymbol{\Lambda}^{\prime})\hspace{-0.05cm}-\hspace{-0.05cm}2\mathcal{R}\lbrace\bar{\mathbf{f}}^{H}_{k,l}\boldsymbol{\Delta}\boldsymbol{\Lambda}^{\prime}\mathbf{p}_{k,l}\rbrace\hspace{-0.08cm}+\hspace{-0.05cm}\bar{\omega}_{k,l}\hspace{-0.05cm}-\hspace{-0.05cm}\bar{d}_{k,l}\right)\hspace{-0.05cm}\right]\hspace{-0.09cm},
		\label{eqn:f_new}
	\end{align}
	where the vector $\boldsymbol{\omega}$ is composed of the elements $\omega^{(m)}_{c,k,l}$ and $\omega^{(m)}_{k,l}$, and the vector $\mathbf{g}$ is composed of the elements $g^{(m)}_{c,k,l}$ and $g^{(m)}_{k,l}$, $\forall k \in \mathcal{K}$, $\forall l \in \mathcal{L}$, and, $\forall m \in \mathcal{M}$. Finally, substituting \eqref{eqn:f_new} in \eqref{eqn:v_update_2}, we can formulate the update step for $\mathbf{v}$ as
	\begin{align}
		\min_{\mathbf{v}_{r}}&(\frac{1}{L}\sum_{l \in \mathcal{L}}\left[ -\bar{C}_{l}+\sum_{k \in \mathcal{K}}\left(\sum_{i \in \mathcal{K}}\mathbf{p}_{i,l}^{H}\boldsymbol{\Lambda}^{\prime}\boldsymbol{\Delta}\bar{\mathbf{A}}_{k,l}\boldsymbol{\Delta}\boldsymbol{\Lambda}^{\prime}\mathbf{p}_{i,l}+\bar{q}_{k,l}\sigma_{n}^{2}\right.\right.\nonumber \\
		&\left.\left.+\tr(\bar{\mathbf{A}}_{k,l}\boldsymbol{\Lambda}^{\prime}\boldsymbol{\Sigma}\boldsymbol{\Lambda}^{\prime})\hspace{-0.05cm}-\hspace{-0.05cm}2\mathcal{R}\lbrace\bar{\mathbf{f}}^{H}_{k,l}\boldsymbol{\Delta}\boldsymbol{\Lambda}^{\prime}\mathbf{p}_{k,l}\rbrace\hspace{-0.08cm}+\hspace{-0.05cm}\bar{\omega}_{k,l}\hspace{-0.05cm}-\hspace{-0.05cm}\bar{d}_{k,l}\right)\hspace{-0.05cm}\right]\nonumber\\
		&+\frac{\zeta}{2}\hspace{-0.1cm}\sum_{l \in \mathcal{L}}\hspace{-0.05cm}||\mathbf{v}_{r,l}-\mathbf{u}^{t}_{r,l}+\mathbf{w}^{t}_{r,l}||^{2}_{2} ),  \label{eqn:v_update_3} 
	\end{align}
	which is convex for given $\boldsymbol{\omega}$ and $\mathbf{g}$, and can be solved by interior-point methods. We note the abuse of notation in \eqref{eqn:v_update_3} , where $\mathbf{v}_{r}$ and $\mathbf{p}_{i,l}$ are used interchangeably for the sake of simplicity, although they both represent the precoder coefficients to be optimized.

\subsubsection{Update Step for $\mathbf{u}$ :}
One can note that for given $\mathbf{B}$ and $\boldsymbol{\Lambda}^{\prime}$, the constraints \eqref{eqn:common_rate_2}, \eqref{eqn:comms}, and \eqref{eqn:radar_power_1} in the update step \eqref{eqn:u_update} are not convex with respect to $\mathbf{P}_{l}$. We first address the non-convexity of the constraints \eqref{eqn:common_rate_2} and \eqref{eqn:comms}. As done in the previous subsection, we express the rate expressions in terms of augmented MSEs, and write the constraints as
\begin{subequations}
		\begin{alignat}{3}
			&1-\bar{C}_{l} \geq \bar{\xi}_{c,k,l}, \  \forall k \in \mathcal{K}, \ \forall l \in \mathcal{L},  \label{eqn:common_rate_1_f}  \\
			&\frac{1}{L}\sum_{l \in \mathcal{L}}\left( - \bar{C}_{l}+\sum_{k \in \mathcal{K}}\bar{\xi}_{k,l} \right)\leq K-R_{th},  \label{eqn:comms_1_f}
		\end{alignat}
	\end{subequations}
	respectively, and denote the corresponding indicator function as $\bar{\Pi}(\mathbf{u}_{r}, \boldsymbol{\Lambda}^{\prime}, \mathbf{B})$.
Then, we rewrite the constraints \eqref{eqn:common_rate_1_f} and \eqref{eqn:comms_1_f} in terms of the terms in \eqref{eqn:allsafs} for the indicator function $\bar{\Pi}(\mathbf{u}_{r}, \boldsymbol{\Lambda}^{\prime}, \boldsymbol{\omega},\mathbf{g}, \mathbf{B})$ as 
\begin{subequations}
	\begin{alignat}{3}
		1-\bar{C}_{l}& \geq \mathbf{p}_{c,l}^{H}\boldsymbol{\Lambda}^{\prime}\boldsymbol{\Delta}\bar{\mathbf{A}}_{c,k,l}\boldsymbol{\Delta}\boldsymbol{\Lambda}^{\prime}\mathbf{p}_{c,l}\hspace{-0.1cm}+\hspace{-0.1cm}\sum_{i \in \mathcal{K}}\mathbf{p}_{i,l}^{H}\boldsymbol{\Lambda}^{\prime}\boldsymbol{\Delta}\bar{\mathbf{A}}_{c,k,l}\boldsymbol{\Delta}\boldsymbol{\Lambda}^{\prime}\mathbf{p}_{i,l}\nonumber\\ 
		&+\bar{q}_{c,k,l}\sigma_{n}^{2}+\tr(\bar{\mathbf{A}}_{c,k,l}\boldsymbol{\Lambda}^{\prime}\boldsymbol{\Sigma}\boldsymbol{\Lambda}^{\prime})-2\mathcal{R}\lbrace\bar{\mathbf{f}}^{H}_{c,k,l}\boldsymbol{\Delta}\boldsymbol{\Lambda}^{\prime}\mathbf{p}_{c,l}\rbrace\nonumber \\
		&+\bar{\omega}_{c,k,l}-\bar{d}_{c,k,l}, \  \forall k \in \mathcal{K}, \ \forall l \in \mathcal{L},  \label{eqn:common_rate_1_f_saf}  \\
		&\hspace{-1.1cm}\frac{1}{L}\sum_{l \in \mathcal{L}}\left[-\bar{C}_{l}\hspace{-0.1cm}+\hspace{-0.1cm}\sum_{k \in \mathcal{K}}\hspace{-0.1cm}\left(\sum_{i \in \mathcal{K}}\hspace{-0.1cm}\mathbf{p}_{i,l}^{H}\boldsymbol{\Lambda}^{\prime}\boldsymbol{\Delta}\bar{\mathbf{A}}_{k,l}\boldsymbol{\Delta}\boldsymbol{\Lambda}^{\prime}\mathbf{p}_{i,l}+\bar{q}_{k,l}\sigma_{n}^{2}\right.\right.\nonumber \\
		&\left.\left.+\tr(\bar{\mathbf{A}}_{k,l}\boldsymbol{\Lambda}^{\prime}\boldsymbol{\Sigma}\boldsymbol{\Lambda}^{\prime})-2\mathcal{R}\lbrace\bar{\mathbf{f}}^{H}_{k,l}\boldsymbol{\Delta}\boldsymbol{\Lambda}^{\prime}\mathbf{p}_{k,l}\rbrace+\bar{\omega}_{k,l}-\bar{d}_{k,l}\right)\right]\nonumber \\
		&\hspace{4.8cm} \leq K-R_{th},  \label{eqn:comms_1_f_saf}
	\end{alignat}
\end{subequations}

\begin{figure}[t!]
	\removelatexerror
	\begin{algorithm}[H]
		\caption{AO-ADMM-Based Algorithm}
		\label{alg:ao_admm}
		$t \gets 1$, $\mathbf{v}_{r}^{0}$, $\mathbf{u}_{r}^{0}$,  $\mathbf{w}_{r}^{0}$, $\boldsymbol{\Lambda}^{\prime}$, $\mathbf{B}$\\
		\While{$|\mathrm{EE}^{t}-\mathrm{EE}^{t-1}|> \epsilon_{ao}$}{
			$\boldsymbol{\omega}^{t} \gets$ updateWeights$\left(\mathbf{u}^{t-1}_{r},  \boldsymbol{\Lambda}^{\prime}\right)$ \\
			$\boldsymbol{g}^{t} \gets$ updateFilters$\left(\mathbf{u}^{t-1}_{r},  \boldsymbol{\Lambda}^{\prime}\right)$ \\
			$[\mathbf{u}^{t}_{r},\mathbf{w}_{r}^{t}]\gets \mathrm{\mathbf{ALG1}}(\boldsymbol{\Lambda}^{\prime},\mathbf{B}, \mathbf{u}_{r}^{t-1}, \mathbf{w}_{r}^{0}, \boldsymbol{\omega}^{t}, \boldsymbol{g}^{t})$\\
			$\mathrm{EE}^{t+1} \gets$ updateEE$\left(\mathbf{u}^{t}_{r}\right)$ \\
			$t \gets t + 1$\\
		}
		\Return $\mathbf{u}^{t}_{r}$
	\end{algorithm}
	\vspace{-0.5cm}
\end{figure}
Next, we propose the SDR method to deal with the non-convex constraint \eqref{eqn:radar_power_1} \cite{ma_2010}. 
Consider the following formulation for the minimization step for $\mathbf{u}$ expressed in terms of the complex-valued parameters as in \eqref{eqn:problem_umin}, where the defined matrices are
\begin{table*}
\begin{subequations}
	\begin{alignat}{3}
		\min_{\mathbf{u}}&     \quad  \sum_{l \in \mathcal{L}}||\mathbf{v}^{t+1}_{l}-\mathbf{u}_{l}+\mathbf{w}^{t}_{l}||^{2}_{2}         \\
		\text{s.t.}& \quad 1-\mathbf{e}^{T}_{N_{t}(K+1)+1,1}\mathbf{u}_{l}\geq \nonumber \\
		&\mathbf{u}_{l}^{H}\widehat{\mathbf{D}}^{H}_{c}\boldsymbol{\Lambda}^{\prime}\boldsymbol{\Delta}\bar{\mathbf{A}}_{c,k,l}\boldsymbol{\Delta}\boldsymbol{\Lambda}^{\prime}\widehat{\mathbf{D}}_{c}\mathbf{u}_{l}\hspace{-0.1cm}+\hspace{-0.1cm}\sum_{i \in \mathcal{K}}\hspace{-0.1cm}\mathbf{u}_{l}^{H}\widehat{\mathbf{D}}^{H}_{i}\boldsymbol{\Lambda}^{\prime}\boldsymbol{\Delta}\bar{\mathbf{A}}_{c,k,l}\boldsymbol{\Delta}\boldsymbol{\Lambda}^{\prime}\widehat{\mathbf{D}}_{c}\mathbf{u}_{l}\hspace{-0.05cm}+\hspace{-0.05cm}\bar{q}_{c,k,l}\sigma_{n}^{2}\hspace{-0.05cm}+\hspace{-0.05cm}\tr(\bar{\mathbf{A}}_{c,k,l}\boldsymbol{\Lambda}^{\prime}\boldsymbol{\Sigma}\boldsymbol{\Lambda}^{\prime})\hspace{-0.05cm}-\hspace{-0.05cm}2\mathcal{R}\lbrace\bar{\mathbf{f}}^{H}_{c,k,l}\boldsymbol{\Delta}\boldsymbol{\Lambda}^{\prime}\widehat{\mathbf{D}}_{c}\mathbf{u}_{l}\rbrace+\bar{\omega}_{c,k,l}-\bar{d}_{c,k,l}, \label{eqn:u_upd_1_domain}\\
		& \quad ||\sum_{l \in \mathcal{L}}\widehat{\mathbf{D}}_{c}\widehat{\boldsymbol{\Delta}}\widehat{\mathbf{\Lambda}}^{\prime}\mathbf{u}_{l}\mathbf{u}_{l}^{H}\widehat{\mathbf{\Lambda}}^{\prime}\widehat{\boldsymbol{\Delta}}\widehat{\mathbf{D}}_{c}^{H}+\sum_{l \in \mathcal{L}}\sum_{k \in \mathcal{K}}\widehat{\mathbf{D}}_{k}\widehat{\boldsymbol{\Delta}}\widehat{\mathbf{\Lambda}}^{\prime}\mathbf{u}_{l}\mathbf{u}_{l}^{H}\widehat{\mathbf{\Lambda}}^{\prime}\widehat{\boldsymbol{\Delta}}\widehat{\mathbf{D}}_{k}^{H}+L\boldsymbol{\Lambda}^{\prime}\boldsymbol{\Sigma}\boldsymbol{\Lambda}^{\prime}-\mathbf{U} ||^{2}_{2} \leq \tau, \label{eqn:4_domain} \\
		& \frac{1}{L}\sum_{l \in \mathcal{L}}\left[-\mathbf{e}^{T}_{N_{t}(K+1)+1,1}\mathbf{u}_{l}\hspace{-0.1cm}+\hspace{-0.1cm}\sum_{k \in \mathcal{K}}\hspace{-0.1cm}\left(\sum_{i \in \mathcal{K}}\hspace{-0.1cm}\boldsymbol{\Lambda}^{\prime}\boldsymbol{\Delta}\mathbf{u}_{l}^{H}\widehat{\mathbf{D}}^{H}_{i}\bar{\mathbf{A}}_{k,l}\boldsymbol{\Delta}\boldsymbol{\Lambda}^{\prime}\widehat{\mathbf{D}}_{i}\mathbf{u}_{l})+\bar{q}_{k,l}\sigma_{n}^{2}+\tr(\bar{\mathbf{A}}_{k,l}\boldsymbol{\Lambda}^{\prime}\boldsymbol{\Sigma}\boldsymbol{\Lambda}^{\prime})-2\mathcal{R}\lbrace\bar{\mathbf{f}}^{H}_{k,l}\boldsymbol{\Delta}\boldsymbol{\Lambda}^{\prime}\widehat{\mathbf{D}}_{k}\mathbf{u}_{l}\rbrace+\bar{\omega}_{k,l}-\bar{d}_{k,l}\right)\right] \nonumber \\
		& \hspace{15cm}\leq K-R_{th}, \label{eqn:45_domain} \\
		&   \mathrm{tr}\left(\mathbf{E}_{N_{t},i}\left((\widehat{\mathbf{D}}_{c}\widehat{\boldsymbol{\Delta}}\mathbf{u}_{l})(\widehat{\mathbf{D}}_{c}\widehat{\boldsymbol{\Delta}}\mathbf{u}_{l})^{H}+\sum_{k \in \mathcal{K}}(\widehat{\mathbf{D}}_{k}\widehat{\boldsymbol{\Delta}}\mathbf{u}_{l})(\widehat{\mathbf{D}}_{k}\widehat{\boldsymbol{\Delta}}\mathbf{u}_{l})^{H}+\boldsymbol{\Sigma}\right)\right) = P_{ant},  \quad    i \in \left\lbrace 1,2,\ldots, N_{t} \right\rbrace, \  \forall l \in \mathcal{L}\label{eqn:5_domain}.
	\end{alignat}
	\label{eqn:problem_umin}
\end{subequations}
\hrule
\end{table*}
\begin{align}
	\widehat{\boldsymbol{\Lambda}}^{\prime}&=\mathrm{Diag}([0, \underbrace{\lambda^{\prime}_{1}, \lambda^{\prime}_{2}, \ldots, \lambda^{\prime}_{N_{t}}, \ldots, \lambda^{\prime}_{1}, \lambda^{\prime}_{2}, \ldots, \lambda^{\prime}_{N_{t}}}_{(K+1)N_{t}}]) \nonumber \\
	\widehat{\boldsymbol{\Delta}}&=\mathrm{Diag}([0, \underbrace{\delta_{1}, \delta_{2}, \ldots, \delta_{N_{t}}, \ldots, \delta_{1}, \delta_{2}, \ldots, \delta_{N_{t}}}_{(K+1)N_{t}}]) \nonumber \\
	\widehat{\mathbf{D}}_{c}&=[\mathbf{0}^{N_{t}\times 1}, \ \mathbf{I}_{N_{t}}, \ \mathbf{0}^{N_{t}\times KN_{t}}], \nonumber \\
	\widehat{\mathbf{D}}_{k}&=[\mathbf{0}^{N_{t}\times(1+kN_{t})}, \ \mathbf{I}_{N_{t}}, \ \mathbf{0}^{N_{t}\times (K-k)N_{t}}]. \nonumber
\end{align}
and $\mathbf{e}_{N_{t}(K+1)+1,k}$ is the $k$-th standard basis vector of length $N_{t}(K+1)+1$. The inhomogeneous Quadratically Constrained Quadratic
Program (QCQP) formulation \eqref{eqn:problem_umin} can be transformed into an equivalent homogeneous QCQP. 
Accordingly, we define the matrices, 
\begin{align}
	\mathbf{O}_{l}&=[\mathbf{u}^{T}_{l}, a_{l}]^{T}([\mathbf{u}^{T}_{l}, a_{l}]^{T})^{H}, \nonumber \\
	\mathbf{J}_{l}&=
	\begin{bmatrix}
		\mathbf{I}_{N_{t}(K+1)+1} & -(\mathbf{v}^{t+1}_{l}+\mathbf{w}^{t}_{l})\\
		-(\mathbf{v}^{t+1}_{l}+\mathbf{w}^{t}_{l})^{H} & ||\mathbf{v}^{t+1}_{l}+\mathbf{w}^{t}_{l}||^{2} 
	\end{bmatrix} \nonumber \\
	\widehat{\mathbf{T}}_{c,k,l}&=\mathrm{Diag}([0, \underbrace{\boldsymbol{\Lambda}^{\prime}\boldsymbol{\Delta}\bar{\mathbf{A}}_{c,k,l}\boldsymbol{\Delta}\boldsymbol{\Lambda}^{\prime}, \ldots, \boldsymbol{\Lambda}^{\prime}\boldsymbol{\Delta}\bar{\mathbf{A}}_{c,k,l}\boldsymbol{\Delta}\boldsymbol{\Lambda}^{\prime}}_{K+1}]), \nonumber \\
	\mathbf{T}_{c,k,l}&=
	\begin{bmatrix}
		 \hspace{-3.9cm}\widehat{\mathbf{T}}_{c,k,l}&\hspace{-3.7cm} -(\widehat{\mathbf{D}}_{c}^{H}\boldsymbol{\Lambda}^{\prime}\boldsymbol{\Delta}\bar{\mathbf{f}}_{c,k,l}-0.5\mathbf{e}_{N_{t}(K+1)+1,1})\\
		-(\bar{\mathbf{f}}^{H}_{c,k,l}\boldsymbol{\Delta}\boldsymbol{\Lambda}^{\prime}\widehat{\mathbf{D}}_{c}-0.5\mathbf{e}^{T}_{N_{t}(K+1)+1,1}) & 0 
	\end{bmatrix} \nonumber \\
	\widehat{\mathbf{T}}_{k,l}&=\mathrm{Diag}([\underbrace{0,\ldots,0}_{N_{t}+1}, \underbrace{\boldsymbol{\Lambda}^{\prime}\boldsymbol{\Delta}\bar{\mathbf{A}}_{k,l}\boldsymbol{\Delta}\boldsymbol{\Lambda}^{\prime}, \ldots, \boldsymbol{\Lambda}^{\prime}\boldsymbol{\Delta}\bar{\mathbf{A}}_{k,l}\boldsymbol{\Delta}\boldsymbol{\Lambda}^{\prime}}_{K}]), \nonumber \\
	\mathbf{T}_{k,l}&=
	\begin{bmatrix}
		\hspace{-3.9cm}\widehat{\mathbf{T}}_{k,l}&\hspace{-3.7cm} -(\widehat{\mathbf{D}}_{k}^{H}\boldsymbol{\Lambda}^{\prime}\boldsymbol{\Delta}\bar{\mathbf{f}}_{k,l}+\frac{0.5}{K}\mathbf{e}_{N_{t}(K+1)+1,1})\\
		-(\bar{\mathbf{f}}^{H}_{k,l}\boldsymbol{\Delta}\boldsymbol{\Lambda}^{\prime}\widehat{\mathbf{D}}_{k}+\frac{0.5}{K}\mathbf{e}^{T}_{N_{t}(K+1)+1,1}) & 0 
	\end{bmatrix}, \nonumber \\
	\breve{\boldsymbol{\Lambda}}^{\prime}&=
	\begin{bmatrix}
		\widehat{\boldsymbol{\Lambda}}^{\prime}&\mathbf{0}^{N_{t}(K+1)+1 \times 1}\\
		\mathbf{0}^{ 1 \times N_{t}(K+1)+1} & 0 
	\end{bmatrix}, \nonumber \\
	\breve{\boldsymbol{\Delta}}&=
	\begin{bmatrix}
		\widehat{\boldsymbol{\Delta}}&\mathbf{0}^{N_{t}(K+1)+1 \times 1}\\
		\mathbf{0}^{ 1 \times N_{t}(K+1)+1} & 0 
	\end{bmatrix}, \nonumber \\
	\mathbf{D}_{c}&=[\widehat{\mathbf{D}}_{c}, \  \mathbf{0}^{N_{t} \times 1}], \nonumber \\
	\mathbf{D}_{k}&=[\widehat{\mathbf{D}}_{k}, \  \mathbf{0}^{N_{t} \times 1}].  \nonumber
\end{align}	
Then, the formulation in \eqref{eqn:problem_umin} can be rewritten as
\begin{subequations}
	\begin{alignat}{3}
		&\min_{\mathbf{O}_{1}, \ldots, \mathbf{O}_{L}}     \quad  \sum_{l \in \mathcal{L}}\tr{(\mathbf{J}_{l}\mathbf{O}_{l})}         \\
		&\quad \text{s.t.} \ \ \ \mathbf{O}_{l} \succcurlyeq 0,  \ \forall l \in \mathcal{L},\\
		&\quad\quad\quad\tr{(\mathbf{T}_{c,k,l}\mathbf{O}_{l})}+\bar{q}_{c,k,l}\sigma_{n}^{2}+\tr(\bar{\mathbf{A}}_{c,k,l}\breve{\boldsymbol{\Lambda}}^{\prime}\boldsymbol{\Sigma}\breve{\boldsymbol{\Lambda}}^{\prime}) \nonumber \\
		&\quad\quad\quad\quad+\bar{\omega}_{c,k,l}-\bar{d}_{c,k,l}  \leq  1, \ \forall k \in \mathcal{K}, \ \forall l \in \mathcal{L}, \label{eqn:u_upd_1_domain_sdr}\\
		& \quad\quad\quad ||\sum_{l \in \mathcal{L}}\mathbf{D}_{c}\breve{\boldsymbol{\Delta}}\breve{\mathbf{\Lambda}}^{\prime}\mathbf{O}_{l}\breve{\mathbf{\Lambda}}^{\prime}\breve{\boldsymbol{\Delta}}\mathbf{D}_{c}^{H}+L\breve{\boldsymbol{\Lambda}}^{\prime}\boldsymbol{\Sigma}\breve{\boldsymbol{\Lambda}}^{\prime}\nonumber \\
		&\quad\quad\quad +\sum_{l \in \mathcal{L}}\sum_{k \in \mathcal{K}}\mathbf{D}_{k}\breve{\boldsymbol{\Delta}}\breve{\mathbf{\Lambda}}^{\prime}\mathbf{O}_{l}\breve{\mathbf{\Lambda}}^{\prime}\breve{\boldsymbol{\Delta}}\mathbf{D}_{k}^{H}\hspace{-0.1cm}-\hspace{-0.1cm}\mathbf{U}||^{2}_{2} \leq \tau, \label{eqn:4_domain_sdr} \\
		& \quad\quad\quad \frac{1}{L}\sum_{l \in \mathcal{L}} \sum_{k \in \mathcal{K}}(\tr{(\mathbf{T}_{k,l}\mathbf{O}_{l})}+\bar{q}_{k,l}\sigma_{n}^{2}+\tr(\bar{\mathbf{A}}_{k,l}\breve{\boldsymbol{\Lambda}}^{\prime}\boldsymbol{\Sigma}\breve{\boldsymbol{\Lambda}}^{\prime}) \nonumber \\
		& \quad\quad\quad\quad\quad\quad\quad\quad\quad\quad+\bar{\omega}_{k,l}-\bar{d}_{k,l})
		\leq K-R_{th}, \label{eqn:45_domain_sdr} \\
		&  \quad\quad\quad \mathrm{tr}(\mathbf{E}_{N_{t},i}(\mathbf{D}_{c}\breve{\boldsymbol{\Delta}}\mathbf{O}_{l}\breve{\boldsymbol{\Delta}}\mathbf{D}_{c}^{H}+\sum_{k \in \mathcal{K}}\mathbf{D}_{k}\breve{\boldsymbol{\Delta}}\mathbf{O}_{l}\breve{\boldsymbol{\Delta}}\mathbf{D}_{k}^{H}+\boldsymbol{\Sigma})) \nonumber \\ 
		&\quad\quad\quad\quad\quad  = P_{ant}, \ \  \forall i \in \left\lbrace 1,2,\ldots, N_{t} \right\rbrace, \  \forall l \in \mathcal{L}\label{eqn:5_domain_sdr} ,\\
		&\quad\quad\quad \tr{(\mathbf{O}_{l}\mathbf{E}_{N_{t}(K+1)+2,N_{t}(K+1)+2})}=1, \ \forall l \in \mathcal{L}.
	\end{alignat}
	\label{eqn:problem_umin_sdr}
\end{subequations}

Note that the rank-1 constraint for $\mathbf{O}_{l}$ is omitted in \eqref{eqn:problem_umin_sdr} in accordance with the SDR method. The formulation in \eqref{eqn:problem_umin_sdr} is a Semi-Definite Problem (SDP) and can be solved by interior-point methods. We describe the ADMM-based algorithm to perform the update steps for $\mathbf{v}$, $\mathbf{u}$, and $\mathbf{w}$ for given ${\boldsymbol{\omega}}$ and $\mathbf{g}$ in Alg.~\ref{alg:admm}.
Alg.~\ref{alg:ao_admm} describes the AO-ADMM-based algorithm to solve the updates using the ADMM-based in Alg.~\ref{alg:admm} and $\boldsymbol{\omega}$ and $\mathbf{g}$ calculations. 
The outer iterations of the AO algorithm serve to update the MSE weights $\boldsymbol{\omega}^{t}$ and equalizers $\mathbf{g}^{t}$ based on the precoders calculated by the ADMM-based algorithm at iteration-$t-1$. By the calculated $\boldsymbol{\omega}^{t}$ and $\mathbf{g}^{t}$, the ADMM-based algorithm performs the update steps for $\mathbf{v}$, $\mathbf{u}$, and $\mathbf{w}$ at iteration-$t$. 

\textit{Proposition 2:}
For given $\boldsymbol{\Lambda}^{\prime}$, the proposed algorithm in Alg.~\eqref{alg:ao_admm} converges to a stationary point of the problem \eqref{eqn:problem3_v2}.

\textit{Proof:}
We prove the convergence of the proposed AO-ADMM-based algorithm in Alg.~\eqref{alg:ao_admm} over the real-valued equivalent definitions of the functions. First, we consider the inner problem in Alg.~\eqref{alg:ao_admm} at iteration-$t$, which is solved by the ADMM-based algorithm in Alg.~\eqref{alg:admm} and can be written as, 
	\begin{alignat}{3}
		\min_{\mathbf{v}_{r}}&  \quad  \bar{f}(\mathbf{v_{r}}, \boldsymbol{\Lambda}^{\prime},\boldsymbol{\omega}^{t},\mathbf{g}^{t}, \mathbf{B})  \nonumber \\
		\text{s.t.}& \quad  \mathbf{v}_{r} \in \mathcal{D}^{t}, \nonumber
	\end{alignat}
\hspace{-0.2cm}where $\mathcal{D}^{t}$ represents the problem domain at iteration-$t$ of the AO algorithm. It can be shown that the function $\bar{f}(\mathbf{v_{r}}, \boldsymbol{\Lambda}^{\prime},\boldsymbol{\omega}^{t},\mathbf{g}^{t}, \mathbf{B})$ is Lipschitz differentiable \cite{wang_2019} and $\mathcal{D}^{t}$ is a compact set. Then, the solution sequence
($\mathbf{v}^{t}_{r}$ , $\mathbf{u}^{t}_{r}$, $\mathbf{w}^{t}_{r}$) has at least one limit point, and each limit point is a stationary point of $\mathcal{L}_{\zeta}$ for any sufficiently large $\zeta$ \cite[Cor. 2]{wang_2019}. This is also valid when the subproblems are solved inexactly with summable errors, which is a condition satisfied by the SDR method \cite{zuo_2010}, assuming a rank-1 solution exists.

Since the solution $\mathbf{u}^{t}_{r}$ at iteration-$t$ is also a feasible solution at iteration $t+1$, the $EE^{t}$ is non-decreasing with $t$ and is bounded above due to the per antenna power constraints, per antenna RF chain power consumption and the constant terms $P_{BB}$ and $P_{syn}$ in $\mathcal{E}(\mathbf{P}, \boldsymbol{\Lambda}^{\prime}, \mathbf{B})$ \cite{bjornson_2015} \cite[Sec. IV-F]{sun_2021}\cite{yi_2011}.
Given the convergence of the ADMM part for the inner iterations and $EE^{t}$ being bounded, the convergence of the AO algorithm for the outer iterations follow from \cite[Prop. 1]{clerckxTC2016}.\hspace{1.95cm}$\blacksquare$

We note that due to the non-convex nature of the problem, the global optimality of the solutions cannot be guaranteed and is dependent on the initializations of the precoders.   

\subsection{Proposed SCA/SDR-Based Algorithm for Optimal RF Chain Selection}
\label{sec:scasdr}
The subproblem of finding the optimal RF chain selection matrix $\boldsymbol{\Lambda}$ for given precoders, $\mathbf{P}^{\prime}_{l}$, $\forall l \in \mathcal{L}$, can be written as  
\begin{subequations}
	\begin{alignat}{3}
		&\max_{ \bar{\mathbf{C}}, \boldsymbol{\Lambda}}   \   \frac{1}{L}\sum_{l \in \mathcal{L}}\frac{(\bar{C}_{l}(\mathbf{P}^{\prime}_{l}, \boldsymbol{\Lambda}, \mathbf{B}
			)+\sum_{k \in \mathcal{K}}\bar{R}_{k,l}(\mathbf{P}^{\prime}_{l}, \boldsymbol{\Lambda}, \mathbf{B}))}{P_{tot}(\boldsymbol{\Lambda},  \mathbf{B})}    \\
		&\  \text{s.t.}  \  \bar{C}_{l}(\mathbf{P}^{\prime}_{l}, \boldsymbol{\Lambda}, \mathbf{B}) \leq \bar{R}_{c,k,l}(\mathbf{P}^{\prime}_{l}, \boldsymbol{\Lambda}, \mathbf{B}),  k \in \mathcal{K},  \forall l \in \mathcal{L} \label{eqn:common_rate_25} \\
		& \quad\quad \ \ \ \lambda_{i} \in \left\{0,1 \right\},\    \forall i \in \left\lbrace 1,2,\ldots, N_{t} \right\rbrace,  \label{eqn:3_nonconvex} \\
		& \quad\quad \quad\Bigg|\Bigg|\widehat{\mathbf{R}}_{\tilde{\mathbf{x}}}\left(\mathbf{P}^{\prime}, \boldsymbol{\Lambda}, \mathbf{B}\right)-\mathbf{U} \Bigg|\Bigg|^{2}_{2} \leq \tau
		\label{eqn:4} \\
		& \quad\quad\quad \bar{R}_{sum}(\mathbf{P}^{\prime}, \boldsymbol{\Lambda}, \mathbf{B}) \geq R_{th}. \label{eqn:45}
	\end{alignat}
	\label{eqn:problem4}
\end{subequations}

The problem \eqref{eqn:problem4} is non-convex due to the non-convex objective function and constraints \eqref{eqn:common_rate_25}, \eqref{eqn:3_nonconvex}, and \eqref{eqn:45} with respect to $\boldsymbol{\Lambda}$. Moreover, it is stochastic in nature, as discussed in the previous subsection. 
First, we relax the constraint \eqref{eqn:3_nonconvex} as 
\begin{align}
	0 \leq \tilde{\lambda}_{i} \leq 1, 
	\label{eqn:lambdarelaxed}
\end{align}
and denote the corresponding vectors and matrices as $\tilde{\boldsymbol{\lambda}}$ and $\tilde{\boldsymbol{\Lambda}}$, respectively. 
Before moving to obtain a convex and deterministic equivalent formulation for the considered problem, we introduce the following lemma.

\textit{Lemma 1:} Define the vectors $\mathbf{a}$, $\mathbf{b}$, $\mathbf{c}$ $\in \mathbb{C}^{M\times1}$ and the matrix $\mathbf{D}_{\mathbf{b}}=\mathrm{Diag}(\mathbf{b}^{T})$. Then, the following holds
\begin{align}
	\mathbf{a}^{H}\mathbf{D}_{\mathbf{b}}\mathbf{c}=\mathbf{b}^{T}(\mathbf{a}^{*}\circ\mathbf{c}),
\end{align}
where $\circ$ represents the Hadamard product.

\textit{Proof:} We start by writing $\mathbf{a}^{H}\mathbf{D}_{\mathbf{b}}\mathbf{c}$ by using the Hadamard product as
\begin{align}
	\mathbf{a}^{H}\mathbf{D}_{\mathbf{b}}\mathbf{c}=\mathbf{a}^{H}(\mathbf{b}\circ\mathbf{c}). 
	\label{eqn:hadamard}
\end{align}
Using the property $\mathbf{x}^{H}(\mathbf{X}\circ\mathbf{Y})\mathbf{y}=\mathrm{tr}(\mathbf{D}^{H}_{\mathbf{x}}\mathbf{X}\mathbf{D}_{\mathbf{y}}\mathbf{Y}^{T})$, we write
\begin{align}
	\mathbf{a}^{H}(\mathbf{b}\circ\mathbf{c})=\mathrm{tr}(\mathbf{D}^{H}_{\mathbf{a}}\mathbf{b}\mathbf{c}^{T})=\mathbf{c}^{T}\mathbf{D}^{H}_{\mathbf{a}}\mathbf{b}=\mathbf{b}^{T}\mathbf{D}^{H}_{\mathbf{a}}\mathbf{c}. \nonumber
\end{align}
Finally, using the property in \eqref{eqn:hadamard} once again, we obtain
\begin{align}
	\mathbf{b}^{T}\mathbf{D}^{H}_{\mathbf{a}}\mathbf{c}=\mathbf{b}^{T}(\mathbf{a}^{*}\circ\mathbf{c}), 
	\label{eqn:hadamard_2}
\end{align}
which completes the proof. \hspace{4.5cm}$\blacksquare$

Now, we move to replace the non-convex and stochastic expressions in \eqref{eqn:problem4} with their convex and deterministic counterparts. We benefit from Lemma 1 to write the SAFs of rate expressions as given below.
		\begin{alignat}{3}
		&R^{(m)}_{k,l}(\mathbf{P}^{\prime}_{l}, \tilde{\boldsymbol{\Lambda}}, \mathbf{B}, \mathbf{h}^{(m)}_{k}
		) \nonumber \\
		&=\log\hspace{-0.1cm}\left(\hspace{-0.1cm}1\hspace{-0.1cm}+\hspace{-0.1cm}\frac{|\tilde{\boldsymbol{\lambda}}^{T}(\boldsymbol{\Delta}(\mathbf{h}^{(m)}_{k})^{*}\circ\mathbf{p}^{\prime}_{k,l})|^{2}}{\sigma_{n}^{2}\hspace{-0.1cm}+\hspace{-0.1cm}\tilde{\boldsymbol{\lambda}}^{T}\boldsymbol{\Phi}_{k}^{(m)}\tilde{\boldsymbol{\lambda}}\hspace{-0.1cm}+\hspace{-0.1cm}\sum_{\substack{k^{\prime}\in\mathcal{K}, \\ k^{\prime} \neq k}}|\tilde{\boldsymbol{\lambda}}^{T}(\boldsymbol{\Delta}(\mathbf{h}^{(m)}_{k})^{*}\circ\mathbf{p}^{\prime}_{k^{\prime},l})|^{2}}\hspace{-0.1cm}\right),  \nonumber \\
		&R^{(m)}_{c,k,l}(\mathbf{P}^{\prime}_{l}, \tilde{\boldsymbol{\Lambda}}, \mathbf{B}, \mathbf{h}^{(m)}_{k}
		) \nonumber \\
		&=\log\hspace{-0.1cm}\left(\hspace{-0.1cm}1\hspace{-0.1cm}+\hspace{-0.1cm}\frac{|\tilde{\boldsymbol{\lambda}}^{T}(\boldsymbol{\Delta}(\mathbf{h}^{(m)}_{k})^{*}\circ\mathbf{p}^{\prime}_{c,l})|^{2}}{\sigma_{n}^{2}\hspace{-0.1cm}+\hspace{-0.1cm}\tilde{\boldsymbol{\lambda}}^{T}\boldsymbol{\Phi}_{k}^{(m)}\tilde{\boldsymbol{\lambda}}\hspace{-0.1cm}+\hspace{-0.1cm}\sum_{k^{\prime}\in\mathcal{K}}|\tilde{\boldsymbol{\lambda}}^{T}(\boldsymbol{\Delta}(\mathbf{h}^{(m)}_{k})^{*}\circ\mathbf{p}^{\prime}_{k^{\prime},l})|^{2}}	\hspace{-0.1cm}\right),\nonumber
	\end{alignat}
\begin{table*}
\begin{subequations}
		\begin{alignat}{3}
			&\max_{\mathbf{C} , \tilde{\boldsymbol{\Lambda}}, t}   \   t    \\
			\quad \text{s.t.} &\  t \leq  \frac{1}{L}\sum_{l \in \mathcal{L}}\frac{\bar{C}_{l}}{P_{tot}(\tilde{\boldsymbol{\Lambda}},  \mathbf{B})}+\frac{1}{LM}\sum_{l \in \mathcal{L}}\sum_{m \in \mathcal{M}}\sum_{k \in \mathcal{K}}\frac{\ln\left(\sigma_{n}^{2}+\tilde{\boldsymbol{\lambda}}^{T}\boldsymbol{\Phi}_{k}^{(m)}\tilde{\boldsymbol{\lambda}}+\sum_{k^{\prime}\in\mathcal{K}}|\tilde{\boldsymbol{\lambda}}^{T}(\boldsymbol{\Delta}(\mathbf{h}^{(m)}_{k})^{*}\circ\mathbf{p}^{\prime}_{k^{\prime},l})|^{2}\right)}{(\ln2)P_{tot}(\tilde{\boldsymbol{\Lambda}},  \mathbf{B})} \nonumber \\
			&\hspace{6cm}-\frac{1}{LM}\sum_{l \in \mathcal{L}}\sum_{m \in \mathcal{M}}\sum_{k \in \mathcal{K}}\frac{\ln(\sigma_{n}^{2}+\tilde{\boldsymbol{\lambda}}^{T}\boldsymbol{\Phi}_{k}^{(m)}\tilde{\boldsymbol{\lambda}}+\sum_{\substack{k^{\prime}\in\mathcal{K}, \\ k^{\prime} \neq k}}|\tilde{\boldsymbol{\lambda}}^{T}(\boldsymbol{\Delta}(\mathbf{h}^{(m)}_{k})^{*}\circ\mathbf{p}^{\prime}_{k^{\prime},l})|^{2})}{(\ln2)P_{tot}(\tilde{\boldsymbol{\Lambda}},  \mathbf{B})},\label{eqn:new1} \\
			&\ (\ln2)\bar{C}_{l} \leq \frac{1}{M}\sum_{m \in \mathcal{M}}\hspace{-0.1cm}\ln\left(\sigma_{n}^{2}+\tilde{\boldsymbol{\lambda}}^{T}\boldsymbol{\Phi}_{k}^{(m)}\tilde{\boldsymbol{\lambda}}+\hspace{-0.4cm}\sum_{k^{\prime}\in\left\lbrace\mathcal{K},c\right\rbrace}\hspace{-0.3cm}|\tilde{\boldsymbol{\lambda}}^{T}(\boldsymbol{\Delta}(\mathbf{h}^{(m)}_{k})^{*}\circ\mathbf{p}^{\prime}_{k^{\prime},l})|^{2}\right)-\frac{1}{M}\sum_{l \in \mathcal{M}}\ln\hspace{-0.1cm}\left(\sigma_{n}^{2}+\tilde{\boldsymbol{\lambda}}^{T}\boldsymbol{\Phi}_{k}^{(m)}\tilde{\boldsymbol{\lambda}}+\hspace{-0.2cm}\sum_{k^{\prime}\in\mathcal{K}}|\tilde{\boldsymbol{\lambda}}^{T}(\boldsymbol{\Delta}(\mathbf{h}^{(m)}_{k})^{*}\circ\mathbf{p}^{\prime}_{k^{\prime},l})|^{2}\right)\hspace{-0.1cm}, \nonumber \\
			&\hspace{14.5cm} \forall k \in \mathcal{K}, \ \forall l \in \mathcal{L},   \label{eqn:common_rate_3} \\
			& \   \frac{1}{L}\sum_{l \in \mathcal{L}}(\ln2)\bar{C}_{l}+\frac{1}{LM}\sum_{l \in \mathcal{L}}\sum_{m \in \mathcal{M}}\sum_{k \in \mathcal{K}}\ln(\sigma_{n}^{2}+\tilde{\boldsymbol{\lambda}}^{T}\boldsymbol{\Phi}_{k}^{(m)}\tilde{\boldsymbol{\lambda}}+\sum_{k^{\prime}\in\mathcal{K}}|\tilde{\boldsymbol{\lambda}}^{T}(\boldsymbol{\Delta}(\mathbf{h}^{(m)}_{k})^{*}\circ\mathbf{p}^{\prime}_{k^{\prime},l})|^{2}) \nonumber \\
			&\hspace{5cm} -\frac{1}{LM}\sum_{l \in \mathcal{L}}\sum_{m \in \mathcal{M}}\sum_{k \in \mathcal{K}}\ln(\sigma_{n}^{2}+\tilde{\boldsymbol{\lambda}}^{T}\boldsymbol{\Phi}_{k}^{(m)}\tilde{\boldsymbol{\lambda}}+\sum_{\substack{k^{\prime}\in\mathcal{K}, \\ k^{\prime} \neq k}}|\tilde{\boldsymbol{\lambda}}^{T}(\boldsymbol{\Delta}(\mathbf{h}^{(m)}_{k})^{*}\circ\mathbf{p}^{\prime}_{k^{\prime},l})|^{2}) \geq (\ln2)R_{th},\label{eqn:rate_const_3} \\
			& \quad\quad\ \  \eqref{eqn:4}, \ \eqref{eqn:lambdarelaxed}.   \nonumber  
		\end{alignat}
		\label{eqn:problem5_eqv}
	\end{subequations}
	\hrule
\end{table*}
where $\boldsymbol{\Phi}_{k}^{(m)}=\mathrm{Diag}(\sigma^{2}_{e,1}|h^{(m)}_{k,1}|^{2}, \ldots, \sigma^{2}_{e,N_{t}}|h^{(m)}_{k,N_{t}}|^{2})$
and the second term in the denominator follows from the fact that $\mathbf{D}_{\mathbf{a}}\mathbf{b}=\mathbf{D}_{\mathbf{b}}\mathbf{a}$.

First, we write an equivalent problem as in \eqref{eqn:problem5_eqv}. 
Then, we introduce the slack variables $\varepsilon^{(m)}_{k,l}$ and $\nu^{(m)}_{k,l}$ and rewrite the constraint \eqref{eqn:common_rate_3} in 3 new constraints as
\begin{subequations}
	\begin{alignat}{3}
		&\ln\hspace{-0.1cm}\left(\sigma_{n}^{2}+\tilde{\boldsymbol{\lambda}}^{T}\boldsymbol{\Phi}_{k}^{(m)}\tilde{\boldsymbol{\lambda}}+\sum_{k^{\prime}\in\mathcal{K}}|\tilde{\boldsymbol{\lambda}}^{T}(\boldsymbol{\Delta}(\mathbf{h}^{(m)}_{k})^{*}\circ\mathbf{p}^{\prime}_{k^{\prime},l})|^{2}\hspace{-0.1cm}\right) \hspace{-0.1cm}\leq \varepsilon^{(m)}_{k,l},\nonumber \\
		&\hspace{3.3cm}\forall k \in \mathcal{K}, \ \forall l \in \mathcal{L}, \ \forall m \in \mathcal{M}, \label{eqn:1b_1}\\
		&\ln\hspace{-0.1cm}\left(\sigma_{n}^{2}+\tilde{\boldsymbol{\lambda}}^{T}\boldsymbol{\Phi}_{k}^{(m)}\tilde{\boldsymbol{\lambda}}+\hspace{-0.2cm}\sum_{k^{\prime}\in\left\lbrace\mathcal{K},c\right\rbrace}\hspace{-0.2cm}|\tilde{\boldsymbol{\lambda}}^{T}(\boldsymbol{\Delta}(\mathbf{h}^{(m)}_{k})^{*}\circ\mathbf{p}^{\prime}_{k^{\prime},l})|^{2}\hspace{-0.1cm}\right) \hspace{-0.1cm}\geq \nu^{(m)}_{k,l}, \nonumber \\
		&\hspace{3.3cm}\forall k \in \mathcal{K},\  \forall l \in \mathcal{L}, \ \forall m \in \mathcal{M}, \label{eqn:1c_1}\\
		&(\ln2)\bar{C}_{l} - \frac{1}{M}\hspace{-0.2cm}\sum_{m \in \mathcal{M}}\hspace{-0.15cm}(\nu^{(m)}_{k,l}\hspace{-0.1cm}+\hspace{-0.1cm}\varepsilon^{(m)}_{k,l}) \leq 0 , \  \forall k \in \mathcal{K}, \ \forall l \in \mathcal{L}. \label{eqn:1a}
	\end{alignat}
	\label{eqn:1}
\end{subequations}
Next, we simplify the expressions using a stacking method employed in \cite{choi_2020, an_2020}. Define the matrices
\begin{align}
	\boldsymbol{\Gamma}^{(m)}_{k,l}&=\sum_{k^{\prime}\in\mathcal{K}}(\boldsymbol{\Delta}(\mathbf{h}^{(m)}_{k})^{*}\circ\mathbf{p}^{\prime}_{k^{\prime},l})(\boldsymbol{\Delta}(\mathbf{h}^{(m)}_{k})^{*}\circ\mathbf{p}^{\prime}_{k^{\prime},l})^{H}\nonumber \\
	&+\sigma_{n}^{2}\mathbf{I}_{N_{t}}+\boldsymbol{\Phi}_{k}^{(m)}, \nonumber \\ \boldsymbol{\Xi}^{(m)}_{k,l}&=\sum_{k^{\prime}\in\left\lbrace\mathcal{K},c\right\rbrace}(\boldsymbol{\Delta}(\mathbf{h}^{(m)}_{k})^{*}\circ\mathbf{p}^{\prime}_{k^{\prime},l})(\boldsymbol{\Delta}(\mathbf{h}^{(m)}_{k})^{*}\circ\mathbf{p}^{\prime}_{k^{\prime},l})^{H}\nonumber \\
	&+\frac{\sigma_{n}^{2}}{N_{t}}\mathbf{I}_{N_{t}}+\boldsymbol{\Phi}_{k}^{(m)}. \nonumber
\end{align}
In order to progress, we make use of the boundaries for $||\tilde{\boldsymbol{\lambda}}||^{2}$. One can immediately note from \eqref{eqn:lambdarelaxed} that  $||\tilde{\boldsymbol{\lambda}}||^{2} \leq N_{t}$. We add the additional constraint 
\begin{align}
	||\tilde{\boldsymbol{\lambda}}||^{2} \geq 1
	\label{eqn:lambdalowerbound}
\end{align}
to serve as a lower bound for the optimization problem\footnote{Such a constraint is naturally satisfied by $||\boldsymbol{\lambda}||^{2}$, since at least one active RF chain is required for $\mathcal{E}(\mathbf{P},  \boldsymbol{\Lambda}, \mathbf{B})>0$.}. Accordingly, we can write
\begin{align}
	&\sigma_{n}^{2}\hspace{-0.1cm}+\hspace{-0.1cm}\tilde{\boldsymbol{\lambda}}^{T}\boldsymbol{\Phi}_{k}^{(m)}\tilde{\boldsymbol{\lambda}}\hspace{-0.1cm}+\hspace{-0.2cm}\sum_{k^{\prime}\in\mathcal{K}}\hspace{-0.15cm}|\tilde{\boldsymbol{\lambda}}^{T}(\boldsymbol{\Delta}\mathbf{h}_{k}^{*}\circ\mathbf{p}^{\prime}_{k^{\prime},l})|^{2} \leq \tilde{\boldsymbol{\lambda}}^{T}\boldsymbol{\Gamma}^{(m)}_{k,l}\tilde{\boldsymbol{\lambda}} \nonumber \\
	&\hspace{1.8cm}\leq e^{\varepsilon^{(m)}_{k,l}},\ \forall k \in \mathcal{K}, \ \forall l \in \mathcal{L},\ \forall m \in \mathcal{M},\label{eqn:nonconvex_11} \\
	&\sigma_{n}^{2}\hspace{-0.1cm}+\hspace{-0.1cm}\tilde{\boldsymbol{\lambda}}^{T}\boldsymbol{\Phi}_{k}^{(m)}\tilde{\boldsymbol{\lambda}}\hspace{-0.1cm}+\hspace{-0.5cm}\sum_{k^{\prime}\in\left\lbrace\mathcal{K},c\right\rbrace}\hspace{-0.35cm}|\tilde{\boldsymbol{\lambda}}^{T}(\boldsymbol{\Delta}\mathbf{h}_{k}^{*}\circ\mathbf{p}^{\prime}_{k^{\prime},l})|^{2}\geq \tilde{\boldsymbol{\lambda}}^{T}\boldsymbol{\Xi}^{(m)}_{k,l}\tilde{\boldsymbol{\lambda}}\nonumber \\ 
	&\hspace{1.8cm}\geq e^{\nu^{(m)}_{k,l}},\ \forall k \in \mathcal{K}, \ \forall l \in \mathcal{L},\ \forall m \in \mathcal{M}. \label{eqn:nonconvex_12}
\end{align}
One can observe that the constraints \eqref{eqn:nonconvex_11} and \eqref{eqn:nonconvex_12} are not convex. In order to obtain convex approximations of the considered constraints, we substitute the matrix $\boldsymbol{\Upsilon}=\tilde{\boldsymbol{\lambda}}\tilde{\boldsymbol{\lambda}}^{T}$ into the expressions and apply first order Taylor approximation on \eqref{eqn:nonconvex_11} around the point $(\varepsilon^{(m)}_{k,l})^{t-1}$ to get
\begin{subequations}
	\begin{alignat}{3}
		\mathrm{tr}(\boldsymbol{\Upsilon}\boldsymbol{\Gamma}^{(m)}_{k,l})&\hspace{-0.05cm}\leq\hspace{-0.05cm} e^{(\varepsilon^{(m)}_{k,l})^{t-1}}\hspace{-0.15cm}(\varepsilon^{(m)}_{k,l}\hspace{-0.1cm}-\hspace{-0.1cm}(\varepsilon^{(m)}_{k,l})^{t-1}+1), \nonumber \\
		&\hspace{2.0cm} \forall k \in \mathcal{K}, \ \forall l \in \mathcal{L},\ \forall m \in \mathcal{M},\label{eqn:1b} \\
		\mathrm{tr}(\boldsymbol{\Upsilon}\boldsymbol{\Xi}^{(m)}_{k,l})&\hspace{-0.05cm}\geq\hspace{-0.05cm} e^{\nu^{(m)}_{k,l}}, \forall k \in \mathcal{K}, \ \forall l \in \mathcal{L},\ \forall m \in \mathcal{M}\label{eqn:1c_2}.
	\end{alignat}
\end{subequations}
\begin{figure}[t!]
	\removelatexerror
	\begin{algorithm}[H]
		\caption{SCA/SDR-based algorithm}
		\label{alg:sca}
		$t \gets 1$, $\mathbf{P}^{\prime}$, $\mathbf{B}$   \\
		\While{$|\mathrm{EE}^{t}-\mathrm{EE}^{t-1}|> \epsilon_{sdr}$}{
			$\left(\boldsymbol{\Upsilon},  \bar{\mathbf{C}}, \boldsymbol{\alpha}^{t},  \boldsymbol{\beta}^{t}, \boldsymbol{\gamma}^{t},
			\boldsymbol{\varepsilon}^{t}, \boldsymbol{\nu}^{t},
			\boldsymbol{\kappa}^{t}\right)$ $\gets$    by solving \eqref{eqn:problem_sca} via interior-point methods  \\
			$\tilde{\boldsymbol{\Lambda}} \gets$ updateInd$\left(\boldsymbol{\Upsilon}\right)$\\
			$\mathrm{EE}^{t+1} \gets$ updateEE$\left(\tilde{\boldsymbol{\Lambda}}\right)$ \\
			$t \gets t + 1$\\
		}
		\Return $\tilde{\boldsymbol{\Lambda}}^{t}$
	\end{algorithm}
	\vspace{-0.5cm}
\end{figure}

Next, we define the auxiliary variables $\alpha_{l}$, $\beta$, $\kappa^{(m)}_{k,l}$ and $\gamma^{(m)}_{k,l}$ and write the constraint \eqref{eqn:new1} by means of 5 new constraints 
\begin{subequations}
	\begin{alignat}{3}
		&t \leq \frac{1}{L}\sum_{l \in \mathcal{L}}\frac{\alpha_{l}^{2}}{\beta} \label{eqn:0a_1}, \\
		&\ln\hspace{-0.1cm}\left(\sigma_{n}^{2}+\tilde{\boldsymbol{\lambda}}^{T}\boldsymbol{\Phi}_{k}^{(m)}\tilde{\boldsymbol{\lambda}}+\hspace{-0.2cm}\sum_{\substack{k^{\prime}\in\mathcal{K}, \\ k^{\prime} \neq k}}|\tilde{\boldsymbol{\lambda}}^{T}(\boldsymbol{\Delta}(\mathbf{h}^{(m)}_{k})^{*}\circ\mathbf{p}^{\prime}_{k^{\prime},l})|^{2}\right) \leq \gamma^{(m)}_{k,l}, \nonumber \\
		&\hspace{3.5cm}\forall k \in \mathcal{K}, \ \forall l \in \mathcal{L},\ \forall m \in \mathcal{M}, \label{eqn:0b_1} \\
		&\ln\hspace{-0.1cm}\left(\sigma_{n}^{2}+\tilde{\boldsymbol{\lambda}}^{T}\boldsymbol{\Phi}_{k}^{(m)}\tilde{\boldsymbol{\lambda}}+\hspace{-0.2cm}\sum_{k^{\prime}\in\mathcal{K}}|\tilde{\boldsymbol{\lambda}}^{T}(\boldsymbol{\Delta}(\mathbf{h}^{(m)}_{k})^{*}\circ\mathbf{p}^{\prime}_{k^{\prime},l})|^{2}\right)\geq \kappa^{(m)}_{k,l}, \nonumber \\
		&\hspace{3.5cm}\forall k \in \mathcal{K}, \ \forall l \in \mathcal{L},\ \forall m \in \mathcal{M},\label{eqn:0c_1} \\
		&\alpha^{2}_{l}\hspace{-0.05cm}-\hspace{-0.05cm}(\ln2)\bar{C}_{l}-\hspace{-0.1cm}\frac{1}{M}\hspace{-0.15cm}\sum_{m \in \mathcal{M}}\sum_{k \in \mathcal{K}}\hspace{-0.05cm}(\kappa^{(m)}_{k,l}\hspace{-0.1cm}-\hspace{-0.1cm}\gamma^{(m)}_{k,l}) \leq 0,  \forall l \in \mathcal{L}, \label{eqn:0d}\\
		&P_{tot}(\tilde{\boldsymbol{\Lambda}},  \mathbf{B}) \leq \beta/\ln2. \label{eqn:0e_2}
	\end{alignat}
\end{subequations}
Similar to the approach taken for \eqref{eqn:1}, we first define the matrices
\begin{align}
	\boldsymbol{\Psi}^{(m)}_{k,l}&=\sum_{\substack{k^{\prime}\in\mathcal{K}, \\ k^{\prime} \neq k}}(\boldsymbol{\Delta}(\mathbf{h}^{(m)}_{k})^{*}\circ\mathbf{p}^{\prime}_{k^{\prime},l})(\boldsymbol{\Delta}(\mathbf{h}^{(m)}_{k})^{*}\circ\mathbf{p}^{\prime}_{k^{\prime},l})^{H}\nonumber \\
	&+\sigma_{n}^{2}\mathbf{I}_{N_{t}}+\boldsymbol{\Phi}_{k}^{(m)}, \nonumber\\ \boldsymbol{\Omega}^{(m)}_{k,l}&=\sum_{k^{\prime}\in\left\lbrace\mathcal{K}\right\rbrace}(\boldsymbol{\Delta}(\mathbf{h}^{(m)}_{k})^{*}\circ\mathbf{p}^{\prime}_{k^{\prime},l})(\boldsymbol{\Delta}(\mathbf{h}^{(m)}_{k})^{*}\circ\mathbf{p}^{\prime}_{k^{\prime},l})^{H}\nonumber \\
	&+\frac{\sigma_{n}^{2}}{N_{t}}\mathbf{I}_{N_{t}}+\boldsymbol{\Phi}_{k}^{(m)}, \nonumber
\end{align}
and apply first order Taylor approximation to obtain the convex approximations for the constraints \eqref{eqn:0a_1}, \eqref{eqn:0b_1} and \eqref{eqn:0c_1}, respectively, as
\begin{subequations}
	\begin{alignat}{3}
		&t \leq \frac{1}{L}\sum_{l \in \mathcal{L}}\left(\frac{2\alpha_{l}^{t-1}}{\beta^{t-1}}\right)\alpha_{l}-\left(\frac{\alpha_{l}^{t-1}}{\beta^{t-1}}\right)^{2}\beta, \label{eqn:0a}\\
		&\mathrm{tr}(\boldsymbol{\Upsilon}\boldsymbol{\Psi}^{(m)}_{k})\leq e^{(\gamma^{(m)}_{k,l})^{t-1}}(\gamma^{(m)}_{k,l}-(\gamma^{(m)}_{k,l})^{t-1}+1), \nonumber \\
		&\hspace{3.5cm}\forall k \in \mathcal{K}, \ \forall l \in \mathcal{L},\ \forall m \in \mathcal{M},\label{eqn:0b}\\
		&\mathrm{tr}(\boldsymbol{\Upsilon}\boldsymbol{\Omega}^{(m)}_{k})\geq e^{\kappa^{(m)}_{k,l}}, \ \forall k \in \mathcal{K}, \ \forall l \in \mathcal{L},\ \forall m \in \mathcal{M}.\label{eqn:0c}
	\end{alignat}
\end{subequations}
We can express the constraint \eqref{eqn:0e_2} in terms of $\boldsymbol{\Upsilon}$ as 
\begin{align}
	P_{tot}(\boldsymbol{\Upsilon},  \mathbf{B})&=\frac{\mathrm{tr}(\boldsymbol{\Upsilon})P_{ant}}{\eta_{PA}}+P_{RF}+\mathrm{tr}(\boldsymbol{\Upsilon}\mathbf{P}_{\boldsymbol{\Delta}})+P_{int}+P_{BB} \nonumber \\
	&\leq \beta/\ln2, \label{eqn:0e}
\end{align}
with $P_{RF}\hspace{-0.08cm}=\hspace{-0.08cm}\mathrm{tr}(\boldsymbol{\Upsilon})P_{circ}\hspace{-0.08cm}+\hspace{-0.08cm}P_{syn}$ and $P_{int}=p_{int}2S_{DAC}\mathrm{tr}(\boldsymbol{\Upsilon}\mathbf{B})$.

For the constraint \eqref{eqn:rate_const_3}, we use the auxiliary variables $\kappa^{(m)}_{k,l}$ and $\gamma^{(m)}_{k,l}$ to write
\begin{align}
	\frac{1}{L}\sum_{l \in \mathcal{L}}\left((\ln2)\bar{C}_{l}+\frac{1}{M}\sum_{m \in \mathcal{M}}\sum_{k \in \mathcal{K}}(\kappa^{(m)}_{k,l}-\gamma^{(m)}_{k,l})\right) \geq (\ln 2)R_{th}. 
	\label{eqn:45a}
\end{align}

Next, we move to rewrite the constraint \eqref{eqn:4}. Proposition 3 gives the expression for $\widehat{\mathbf{R}}_{\tilde{\mathbf{x}}}(\mathbf{P}, \tilde{\boldsymbol{\Lambda}}, \mathbf{B})$ in terms of $\boldsymbol{\Upsilon}$.

\textit{Proposition 3:} The matrix $\widehat{\mathbf{R}}_{\tilde{\mathbf{x}}}(\mathbf{P}, \tilde{\boldsymbol{\Lambda}}, \mathbf{B})$ can be expressed in terms of $\boldsymbol{\Upsilon}$ as 
\begin{align}
	\widehat{\mathbf{R}}_{\tilde{\mathbf{x}}}(\mathbf{P}, \tilde{\boldsymbol{\Lambda}}, \mathbf{B})=\sum_{l \in \mathcal{L}}\sum_{i=1}^{N_{t}}(\mu_{i,l}\mathbf{F}_{i,l}\boldsymbol{\Upsilon}\mathbf{F}_{i,l}^{H}+\boldsymbol{\Sigma}\mathbf{E}_{N_{t},i}\boldsymbol{\Upsilon}\mathbf{E}_{N_{t},i}), 
\end{align}
where $\mathbf{F}_{i,l}=\mathrm{Diag}(\mathbf{f}^{T}_{i,l})$ and $\mu_{i,l}$ and $\mathbf{f}_{i,l}$ are respectively the $i$-th eigenvalue and eigenvector of the matrix $\boldsymbol{\Delta}\mathbf{P}_{l}\mathbf{P}_{l}^{H}\boldsymbol{\Delta}$, $\forall i \in \left\lbrace 1,2,\ldots, N_{t}\right\rbrace$ and $\forall l \in \mathcal{L}$.   

\textit{Proof:} Please see Appendix~\ref{sec:app3}.\hspace{3.2cm}$\blacksquare$

Using Proposition 3, the constraint \eqref{eqn:4} can be written as
\begin{align}
	\Bigg|\Bigg|\sum_{l \in \mathcal{L}}\sum_{i=1}^{N_{t}}(\mu_{i,l}\mathbf{F}_{i,l}\boldsymbol{\Upsilon}\mathbf{F}_{i,l}^{H}+\boldsymbol{\Sigma}\mathbf{E}_{N_{t},i}\boldsymbol{\Upsilon}\mathbf{E}_{N_{t},i})-\mathbf{U} \Bigg|\Bigg|^{2}_{2} \leq \tau.
	\label{eqn:4_2}
\end{align}

Finally, we write
\begin{align}
	& 0 \leq \boldsymbol{\Upsilon}_{l}(i,j) \leq 1, \ i,j \in \left\lbrace 1,2,\ldots, N_{t} \right\rbrace, \label{eqn:3}\\
	& \tr(\boldsymbol{\Upsilon}) \geq 1, \label{eqn:lowerbound_sdr}
\end{align}
where $\boldsymbol{\Upsilon}(i,j)$ represents the element of the matrix $\boldsymbol{\Upsilon}$ at the $i$-th column and $j$-th row, and \eqref{eqn:lowerbound_sdr} is the constraint \eqref{eqn:lambdalowerbound} rewritten in terms of $\boldsymbol{\Upsilon}$.
The resulting problem formulation is written as
\begin{subequations}
	\begin{alignat}{3}
		&\max_{\substack{t,  \bar{\mathbf{C}}, \boldsymbol{\Upsilon},
				\boldsymbol{\alpha}, \beta, \\
				\boldsymbol{\gamma}, \boldsymbol{\varepsilon},
				\boldsymbol{\nu}, \boldsymbol{\kappa}
		}}   \   t  \nonumber  \\
		& \quad \quad \text{s.t.}  \quad \boldsymbol{\Upsilon} \succcurlyeq 0, \label{eqn:sdf_const}  \\
		&\ \quad \quad \quad \ \quad \eqref{eqn:1a}, \eqref{eqn:1b}, \eqref{eqn:1c_2},  \eqref{eqn:0d}, \eqref{eqn:0a}, \eqref{eqn:0b},    \nonumber \\
		& \ \quad \quad \quad \ \quad  \eqref{eqn:0c}, \eqref{eqn:0e}, \eqref{eqn:45a},\eqref{eqn:4_2}, \eqref{eqn:3}, \eqref{eqn:lowerbound_sdr},
	\end{alignat}
	\label{eqn:problem_sca}
\end{subequations}
where the vectors $\boldsymbol{\alpha}$, $\boldsymbol{\gamma}$, $\boldsymbol{\varepsilon}$, $\boldsymbol{\nu}$, and $\boldsymbol{\kappa}$ are composed of the elements $\alpha_{l}$, $\gamma^{(m)}_{k,l}$, $\varepsilon^{(m)}_{k,l}$, $\nu^{(m)}_{k,l}$, and $\kappa^{(m)}_{k,l}$, respectively, $\forall k \in \mathcal{K}$, $\forall l \in \mathcal{L}$, and, $\forall m \in \mathcal{M}$. The rank-$1$ constraint for $\boldsymbol{\Upsilon}$ is removed in accordance with the SDR procedure and the problem is in an iterative fashion by means of interior-point methods. The SCA/SDR-based algorithm to solve the problem \eqref{eqn:problem_sca} is given in Alg.~\ref{alg:sca}.

\textit{Proposition 4:} Consider the problem \eqref{eqn:problem4} with the RF chain selection indicator constraint \eqref{eqn:3_nonconvex} relaxed as $0 \leq \tilde{\lambda}_{i} \leq 1$, $\forall i \in \left\lbrace 1, 2, \ldots, N_{t}\right\rbrace$. For the case a rank-$1$ solution exists for problem \eqref{eqn:problem_sca} and for given $\mathbf{P}^{\prime}_{l}$, $\forall l \in \mathcal{L}$, the proposed algorithm in Alg.~\ref{alg:sca} converges to a stationary point of the problem \eqref{eqn:problem4} with the relaxed RF chain selection indicator constraint. 

\textit{Proof:}
Since the solutions $\bar{\mathbf{C}}$ and  $\boldsymbol{\Upsilon}$ at iteration-$t-1$ are also feasible solutions for \eqref{eqn:problem_sca} at iteration-$t$, the $EE^{t}$ is non-decreasing with $t$ and is bounded above due to the per antenna power constraints, per antenna RF chain power consumption and the constant terms $P_{BB}$ and $P_{syn}$ in $\mathcal{E}(\mathbf{P}^{\prime}, \tilde{\boldsymbol{\Lambda}}, \mathbf{B})$ \cite{yi_2011, bjornson_2015}, which guarantees the convergence of the algorithm \cite[Sec. IV-F]{sun_2021}. 

Assuming a rank-$1$ solution exists, the optimal solution  $\tilde{\boldsymbol{\Lambda}}$ at iteration-$t$ obtained by the SDR algorithm is a stationary point of problem \eqref{eqn:problem4}, since problem \eqref{eqn:problem_sca} is a convex approximation of problem \eqref{eqn:problem4} (for which the stationary points can be obtained by SDR \cite{zuo_2010}) and the KKT conditions of problem \eqref{eqn:problem4} are maintained at the point $\tilde{\boldsymbol{\Lambda}}$. Consequently, we obtain a stationary point of the original problem when the solution point  $\tilde{\boldsymbol{\Lambda}}$ at iteration-$t$ is the same as the solution point at iteration-${t-1}$.  
\hspace{6.2cm}$\blacksquare$

\subsection{Proposed AO-Based Algorithm for Optimal Precoder Calculation and RF Chain Selection}
We can now move to combine the algorithms Alg.~\ref{alg:ao_admm} and \ref{alg:sca} to solve the problem \eqref{eqn:problem3_v2}. 
The proposed algorithm is given in Alg.~\ref{alg:algorithm_prop}. The algorithm alternates between updating the precoders and RF chain selection indicators until a stopping condition is satisfied.  
\begin{figure}[t!]
	\removelatexerror
\begin{algorithm}[H]
		\caption{AO-Based Algorithm} 
		\label{alg:algorithm_prop}
		$t \gets 1$, $\mathbf{v}^{0}$, $\mathbf{u}^{0}$,  $\mathbf{w}^{0}$, $\tilde{\boldsymbol{\Lambda}}^{0}=\mathbf{I}_{N_{t}}$, $\mathbf{B}$     \\
		\While{$|\mathrm{EE}^{t}-\mathrm{EE}^{t-1}|> \epsilon_{ao}$}{
			$\boldsymbol{\omega}^{t} \hspace{-0.1cm}\gets$ updateWeights$\left(\mathbf{u}^{t-1}, \tilde{\boldsymbol{\Lambda}}^{t-1}, \mathbf{B}\right)$ \\
			$\boldsymbol{g}^{t} \hspace{-0.1cm}\gets$ updateFilters$\left(\mathbf{u}^{t-1},\tilde{\boldsymbol{\Lambda}}^{t-1}, \mathbf{B}\right)$ \\
			$[\mathbf{u}^{t},\mathbf{w}^{t}]\gets \mathrm{\mathbf{ALG1}}\left(\tilde{\boldsymbol{\Lambda}}^{t-1}, \mathbf{B}, \mathbf{u}^{t-1},  \mathbf{w}^{0}, \boldsymbol{\omega}^{t}, \boldsymbol{g}^{t}\right)$\\
			$\tilde{\boldsymbol{\Lambda}}^{t}  \gets \mathrm{\mathbf{ALG3}}\left(\mathbf{u}^{t}, \mathbf{B}\right)$ \\
			$\mathrm{EE}^{t+1} \gets$ updateEE$\left(\mathbf{u}^{t}, \tilde{\boldsymbol{\Lambda}}^{t}, \mathbf{B}\right)$ \\
			$t \gets t + 1$\\
		}
		\Return $\mathbf{u}^{t}$, $\tilde{\boldsymbol{\Lambda}}^{t}$.
	\end{algorithm}
	\vspace{-0.5cm}
\end{figure}

\begin{figure*}[t!]
	\begin{subfigure}{.5\textwidth}
		\centerline{\includegraphics[width=3.3in,height=3.3in,keepaspectratio]{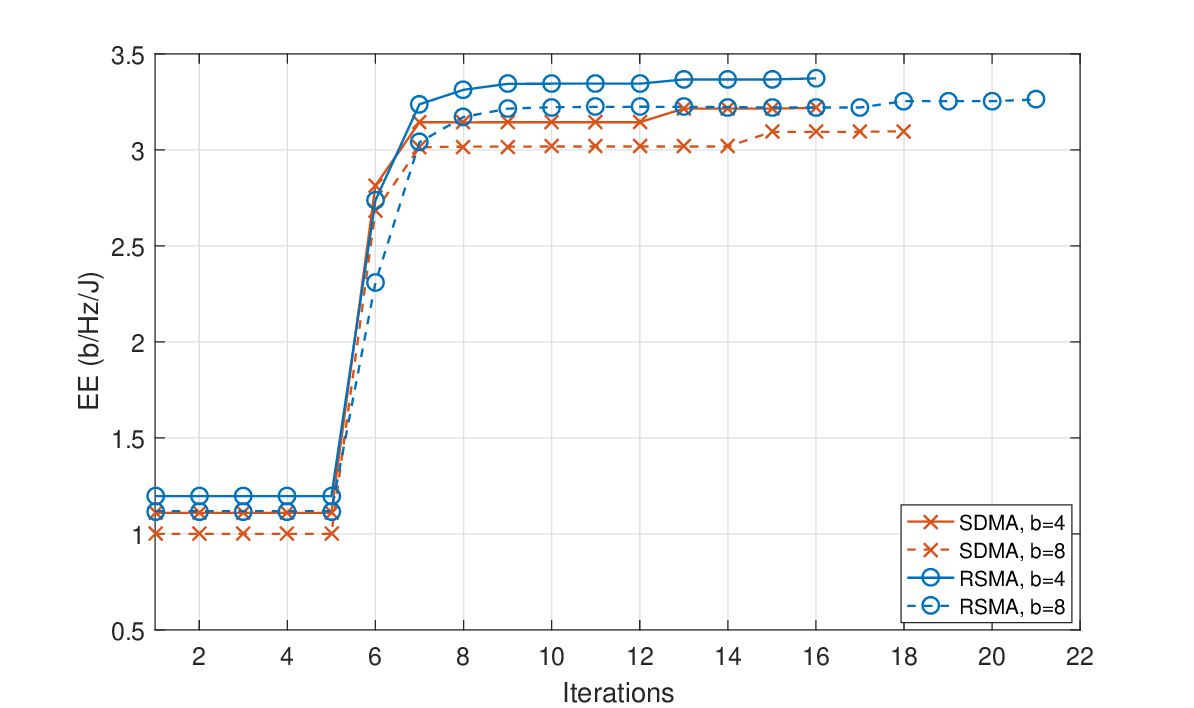}}
		\caption{Optimization for radar detection performance, $\tau=7.0$.}
	\end{subfigure}
	\begin{subfigure}{.5\textwidth}
		\centerline{\includegraphics[width=3.3in,height=3.3in,keepaspectratio]{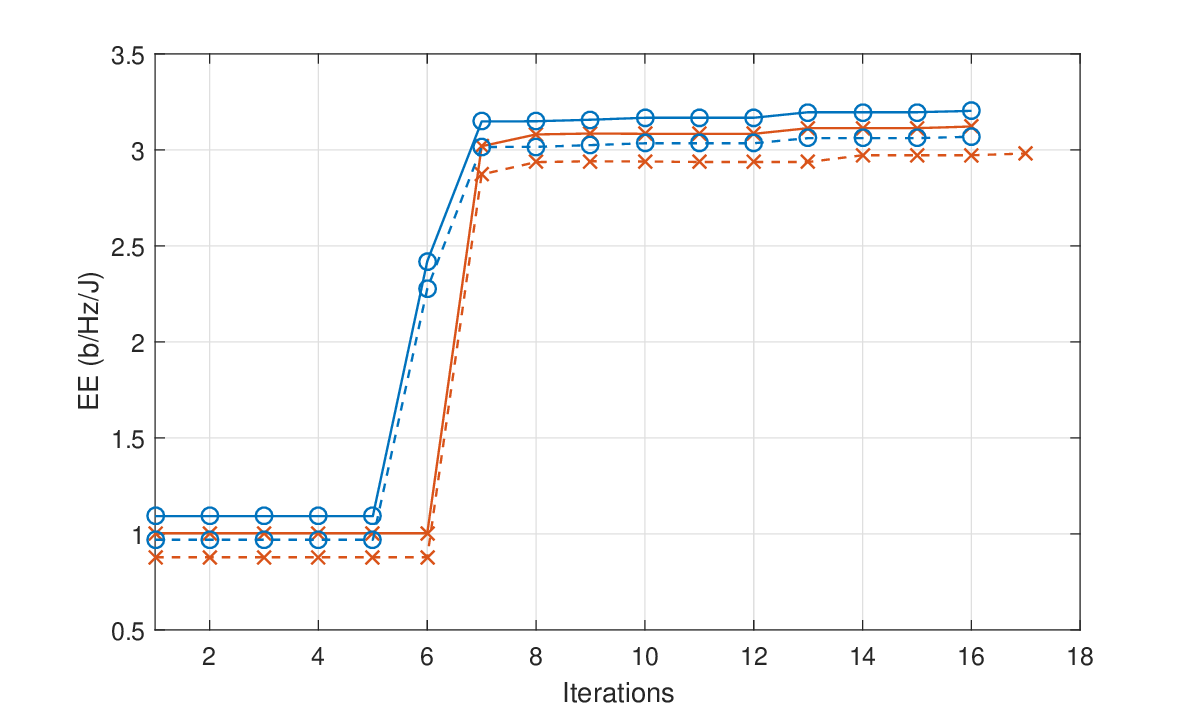}}
		\caption{Optimization for parameter estimation performance, $\tau=65$.}
	\end{subfigure}
	\caption{Convergence behaviour of the proposed algorithm for RSMA and SDMA, imperfect CSIT, single target, $N_{t}=8$, $K=2$, $L=10$, $\theta=\pi/4$. The legend in subfigure (a) is applicable to subfigure (b).}
	\label{fig:conv_ee}
\end{figure*}

\textit{Proposition 5:} Consider the problem \eqref{eqn:problem3_v2} with the constraint \eqref{eqn:s} relaxed as relaxed as $0 \leq \tilde{\lambda}_{i} \leq 1$, $\forall i \in \left\lbrace 1, 2, \ldots, N_{t}\right\rbrace$. For the case a rank-$1$ solution exists for problem \eqref{eqn:problem_sca}, the proposed algorithm in Alg.~\ref{alg:algorithm_prop} converges to a stationary point of the problem \eqref{eqn:problem3_v2} with the relaxed RF chain selection indicator constraint.

\textit{Proof:}
We start the proof by noting that Alg.~\ref{alg:algorithm_prop} is similar to Alg.~\ref{alg:ao_admm} in structure, with Alg.~\ref{alg:sca} run as an  additional step. From Propositions 2 and 4, it is known that the Algorithms \ref{alg:ao_admm} and \ref{alg:sca} converge.
It follows from the convergence of the employed algorithms that $EE^{t}$ increases monotonically and is upper bounded due to the per antenna power constraints, per antenna RF chain power consumption and the constant terms $P_{BB}$ and $P_{syn}$ in $\mathcal{E}(\mathbf{P}, \tilde{\boldsymbol{\Lambda}}, \mathbf{B})$ \cite{yi_2011, bjornson_2015}. Assuming a rank-1 solution exists for problem \eqref{eqn:problem_sca}, the obtained optimal solutions $\mathbf{u}^{t-1}$ and $\tilde{\boldsymbol{\Lambda}}^{t-1}$ are stationary points of problem \eqref{eqn:problem3_v2} with the constraint \eqref{eqn:s} relaxed as $0 \leq \tilde{\lambda}_{i} \leq 1$, $\forall i \in \left\lbrace 1, 2, \ldots, N_{t}\right\rbrace$, since the AO
algorithm is also a special instance of the SCA method \cite{clerckxTC2016} and the KKT conditions of the problem \eqref{eqn:problem3_v2} (with the constraint \eqref{eqn:s} relaxed) are maintained at the points $\mathbf{u}^{t-1}$ and $\tilde{\boldsymbol{\Lambda}}^{t-1}$. Consequently, we obtain a stationary point of the original problem (with the constraint \eqref{eqn:s} relaxed) when the solution points $\mathbf{u}^{t}$ and $\tilde{\boldsymbol{\Lambda}}^{t}$ are the same as the solution points $\mathbf{u}^{t-1}$ and $\tilde{\boldsymbol{\Lambda}}^{t-1}$, respectively. \hspace{4.2cm}$\blacksquare$

We note that the solution $\tilde{\boldsymbol{\Lambda}}^{t}$ is a relaxed version of the actual indicator matrices. Obviously, if the indicators $\tilde{\lambda}_{i} \in \left\lbrace 0, 1\right\rbrace$, $\forall i \in \left\lbrace 1,2,\ldots,N_{t}\right\rbrace$, the obtained solution is also optimal for the non-relaxed problem. Otherwise, we apply a rounding with respect to a threshold to obtain the non-relaxed indicators as given below:
\begin{align}
	\lambda_{i}=
	\left\{
	\begin{array}{ll}
		0,  & \mbox{if} \ \tilde{\lambda}_{i} \leq I_{thr} \\
		1, & \mbox{otherwise}.
	\end{array}
	\right.
	\label{eq:sign}
\end{align}

Finally, we present the complexity of the proposed algorithm in Proposition 6.

\textit{Proposition 6:} For a solution accuracy of $\epsilon > 0$, the worst case complexity of the proposed algorithm at iteration-$t$ is given by    
\begin{align}
	\mathcal{C}_{prop,t}=\mathcal{O}&([T_{1,t}(L(K+1)N_{t})^{4.5} \nonumber \\
	&\quad \quad+T_{3,t}(4KLM+N^{2}_{t})^{4.5}]\log(1/\epsilon)),
	\label{eqn:complexity_1}
\end{align}
where $T_{1,t}$ and $T_{3,t}$ denote the number of iterations of \textbf{ALG1} and \textbf{ALG3} at iteration-$t$, respectively.

\textit{Proof:} Please see Appendix~\ref{sec:app6}.\hspace{3.2cm}$\blacksquare$

\begin{figure*}[t!]
	\begin{subfigure}{.5\textwidth}
		\centerline{\includegraphics[width=3.3in,height=3.3in,keepaspectratio]{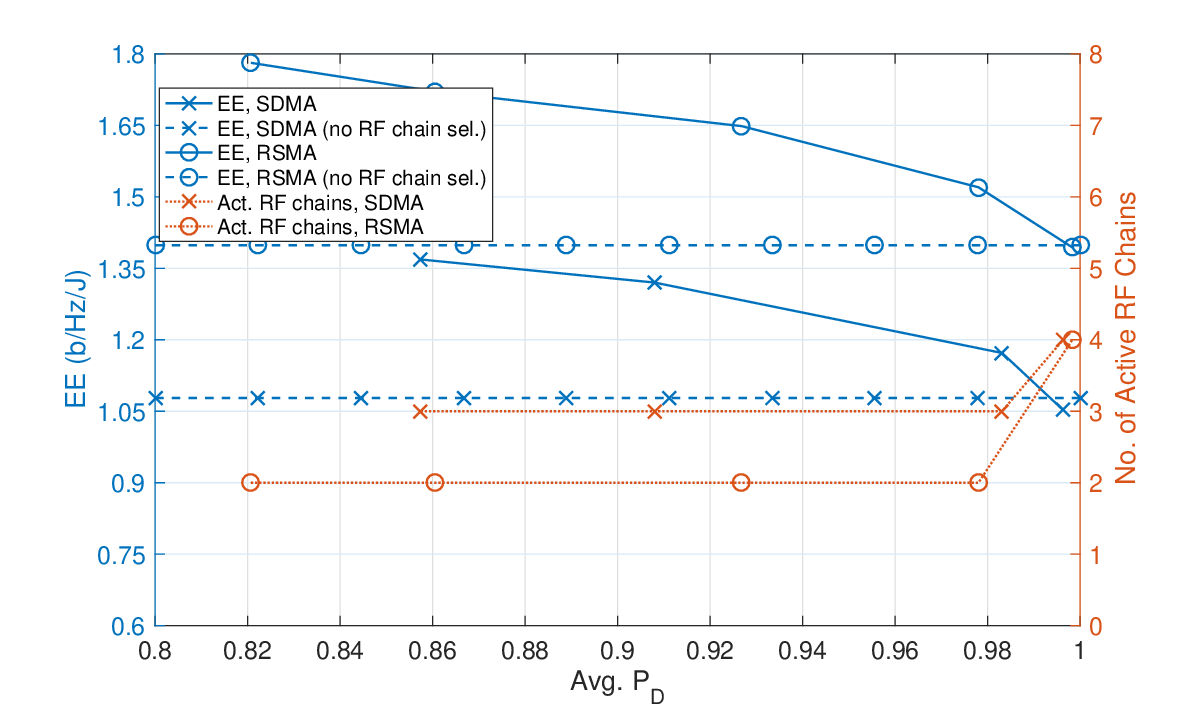}}
		\caption{Perfect CSIT, $b=4$.}
		\label{fig:pdvsee_perf_4}
	\end{subfigure}
	\begin{subfigure}{.5\textwidth}
		\centerline{\includegraphics[width=3.3in,height=3.3in,keepaspectratio]{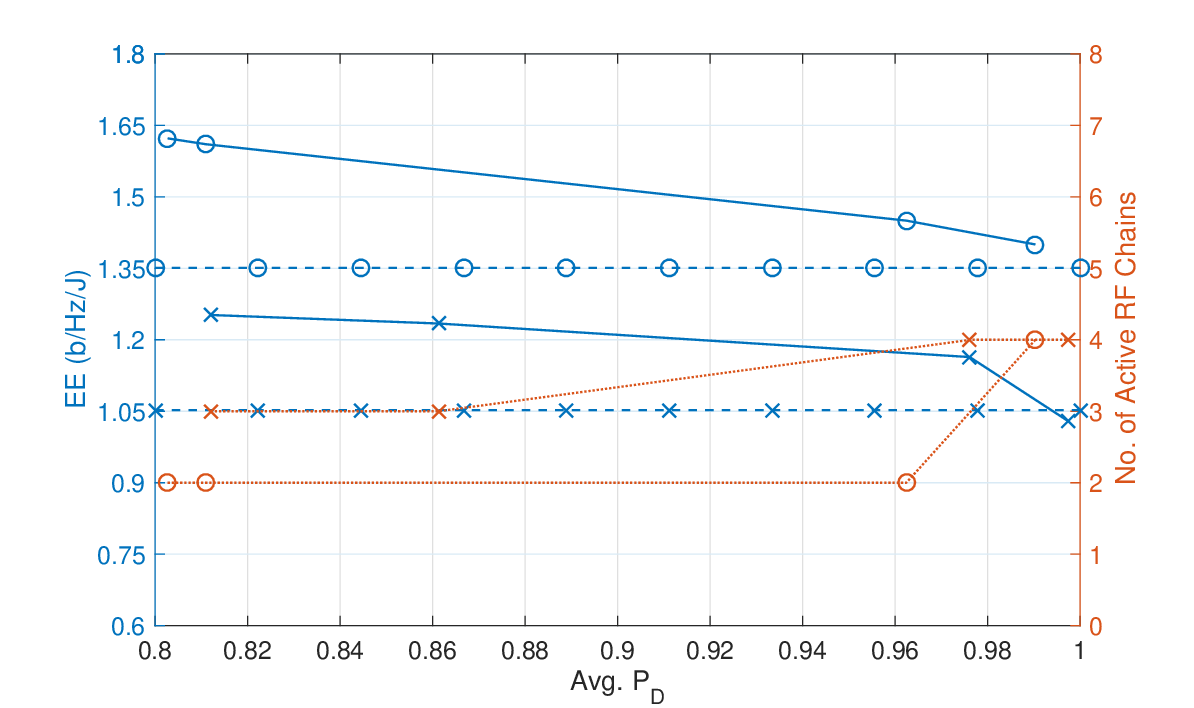}}
		\caption{Perfect CSIT, $b=8$.}
		\label{fig:pdvsee_perf_8}
	\end{subfigure}
	\newline
	\begin{subfigure}{.5\textwidth}
		\centerline{\includegraphics[width=3.3in,height=3.3in,keepaspectratio]{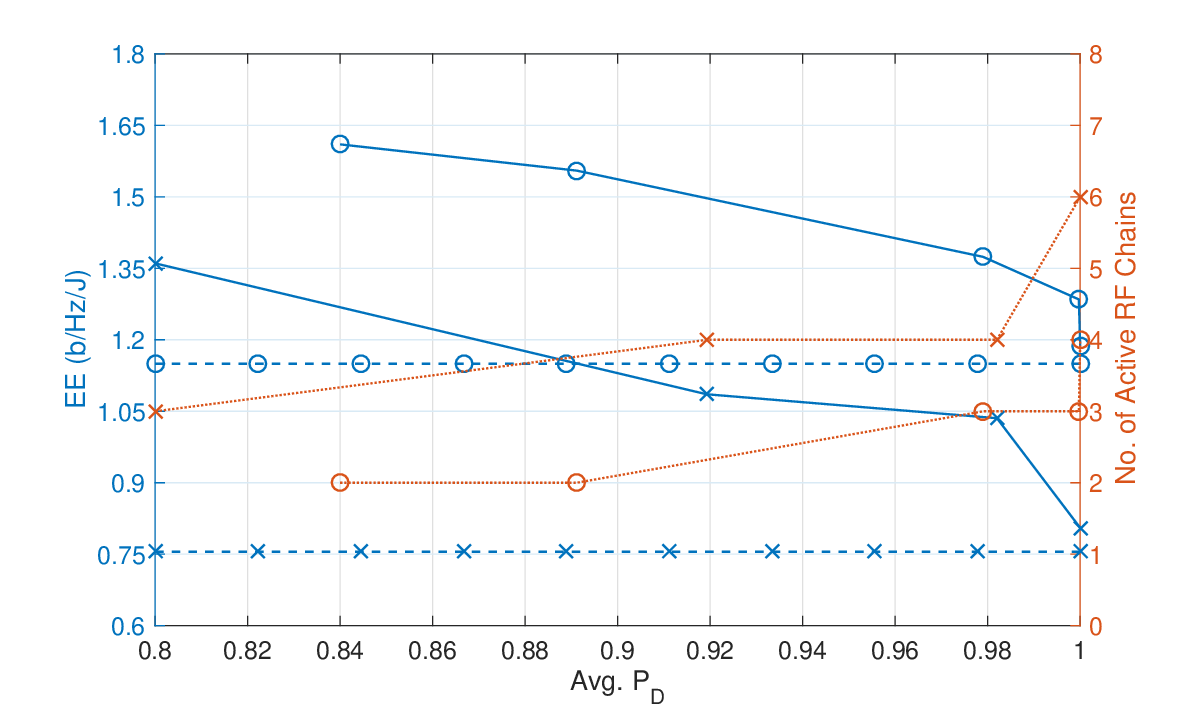}}
		\caption{Imperfect CSIT, $b=4$.}
		\label{fig:pdvsee_imperf_4}
	\end{subfigure}
	\begin{subfigure}{.5\textwidth}
		\centerline{\includegraphics[width=3.3in,height=3.3in,keepaspectratio]{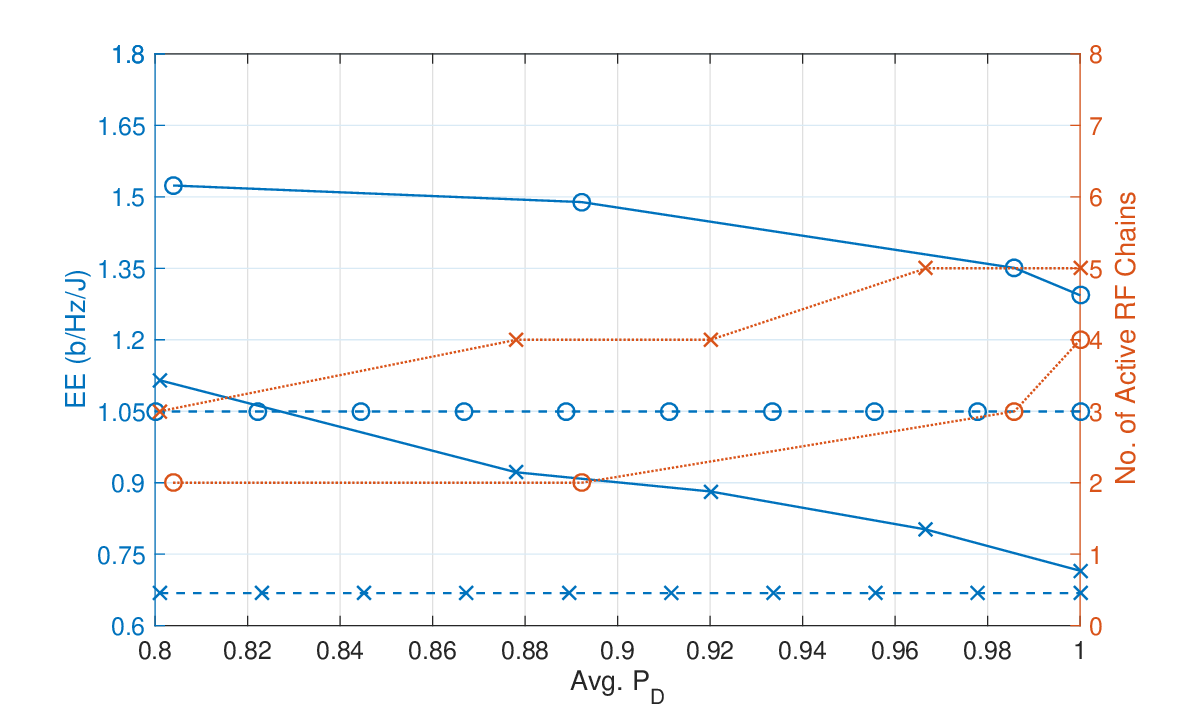}}
		\caption{Imperfect CSIT, $b=8$.}
		\label{fig:pdvsee_imperf_8}
	\end{subfigure}
	\caption{$\bar{P}_{D}$ vs. EE performance of SDMA and RSMA, $SR>8$ b/s/Hz, $P_{F}=10^{-7}$, single target, $N_{t}=8$, $K=2$, $L=10$, $\alpha_{r}=0.1$. The legend in subfigure (a) is applicable to other subfigures (b)-(d).}
	\label{fig:pdvsee}
\end{figure*}

\begin{figure*}[t!]
	\begin{subfigure}{.5\textwidth}
		\centerline{\includegraphics[width=3.3in,height=3.3in,keepaspectratio]{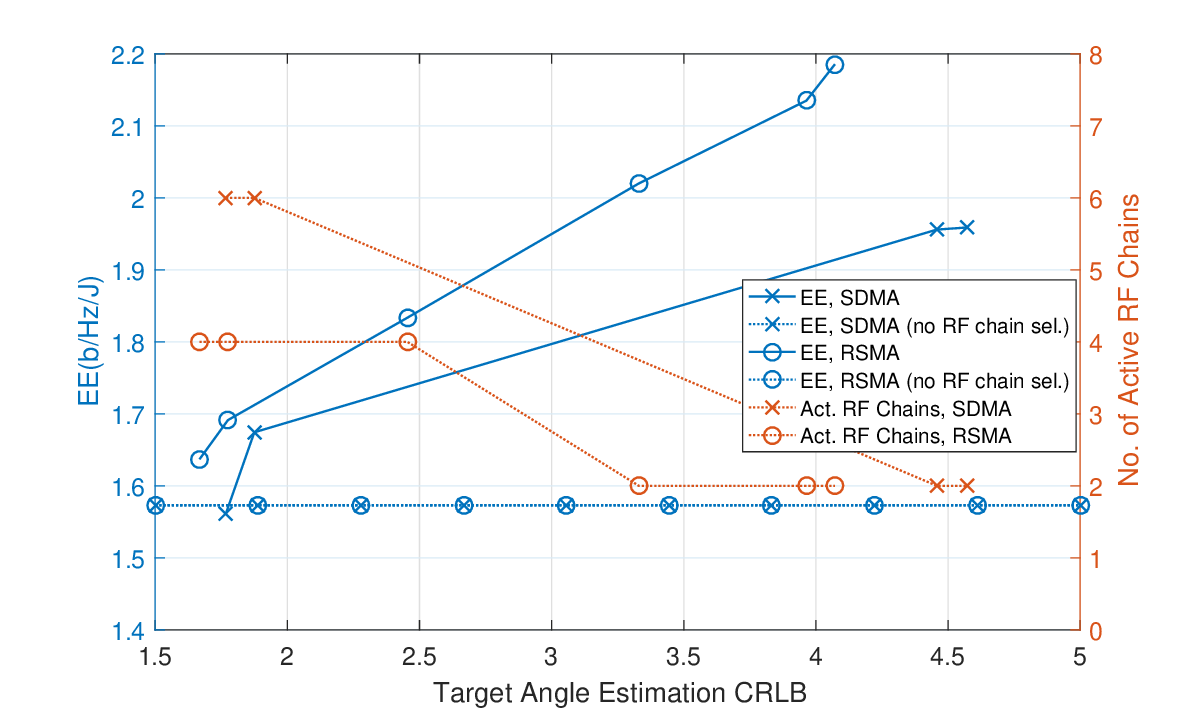}}
		\caption{Perfect CSIT, $b=4$.}
		\label{fig:crlbvsee_perf_4}
	\end{subfigure}
	\begin{subfigure}{.5\textwidth}
		\centerline{\includegraphics[width=3.3in,height=3.3in,keepaspectratio]{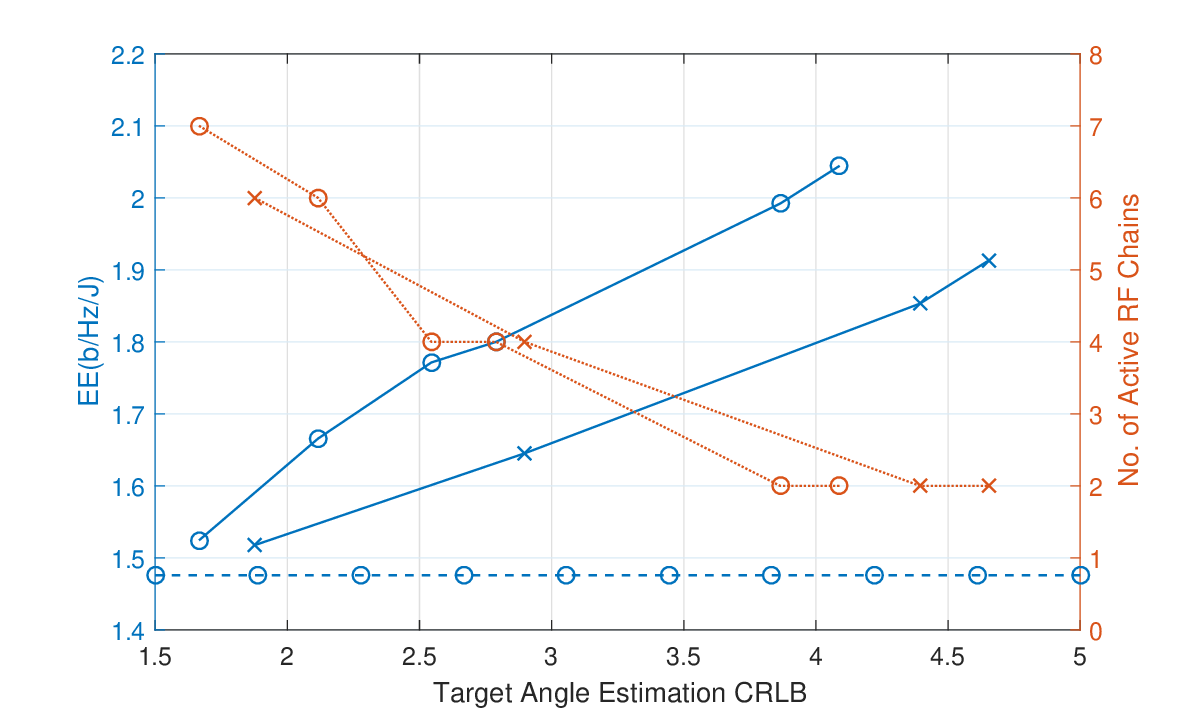}}
		\caption{Perfect CSIT, $b=8$.}
		\label{fig:crlbvsee_perf_8}
	\end{subfigure}
	\newline
	\begin{subfigure}{.5\textwidth}
		\centerline{\includegraphics[width=3.3in,height=3.3in,keepaspectratio]{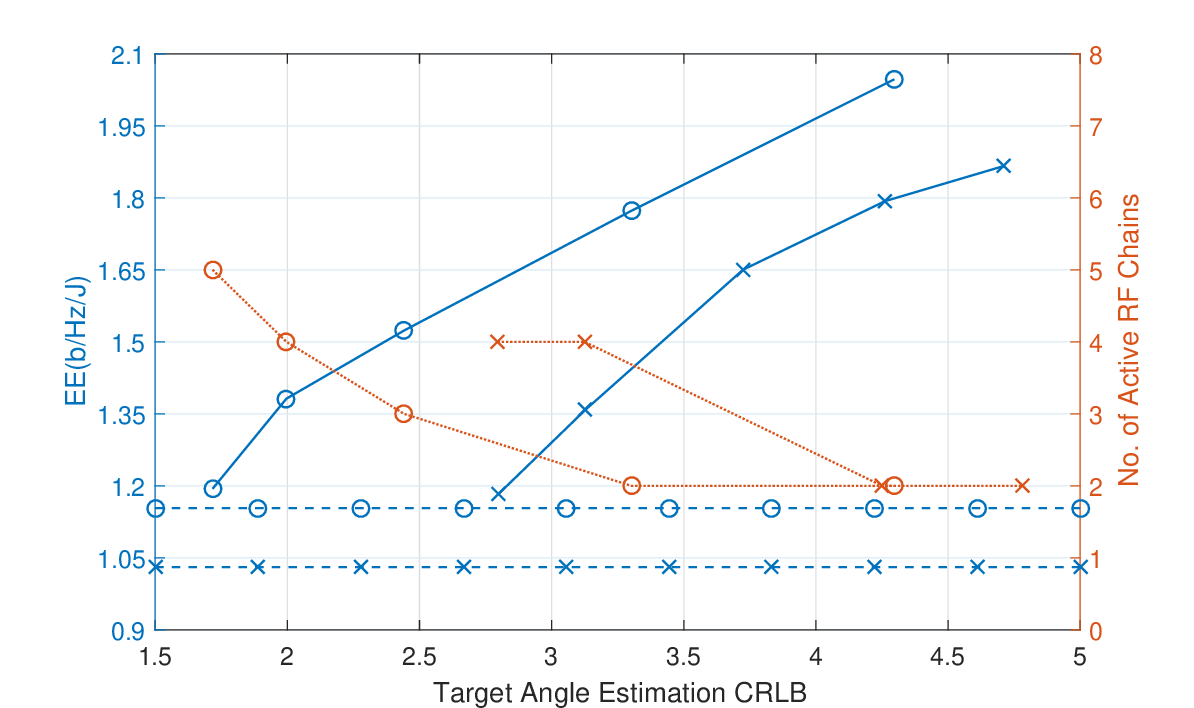}}
		\caption{Imperfect CSIT, $b=4$.}
		\label{fig:crlbvsee_imperf_4}
	\end{subfigure}
	\begin{subfigure}{.5\textwidth}
		\centerline{\includegraphics[width=3.3in,height=3.3in,keepaspectratio]{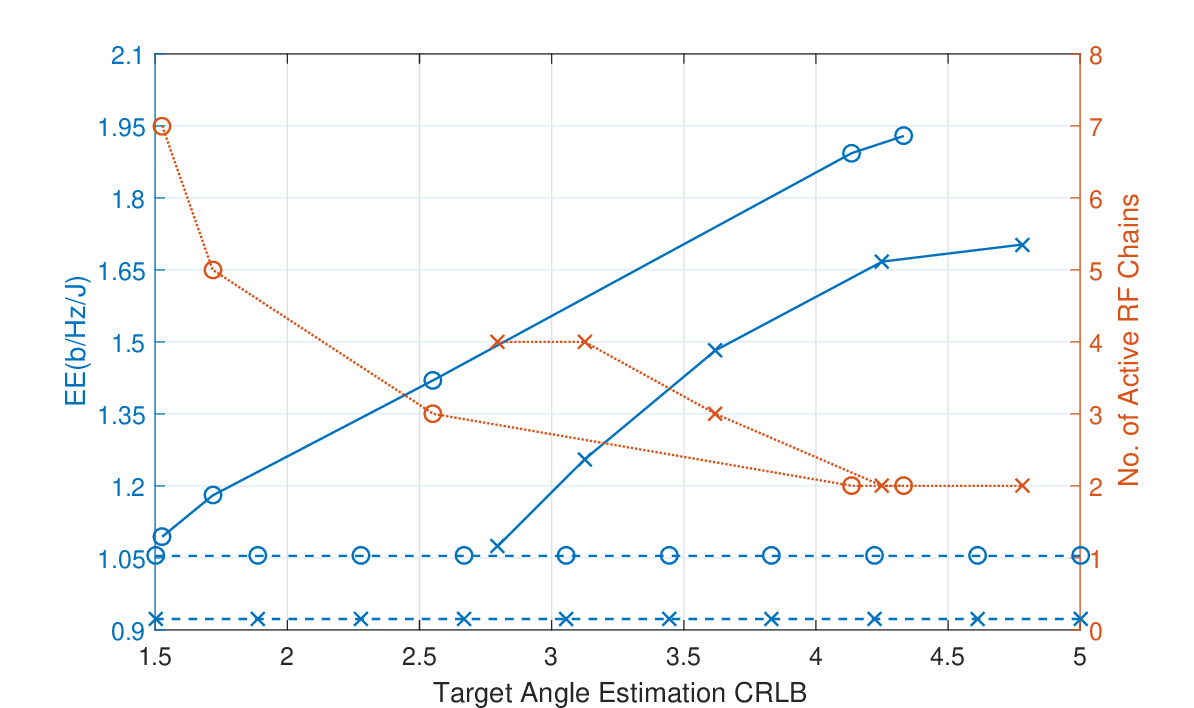}}
		\caption{Imperfect CSIT, $b=8$.}
		\label{fig:crlbvsee_imperf_8}
	\end{subfigure}
	\caption{Target angle estimation CRB vs. EE performance of SDMA and RSMA, $SR>8$ b/s/Hz, $P_{F}=10^{-7}$, single target, $N_{t}=8$, $K=2$, $L=10$, $\alpha_{r}=0.1$, $\theta=\pi/4$. The legend in subfigure (a) is applicable to other subfigures (b)-(d).}
	\label{fig:crlbvsee}
\end{figure*}

\begin{table}[t!]
	\caption{Simulation parameters} 
	\centering 
	\begin{tabular}{| >{\centering\arraybackslash}m{0.6in} | >{\centering\arraybackslash}m{0.4in} | >{\centering\arraybackslash}m{0.6in} | >{\centering\arraybackslash}m{0.4in} | } 
		\hline
		\rule{0pt}{2ex}\textbf{Parameter} & \textbf{Value} & \textbf{Parameter} & \textbf{Value}    \\  
		\hline\hline
		\rule{0pt}{2.2ex} $N_{t}$ & 8 & $P_{BB}[W]$ & 1     \\  
		\hline 
		\rule{0pt}{2ex} $K$ & 2 & $P_{circ}[W]$ & 1   \\   
		\hline	
		\rule{0pt}{2ex}$L$ & 10 & $P_{syn}[W]$ & 2  \\ 
		\hline	
		\rule{0pt}{2ex}$d$ & 0.5 & $P_{DAC}[mW]$ & 1  \\
		\hline	
		\rule{0pt}{2ex}$P_{ant}[mW]$ & 125 & $p_{int}[mW]$ & 25  \\   
		\hline	
		\rule{0pt}{2ex}$N_{0}[mW]$ & 1 & $S_{DAC}[Mbps]$ & 125  \\
		\hline	
		\rule{0pt}{2ex}$\sigma_{ce}$ & 0.2 & $\eta_{PA}$ & 0.39  \\  
		\hline
	\end{tabular}
	\label{table:parameters} 
\end{table} 

\section{Simulation Results}
\label{sec:simulation}
In this section, we present simulation results to investigate the performance of RSMA using the proposed algorithm and perform comparisons with the performance of SDMA. The optimal precoders for
SDMA can be obtained by turning off the common stream in the optimization problem formulation and solving by the proposed algorithm. We analyze the EE performance of the system with the optimized precoders with respect to the communications and radar performance metrics. The parameters used in the simulations are given in Table~\ref{table:parameters}.
\begin{figure*}[t!]
	\begin{subfigure}{.5\textwidth}
		\centerline{\includegraphics[width=3.3in,height=3.3in,keepaspectratio]{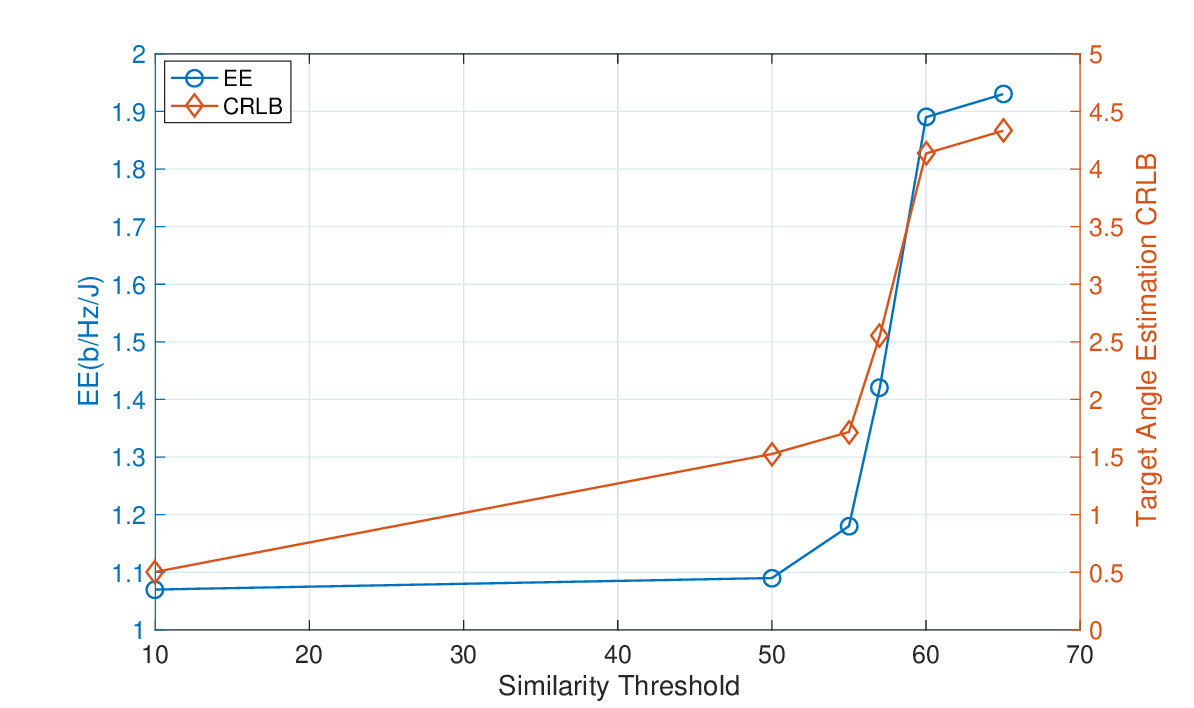}}
		\caption{EE and CRB vs. $\tau$.}
		\label{fig:similarity_metric}
	\end{subfigure}
	\begin{subfigure}{.5\textwidth}
		\centerline{\includegraphics[width=3.3in,height=3.3in,keepaspectratio]{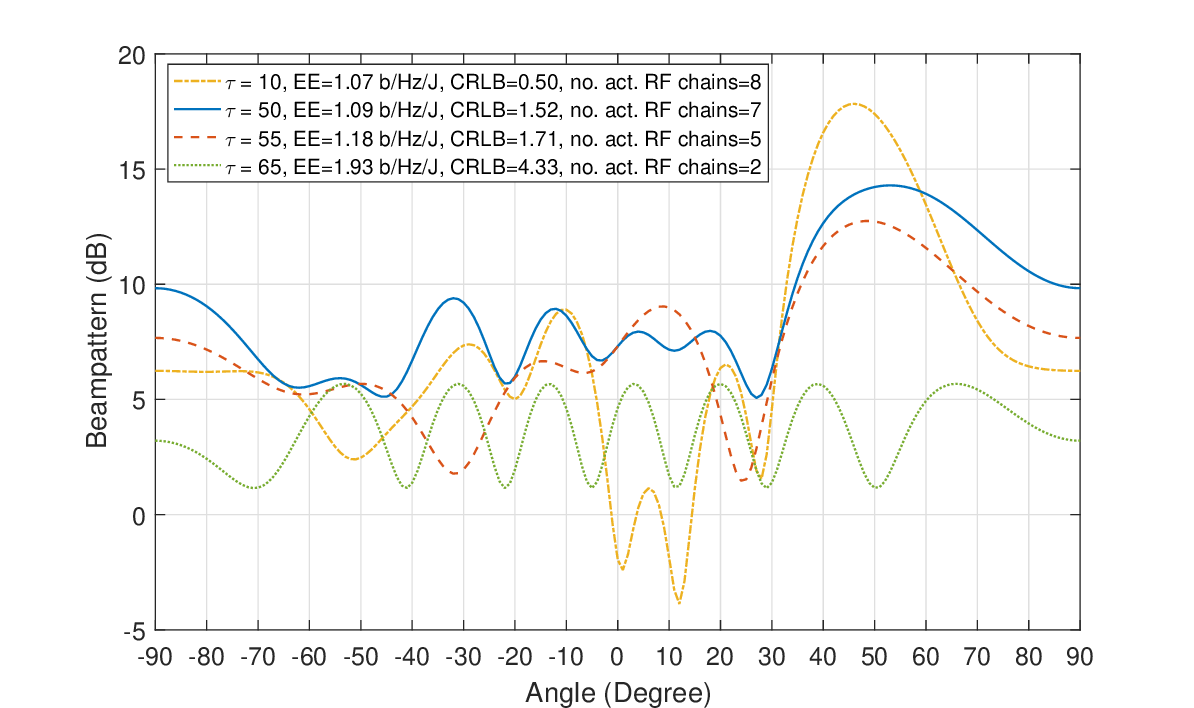}}
		\caption{Beampattern with varying $\tau$.}
		\label{fig:similarity_beam}
	\end{subfigure}
	\caption{Performance metrics vs. $\tau$, RSMA, imperfect CSIT, $SR>8$ b/s/Hz, $P_{F}=10^{-7}$, single target, $N_{t}=8$, $K=2$, $L=10$, $\alpha_{r}=0.1$, $\theta=\pi/4$.}
\end{figure*}

\begin{table}[t!]
	\caption{Run-time results} 
	\centering 
	\begin{tabular}{| >{\centering\arraybackslash}m{1.5in} | >{\centering\arraybackslash}m{0.5in} | >{\centering\arraybackslash}m{0.5in} | } 
		\hline
		\rule{0pt}{2ex} & \textbf{RSMA} & \textbf{SDMA}   \\  
		\hline\hline
		\rule{0pt}{2.2ex} Detection, perfect CSIT & 57s & 37s     \\  
		\hline 
		\rule{0pt}{2ex} Detection, imperfect CSIT  & 372s & 289s   \\
		\hline	
		\rule{0pt}{2.2ex} Estimation, perfect CSIT & 59s & 36s     \\  
		\hline 
		\rule{0pt}{2ex} Estimation, imperfect CSIT  & 376s & 281s   \\
		\hline	
	\end{tabular}
	\label{table:complexity} 
\end{table} 

We investigate the EE performance of the system with respect to radar metrics in detection and tracking modes. We consider target detection probability and Cramer-Rao Bound (CRB) for Direction of Arrival (DoA) estimation. For target detection, we choose the reference covariance matrix to maximize the detection probability as $\mathbf{U}_{det}=P_{ant}L\mathbf{I}_{N_{t}}$ \cite{khawar_2015}. The detection probability for a target at angle $\theta$ is written as 
\begin{align}
	P_{D}(\theta)=1-F_{\chi_{2}^{2}(\rho(\theta))}\left(F^{-1}_{\chi_{2}^{2}}(1-P_{F})\right),
	\label{eqn:detection}
\end{align}
where $P_{F}$ is the false alarm probability, $F^{-1}_{\chi_{2}^{2}}$ is the inverse of central chi-squared distribution function with two degrees of freedom, $F_{\chi_{2}^{2}(\rho(\theta))}$ is the noncentral chi-squared distribution function with two degrees of freedom and noncentrality parameter
$\rho(\theta)\triangleq\frac{|\alpha_{r}|^{2}}{P_{ant}L\sigma_{n}^{2}}|\mathbf{a}^{H}(\theta)\widehat{\mathbf{R}}_{\tilde{\mathbf{x}}}^{T}(\mathbf{P}, \boldsymbol{\Lambda}, \mathbf{B})\mathbf{a}(\theta)|^{2}$ \cite{khawar_2015}. 
\begin{figure*}[hbt]
	\begin{subfigure}{.5\textwidth}
		\centerline{\includegraphics[width=3.3in,height=3.3in,keepaspectratio]{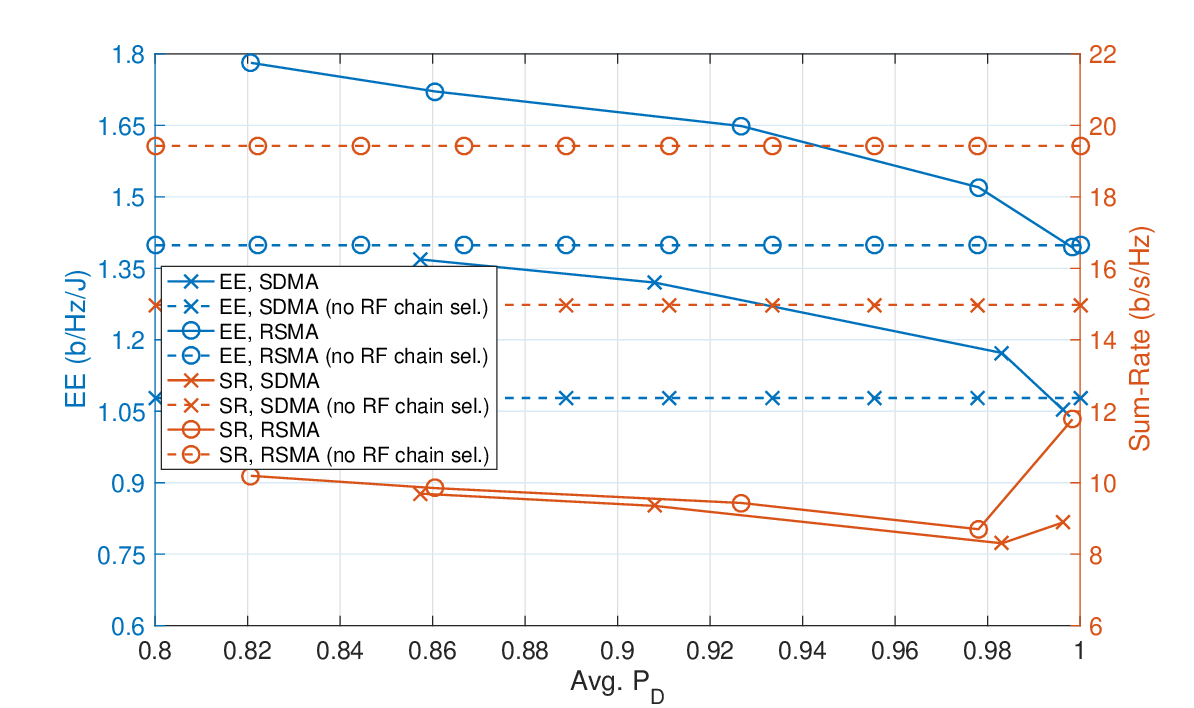}}
		\caption{Perfect CSIT, $b=4$.}
		\label{fig:srvsee_perf_4}
	\end{subfigure}
	\begin{subfigure}{.5\textwidth}
		\centerline{\includegraphics[width=3.3in,height=3.3in,keepaspectratio]{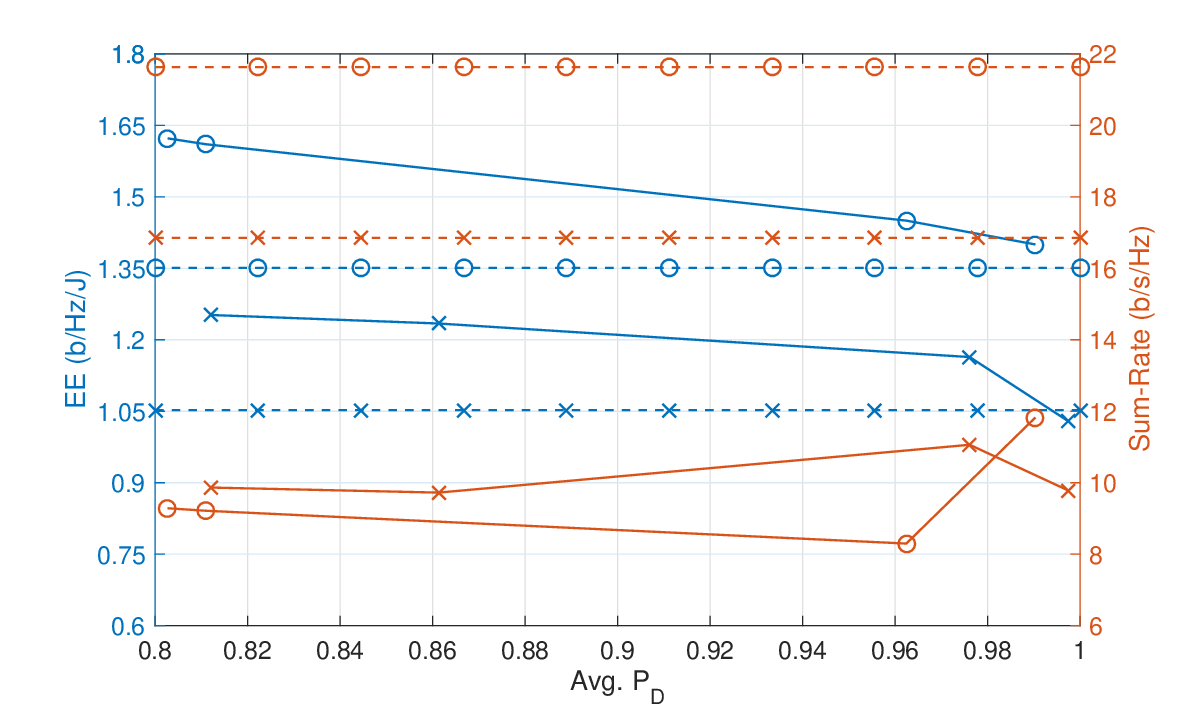}}
		\caption{Perfect CSIT, $b=8$.}
		\label{fig:srvsee_perf_8}
	\end{subfigure}
	\newline
	\begin{subfigure}{.5\textwidth}
		\centerline{\includegraphics[width=3.3in,height=3.3in,keepaspectratio]{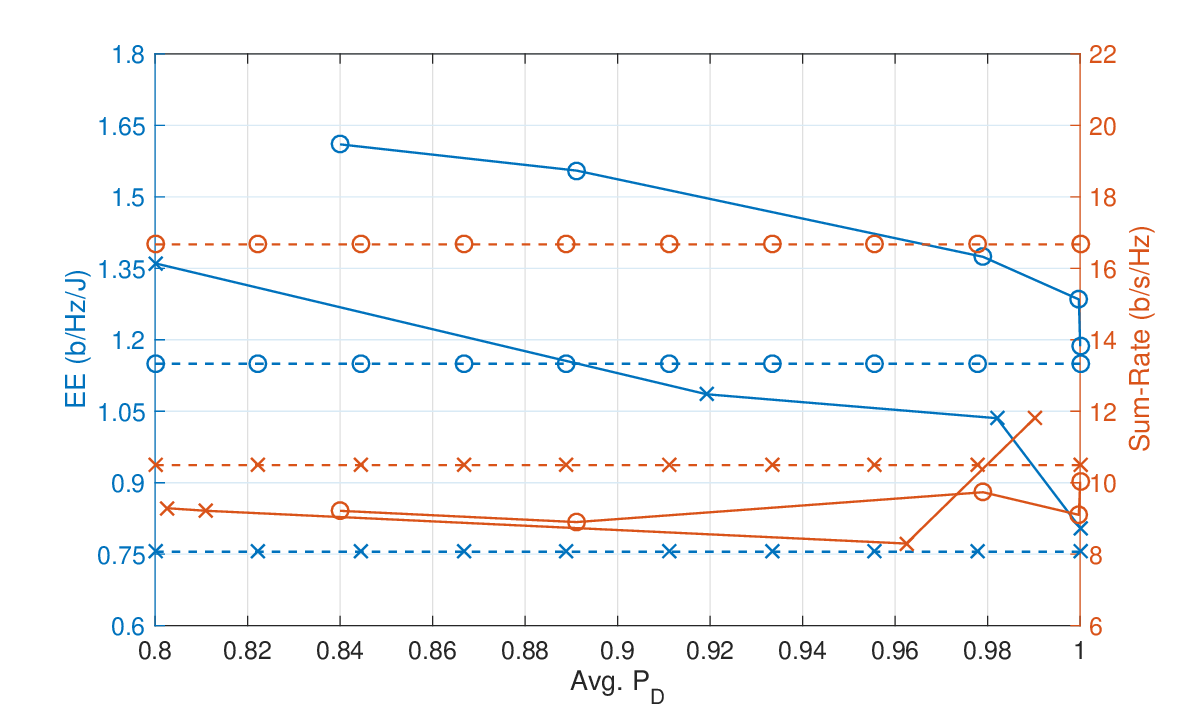}}
		\caption{Imperfect CSIT, $b=4$.}
		\label{fig:srvsee_imperf_4}
	\end{subfigure}
	\begin{subfigure}{.5\textwidth}
		\centerline{\includegraphics[width=3.3in,height=3.3in,keepaspectratio]{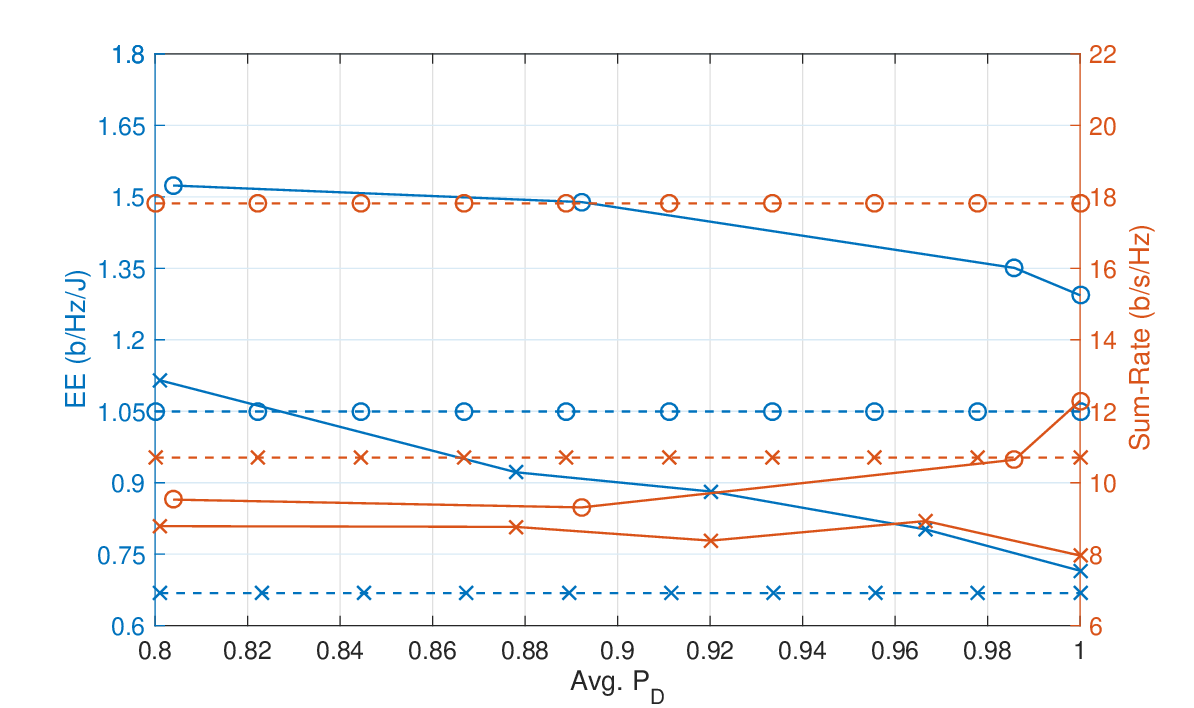}}
		\caption{Imperfect CSIT, $b=8$.}
		\label{fig:srvsee_imperf_8}
	\end{subfigure}
	\caption{$\bar{P}_{D}$ vs. EE and SR performance of SDMA and RSMA, $SR>8$ b/s/Hz, $P_{F}=10^{-7}$, single target, $N_{t}=8$, $K=2$, $L=10$, $\alpha_{r}=0.1$. The legend in subfigure (a) is applicable to other subfigures (b)-(d).}
	\label{fig:srvsee}
\end{figure*}

We note that for similarity threshold $\tau > 0$, the covariance matrix of the designed signal will not be an exact identity matrix. Consequently, detection probabilities at different angles may vary significantly, and thus, result in incorrect conclusions about the performance. Therefore, we consider average detection probability $\bar{P}_{D}$, for which the average is defined over angles between $-90$ and $+90$ degrees, {\sl i.e.,} $\bar{P}_{D}=\mathbb{E}_{\theta}(P_{D}(\theta))$.

For target parameter estimation, typical desired beamformer a number of $N_{tar}$ targets is given as \cite{liu_2019}
\begin{align}
	\mathbf{F}_{rad}=
	\begin{bmatrix}
		\mathbf{v}_{1} & \mathbf{0} & \ldots & \mathbf{0}  \\
		\mathbf{0} & \mathbf{v}_{2}  & \ldots & \mathbf{0}  \\
		\vdots & \vdots & \ddots & \vdots  \\
		\mathbf{0} & \mathbf{0} & \ldots & \mathbf{v}_{N_{tar}}   \\ 
	\end{bmatrix}, 
\end{align}
where $\mathbf{v}_{m} \in \mathbb{C}^{\frac{N_{t}}{N_{tar}}\times1}$ is composed of the entries of $\mathbf{a}(\theta_{m})$, $\forall m \in \left\lbrace1, 2, \ldots, N_{tar}\right\rbrace$, at the corresponding antennas. Then, the corresponding reference covariance matrix is written as $\mathbf{U}_{est}=P_{ant}L\mathbf{F}_{rad}\mathbf{F}^{H}_{rad}$. The CRB for DoA estimation for a single target is given in \eqref{eqn:crb} \cite{bekkerman_2006}. The matrix $\dot{\mathbf{A}}(\theta)$ in \eqref{eqn:crb} is the derivation of $\mathbf{A}(\theta)$ with respect to $\theta$. 
\begin{table*}[t!]
	\begin{align}
		\mathrm{CRB}(\theta)=\frac{\mathrm{tr}(\mathbf{A}(\theta)\widehat{\mathbf{R}}_{\tilde{\mathbf{x}}}\mathbf{A}^{H}(\theta))}{\frac{2|\alpha_{r}|^{2}}{\sigma_{n}^{2}}\left(\mathrm{tr}(\dot{\mathbf{A}}(\theta)\widehat{\mathbf{R}}_{\tilde{\mathbf{x}}}\dot{\mathbf{A}}^{H}(\theta))\mathrm{tr}(\mathbf{A}(\theta)\widehat{\mathbf{R}}_{\tilde{\mathbf{x}}}\mathbf{A}^{H}(\theta))-|\mathrm{tr}(\mathbf{A}(\theta)\widehat{\mathbf{R}}_{\tilde{\mathbf{x}}}\dot{\mathbf{A}}^{H}(\theta))|^{2}\right)}.
		\label{eqn:crb}
	\end{align}
	\hrule
\end{table*}

We start by investigating the convergence behaviour of the proposed algorithm. Fig.~\ref{fig:conv_ee} demonstrates the changes in the optimization variable EE with respect to the total iterations ({\sl i.e.,} $\sum_{t}T_{1,t}+T_{3,t}$) of the proposed algorithm for different radar metrics\footnote{The EE values in the figures are obtained with the relaxed RF chain indicator $\tilde{\mathbf{\lambda}}$, and therefore, may not be reflected in the rest of the EE results in this section, which are calculated after converting $\tilde{\mathbf{\lambda}}$ to $\mathbf{\lambda}$.}. The algorithm is run by setting $\epsilon_{ao}=\epsilon_{sdr}=10^{-2}$, $\epsilon_{admm}=10^{-1}$, and $M=50$. We can verify from the figures that the proposed algorithm converges for different performance targets and number of DAC quantization bits, and the number of required iterations for SDMA and RSMA are similar.
\begin{figure*}[t!]
	\begin{subfigure}{.5\textwidth}
		\centerline{\includegraphics[width=3.3in,height=3.3in,keepaspectratio]{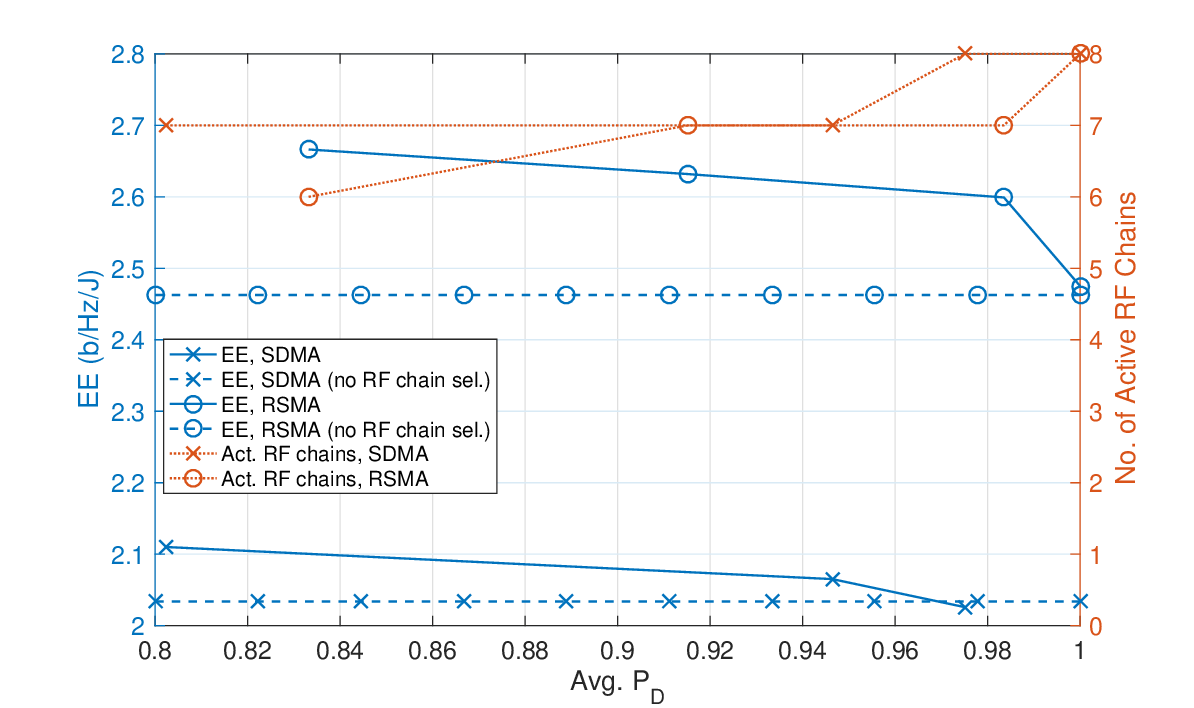}}
		\caption{Perfect CSIT, $b=4$.}
		\label{fig:K4_perfect}
	\end{subfigure}
	\begin{subfigure}{.5\textwidth}
		\centerline{\includegraphics[width=3.3in,height=3.3in,keepaspectratio]{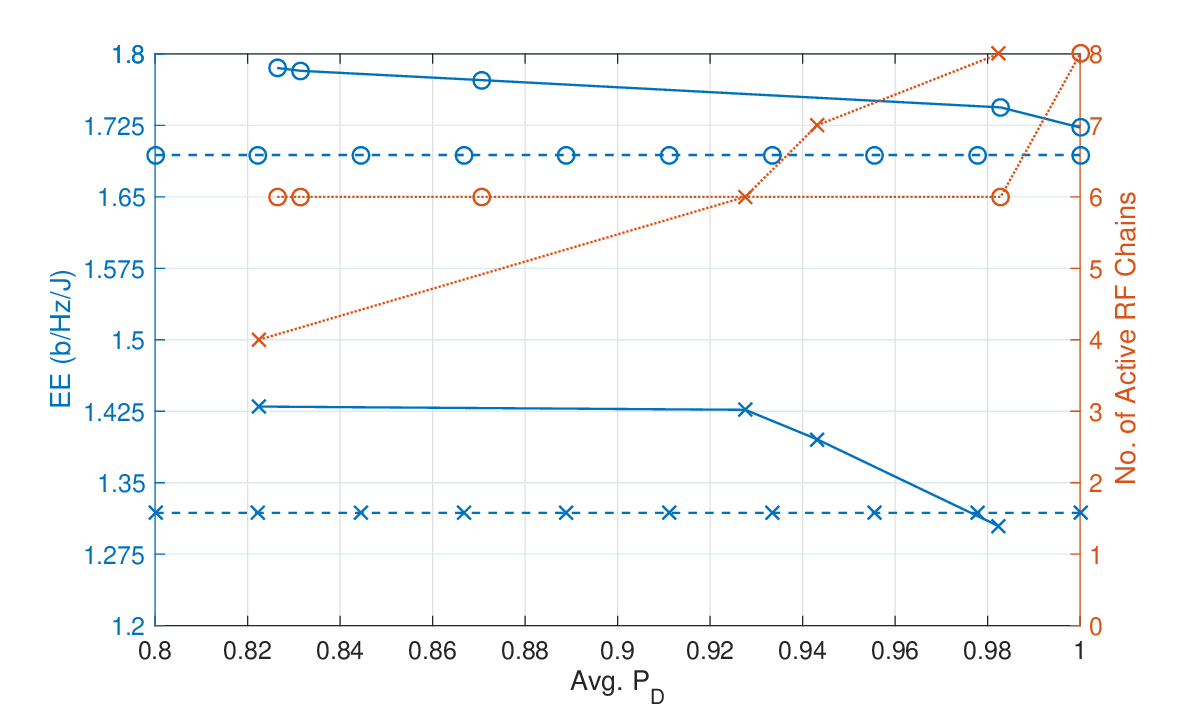}}
		\caption{Imperfect CSIT, $b=4$.}
		\label{fig:K4_imperfect}
	\end{subfigure}
	\caption{$\bar{P}_{D}$ vs. EE performance of SDMA and RSMA, $SR>12$ b/s/Hz, $P_{F}=10^{-7}$, single target, $N_{t}=8$, $K=4$, $L=10$, $\alpha_{r}=0.1$. The legend in subfigure (a) is applicable to subfigure (b).}
	\label{fig:pdvsee_K4}
\end{figure*}

\begin{figure*}[t!]
	\begin{subfigure}{.5\textwidth}
		\centerline{\includegraphics[width=3.3in,height=3.3in,keepaspectratio]{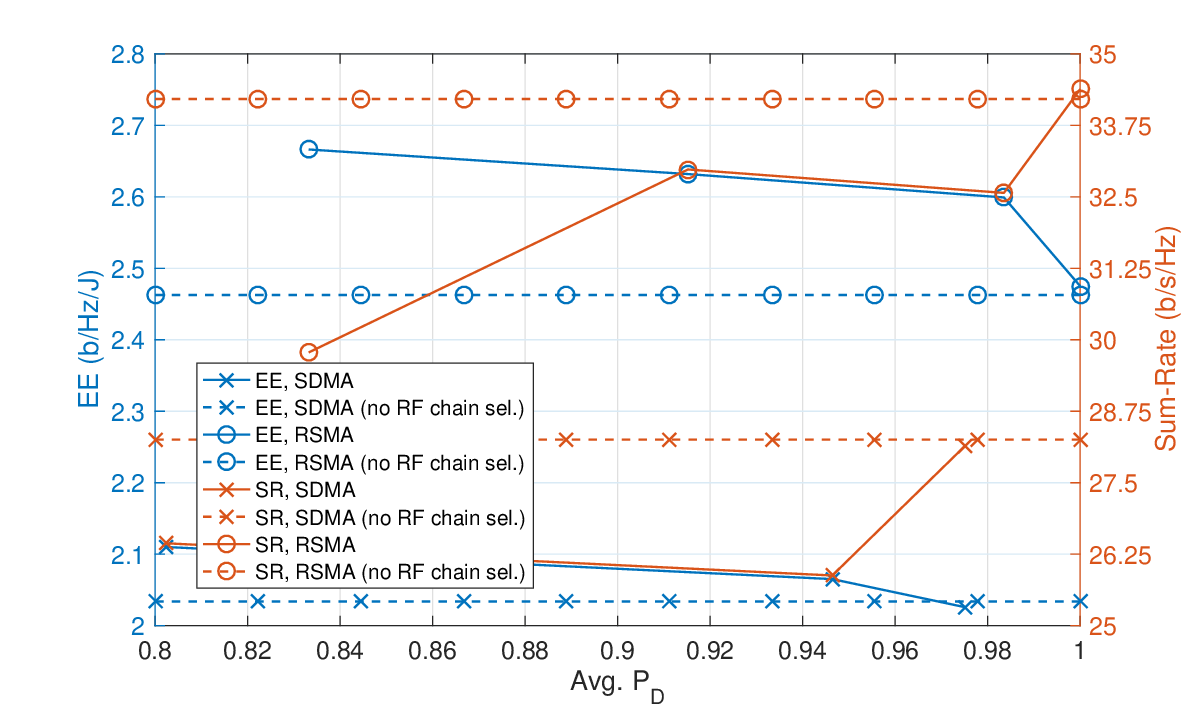}}
		\caption{Perfect CSIT, $b=4$.}
		\label{fig:K4_perfect_EESR}
	\end{subfigure}
	\begin{subfigure}{.5\textwidth}
		\centerline{\includegraphics[width=3.3in,height=3.3in,keepaspectratio]{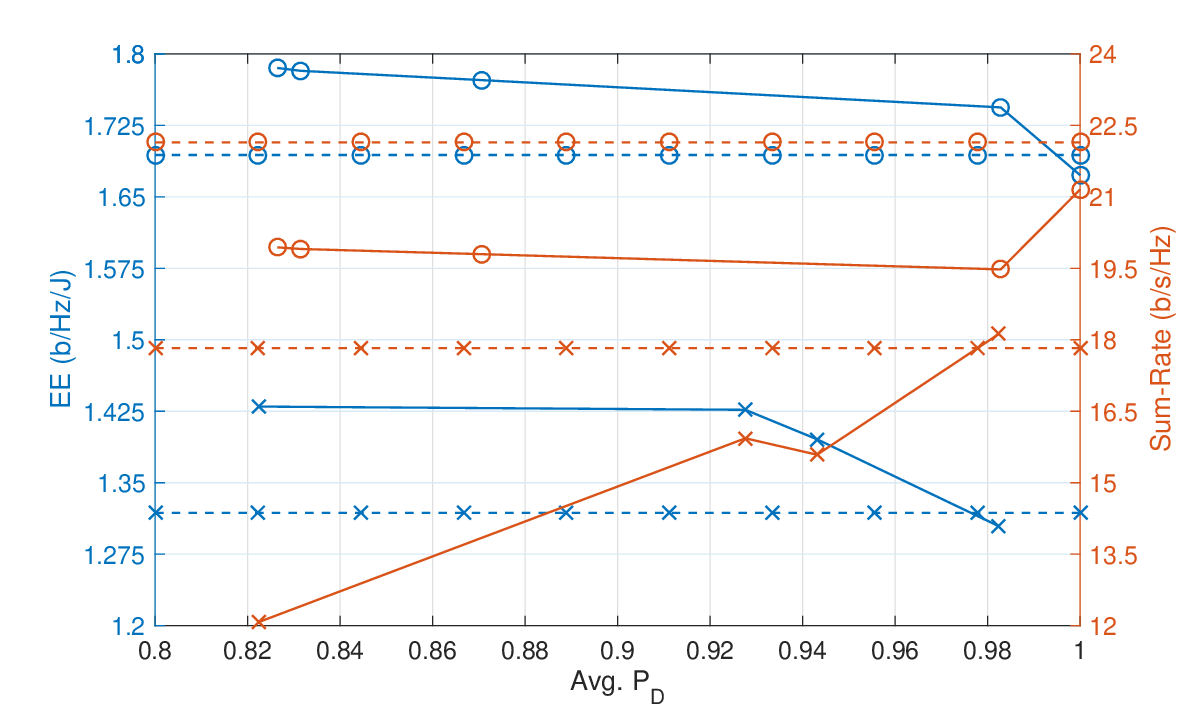}}
		\caption{Imperfect CSIT, $b=4$.}
		\label{fig:K4_imperfect_EESR}
	\end{subfigure}
	\caption{$\bar{P}_{D}$ vs. EE performance of SDMA and RSMA, $SR>12$ b/s/Hz, $P_{F}=10^{-7}$, single target, $N_{t}=8$, $K=4$, $L=10$, $\alpha_{r}=0.1$. The legend in subfigure (a) is applicable to subfigure (b).}
	\label{fig:srvsee_K4}
\end{figure*}

Next, we analyse the complexity of the proposed algorithm and compare the complexities for RSMA and SDMA. Table~\ref{table:complexity} gives the run-time results for the proposed algorithm for RSMA and SDMA under perfect and imperfect CSIT. The algorithms are run with the same settings used to obtain the results in Fig.~\ref{fig:conv_ee} with $b=4$. The results are obtained by MATLAB running CVX \cite{cvx_1} on a 11-th Generation Intel(R) Core(TM) i5-1135G7 processor run at $2.4$Ghz clock speed. The results show that the computational complexity for RSMA is higher than SDMA. In order to analyse the cause, we refer to the complexity expression \eqref{eqn:complexity_1}, which is a function of $N_{t}$, $K$, $L$, $M$, and the number of iterations required for the algorithm to converge. Observing the required number of iterations from Fig.~\ref{fig:conv_ee}, we can see that the required number of iterations is the same for RSMA and SDMA with $b=4$. This implies that the increase in the run-time for RSMA stems from the additional precoder for the common stream to be optimized and the related constraints.

Next, we move to investigate the EE performance of the system with respect to radar and communication metrics. Fig.~\ref{fig:pdvsee} shows the EE performance with respect to average detection probability for $b=\lbrace{4, 8\rbrace}$, $P_{F}=10^{-7}$, and $SR>8$ b/s/Hz under perfect and imperfect CSIT. We study the detection performance for a single target and the reference covariance matrix is set as $\mathbf{I}_{N_{t}}$. We run simulations for $\tau=\lbrace{1, 1.25, \ldots, 8\rbrace}$, $R_{th}=0.1$ b/s/Hz, and $\zeta=\lbrace{1,10,20\rbrace}$. The curves shown in Fig.~\ref{fig:pdvsee} are obtained by first applying \eqref{eq:sign} for $I_{thr}=\lbrace{0, 0.05, \ldots, 1\rbrace}$ to each result obtained with different $\tau$ values. For each result, we determine the optimal $I_{thr}$ value which returns the highest EE, and obtain the corresponding $\boldsymbol{\lambda}$ by \eqref{eq:sign}. Then, we calculate EE, SR and radar metrics for the determined $\boldsymbol{\lambda}$. After all of the results for different $\tau$ values are processed in this manner, we pick the points with highest EE, which satisfy $SR>8$ b/s/Hz. Two different methods are considered to initialize the precoders in order to reduce the dependency of the performance curves on the precoder initialization. The first method of initialization uses the chirp signal in the form
\begin{align}
	X_{0}(n,l)\hspace{-0.1cm}=\hspace{-0.1cm}\sqrt{P_{ant}}\exp\{j2\pi n(l-1)/L\}\exp\{j2\pi (l-1)^{2}/L\}, \nonumber
\end{align}	
$\forall n \in \{1, 2, \ldots, N_{t}\}$ and $\forall l \in \mathcal{L}$ to initialize all precoders \cite{zhao_2021}. The second method uses Zero-Forcing (ZF) precoders to initialize precoders for the private streams and the left-most eigenvector of the channel matrix to initialize precoders for the common stream. We consider equal power allocation among the initial precoders for the private streams and perform simulations by allocating $0.5$ and $0.7$ of the total power to the precoder for the common stream.
The CSIT error vector in \eqref{eqn:error_model} is assumed to have i.i.d. complex Gaussian elements of unit variance. The figures also show the number of active transmit antennas to achieve the demonstrated performance for each scheme. One can immediately observe that RSMA achieves a higher EE than SDMA for all considered scenarios. RSMA satisfies the communications and radar performance metric constraints with fewer number of active antennas, resulting in a higher EE with both perfect and imperfect CSIT. The capability of RSMA to activate fewer antennas than SDMA stems from its higher degrees of freedom in design space due to the additional common stream and its improved SR performance under interference.

Comparing the performance with different numbers of quantization bits, one can notice that EE decreases when the $b$ is increased from $4$ to $8$. Although the SR achieved with $b=8$ is higher due to lower quantization noise and higher multiplicative quantization gain $\delta$, the power consumption is also higher, resulting in a lower EE than that of $b=4$.

Next, we investigate the performance in the tracking mode of the radar system. Fig.~\ref{fig:crlbvsee} shows the EE performance with respect to CRB for $b=\lbrace{4, 8\rbrace}$, $P_{F}=10^{-7}$, and $SR>8$ b/s/Hz under perfect and imperfect CSIT. We consider the angle estimation performance for a single target at $\theta=\pi/4$ and the reference covariance matrix is set as $\mathbf{a}(\pi/4)\mathbf{a}(\pi/4)^{H}$. The curves shown in Fig.~\ref{fig:pdvsee} are obtained by running simulations for $\tau=\lbrace{45, 47.5, \ldots, 80\rbrace}$, $R_{th}=0.1$ b/s/Hz, and $\zeta=\lbrace{1,10,20\rbrace}$, applying the same procedure used for Fig.~\ref{fig:pdvsee}, and picking the points with highest EE which satisfy $SR>8$ b/s/Hz. Two different methods are considered to initialize the precoders in order to reduce the dependency of the performance curves on the precoder initialization. The first method of initialization uses the steering vector for $\theta=\pi/4$. The second method uses Zero-Forcing (ZF) precoders to initialize precoders for the private streams and the left-most eigenvector of the channel matrix to initialize precoders for the common stream. We consider equal power allocation among the initial precoders for the private streams and perform simulations by allocating $0.5$ and $0.7$ of the total power to the precoder for the common stream.

Similar to the case in detection performance, RSMA achieves higher EE than SDMA under both perfect and imperfect CSIT. The number of active antennas used to achieve a given CRB performance are similar with RSMA and SDMA, as opposed to the case with average detection probability. This is due to the reference covariance matrix for minimizing CRB being a non-sparse matrix. This, in turn, requires higher number of active antennas to decrease the distance between the designed and reference covariance matrix and reduce CRB for both RSMA and SDMA.  

We investigate the change in performance metrics with varying similarity threshold values. As an example, Fig.~\ref{fig:similarity_metric} shows the change in target angle estimation CRB achieved by RSMA with respect to $\tau$ for $b=8$ under imperfect CSIT. As expected, the CRB increases with increasing $\tau$. The EE also increases with $\tau$, as increasing  $\tau$ relaxes the similarity constraint and allows the system to deactivate several antennas to improve EE. Fig.~\ref{fig:similarity_beam} shows the beampattern for communicating with $2$ users and tracking a target at $\theta=\pi/4$ for various $\tau$ values. The EE and CRB performance, and the  number of active RF chains can also be observed from the legends in the same figure. One can see that the main lobe of the signal shifts from $\theta=\pi/4$ and the sidelobe power increases as $\tau$ increases. The lowest sidelobe power is achieved when all RF chains are active. On the other hand, increasing $\tau$ allows the system to achieve higher EE by activating smaller number of RF chains with a penalty in estimation performance.

In order to understand the trade-off between sensing and communication performance, Fig.~\ref{fig:srvsee} gives the EE and SR performance with respect to average detection probability. We can see from  Fig.~\ref{fig:srvsee} that the major factor affecting the radar and communication performance is the number of active RF chains. As the active number of RF chains increase, the radar and communication performance improve simultaneously while the EE decreases, implying that the main performance trade-off in the system is between EE and radar-communication performance, which is determined by the number of active RF chains. 

Finally, we investigate the performance with $K=4$ users given in Fig.~\ref{fig:pdvsee_K4}~and~\ref{fig:srvsee_K4}. Comparing Fig.~\ref{fig:pdvsee_K4} and \ref{fig:pdvsee},  one can notice that the required number of active RF chains to achieve high EE increases when the number of communication users in the system is increased. The increase in the EE is due to the increased SR due to the additional users in the system. The increase in the SR can be observed explicitly from Fig.~\ref{fig:srvsee}~and~\ref{fig:srvsee_K4}. Achieving such an SR increase requires larger number of transmit antennas, which leads to the increase in the number of active RF chains. We can also observe that RSMA achieves an even higher EE gain compared to SDMA with $K=4$ both in perfect and imperfect CSIT due to its interference management capability and higher degrees-of-freedom in the design space.


\section{Conclusion}
\label{sec:conclusion}
In this work, we consider RSMA for multi-antenna DFRC systems with low-resolution DACs. We investigate the EE performance of the proposed system with optimal RF chain selection under communications and radar performance constraints and under imperfect CSIT. We formulate a non-convex EE maximization problem and propose an iterative algorithm to solve it. We prove the convergence of the proposed algorithm by analytical means. 
We perform simulations to compare the performance of RSMA and SDMA
in terms of EE, communications, and radar metrics. The results show RSMA can achieve a higher EE both under perfect and imperfect CSIT by means of activating less number of antennas than SDMA, which is enabled by the larger design space and improved interference management capabilities of RSMA.

\section{Appendix}

\subsection{Proof of Proposition 3}
\label{sec:app3}
It is easy to see that the matrix $\mathbf{A}_{l}=\boldsymbol{\Delta}\mathbf{P}_{l}\mathbf{s}_{l}\mathbf{s}_{l}^{H}\mathbf{P}_{l}^{H}\boldsymbol{\Delta}$ is a positive semi-definite matrix, as $\mathbf{b}^{H}\mathbf{A}_{l}\mathbf{b} \geq 0$ for any $\mathbf{b} \in \mathbb{C}^{N_{t}\times1}$.
Consequently, we can express $\mathbf{A}_{l}$ in terms of its eigen-value decomposition, such that
\begin{align}
	\mathbf{A}_{l}=\sum_{i=1}^{N_{t}}\mu_{i,l} \mathbf{f}_{i,l}\mathbf{f}_{i,l}^{H},
\end{align}
where $\mu_{i,l}$ is the $i$-th eigenvalue and  $\mathbf{f}_{i,l}$ is the $i$-th eigenvector of $\mathbf{A}_{l}$, $\forall i \in \left\lbrace 1,2,\ldots, N_{t}\right\rbrace$ and $\forall l \in \mathcal{L}$. 
Then, we write
\begin{subequations}
	\begin{alignat}{3}
		\widehat{\mathbf{R}}_{\tilde{\mathbf{x}}}(&\mathbf{P}, \tilde{\boldsymbol{\Lambda}}, \mathbf{B}) \nonumber \\ 
		&=\sum_{l \in \mathcal{L}}\boldsymbol{\Delta}\tilde{\boldsymbol{\Lambda}}\mathbf{P}_{l}\mathbf{P}_{l}^{H}\tilde{\boldsymbol{\Lambda}}\boldsymbol{\Delta}+L\tilde{\boldsymbol{\Lambda}}\boldsymbol{\Sigma}\tilde{\boldsymbol{\Lambda}} \\
		&=\sum_{l \in \mathcal{L}}\tilde{\boldsymbol{\Lambda}}\mathbf{A}_{l}\tilde{\boldsymbol{\Lambda}}+L\tilde{\boldsymbol{\Lambda}}\boldsymbol{\Sigma}\tilde{\boldsymbol{\Lambda}}  \\
		&=\sum_{l \in \mathcal{L}}\tilde{\boldsymbol{\Lambda}}\left(\sum_{i=1}^{N_{t}}\mu_{i,l} \mathbf{f}_{i,l}\mathbf{f}_{i,l}^{H}\right)\tilde{\boldsymbol{\Lambda}}+L\tilde{\boldsymbol{\Lambda}}\boldsymbol{\Sigma}\tilde{\boldsymbol{\Lambda}} \label{eqn:rx_2} \\
		&=\sum_{l \in \mathcal{L}}\sum_{i=1}^{N_{t}}\mu_{i,l}\mathbf{F}_{i,l}\tilde{\boldsymbol{\lambda}}\tilde{\boldsymbol{\lambda}}^{T} \mathbf{F}_{i,l}^{H}+L\sum_{i=1}^{N_{t}}\boldsymbol{\Sigma}\mathbf{E}_{N_{t},i}\tilde{\boldsymbol{\lambda}}\tilde{\boldsymbol{\lambda}}^{T}\mathbf{E}_{N_{t},i} \label{eqn:rx_3} \\
		&=\sum_{l \in \mathcal{L}}\sum_{i=1}^{N_{t}}(\mu_{i,l}\mathbf{F}_{i,l}\boldsymbol{\Upsilon}\mathbf{F}_{i,l}^{H}+\boldsymbol{\Sigma}\mathbf{E}_{N_{t},i}\boldsymbol{\Upsilon}\mathbf{E}_{N_{t},i}), \label{eqn:rx_3_2} 
	\end{alignat}
\end{subequations}
where $\mathbf{F}_{i,l}=\mathrm{Diag}(\mathbf{f}^{T}_{i,l})$ and \eqref{eqn:rx_2} follows from the fact that $\mathbf{D}_{\mathbf{a}}\mathbf{b}=\mathbf{D}_{\mathbf{b}}\mathbf{a}$.

\subsection{Proof of Proposition 6}
\label{sec:app6}
We calculate the complexity of the proposed algorithm at each iteration by calculating the complexities of the algorithms \textbf{ALG1} and \textbf{ALG3} at iteration-$t$.
\subsubsection{Complexity of \textbf{ALG1}}
The complexity of \textbf{ALG1} at each iteration is analyzed in terms of the complexities of the update steps for $\mathbf{v}_{r}$ and $\mathbf{u}_{r}$ .
\begin{itemize}
	\item Update step for $\mathbf{v}_{r}$: The complexity of this step is equal to the complexity of interior-point methods for solving \eqref{eqn:v_update}, which can be written in the equivalent Quadratically Constrained Quadratic Program (QCQP) form 
	\begin{subequations}
		\begin{alignat}{3}
			\min_{\mathbf{v}_{r}, s}&  \quad  f(\mathbf{v_{r}})+s  \nonumber \\
			\text{s.t.}& \quad  \frac{\zeta}{2}\hspace{-0.1cm}\sum_{n\in\mathcal{S}}||\mathbf{v}_{r,n}-\mathbf{u}^{t}_{r,n}+\mathbf{w}^{t}_{r,n}||^{2} \leq s, \nonumber
		\end{alignat}
	\end{subequations}
	where $s$ is a slack variable. The complexity of solving such a problem by interior point methods is $\mathcal{O}((L(K+1)N_{t}+2L)^{3}\log(1/\epsilon))=\mathcal{O}((L(K+1)N_{t})^{3}\log(1/\epsilon))$ for a given solution accuracy $\epsilon > 0$, with $\mathcal{O}(.)$ denoting the big O function.
	\item Update step for $\mathbf{u}_{r}$: The worst case complexity of this step is approximately given by the complexity of solving an SDP problem using interior point methods with $L(K+1)N_{t}+2L$ optimization variables and $L(K+N_{t}+1)+2$ constraints, which is written as $\mathcal{O}((L(K+1)N_{t}+2L)^{4.5}\log(1/\epsilon))=\mathcal{O}((L(K+1)N_{t})^{4.5}\log(1/\epsilon))$ \cite{ma_2010}.
\end{itemize}
Therefore, the complexity of a single iteration of \textbf{ALG1} is dominated by that of SDR method, and the overall worst case  complexity is written as
\begin{align}
	\mathcal{C}_{ALG1}=\mathcal{O}(T_{1,t}(L(K+1)N_{t})^{4.5}\log(1/\epsilon)), 
\end{align}
where $T_{1,t}$ denotes the number of iterations of \textbf{ALG1} at iteration-$t$.
\subsubsection{Complexity of \textbf{ALG3}}
The worst case complexity of \textbf{ALG3} is approximately given by the complexity of solving an SDP problem using interior point methods with $4KLM+2L+N_{t}+2$ optimization variables and $4KLM+KL+L+N^{2}_{t}+5$ constraints, which is written as
\begin{align}
	\mathcal{C}_{ALG3}&=\mathcal{O}(T_{3,t}(4KLM+KL+L+N^{2}_{t}+5)^{4.5}\log(1/\epsilon))\nonumber\\
	&=\mathcal{O}(T_{3,t}(4KLM+N^{2}_{t})^{4.5}\log(1/\epsilon)), 
\end{align}
where $T_{3,t}$ denotes the number of iterations of \textbf{ALG3} at iteration-$t$ \cite{ma_2010}.
\subsubsection{Complexity of the Proposed Algorithm}
Assuming the same solution accuracy for \textbf{ALG1} and \textbf{ALG3}, the overall complexity of the proposed algorithm at iteration-$t$ can be approximately written as
\begin{align}
	\mathcal{C}_{prop,t}=\mathcal{O}&([T_{1,t}(L(K+1)N_{t})^{4.5} \nonumber \\
	&\quad \quad+T_{3,t}(4KLM+N^{2}_{t})^{4.5}]\log(1/\epsilon)).
	\label{eqn:complexity}
\end{align}

\end{document}